\documentclass[11pt,a4paper]{article}

\usepackage{bm}

\usepackage[top=15truemm,bottom=20truemm,left=20truemm,right=20truemm]{geometry}
\usepackage[colorlinks=true,urlcolor=blue,anchorcolor=black,citecolor=blue,linkcolor=black,filecolor=black,menucolor=black,linktocpage=true,pdfproducer=medialab,pdfa=true]{hyperref}
\usepackage{graphicx}
\usepackage{subcaption}
\usepackage{cite}
\usepackage{amsmath,latexsym,amssymb,mathrsfs,ascmac,physics,mathtools,slashed,xcolor}
\numberwithin{equation}{section} 
\usepackage[affil-it]{authblk}
\newcommand{\email}[1]{\thanks{\href{mailto:#1}{\texttt{#1}}}}

\usepackage[Euler]{upgreek} 

\usepackage{eso-pic}

\newcommand{\preprintnumber}[1]{%
  \AddToShipoutPictureFG*{%
    \AtPageUpperLeft{%
      \put(
        \LenToUnit{\dimexpr\paperwidth-20truemm\relax},
        \LenToUnit{-20truemm}
      ){%
        \makebox[0pt][r]{\small #1}%
      }%
    }%
  }%
}

\usepackage{tabularx}
\usepackage{tikz}
\usetikzlibrary{calc,decorations.pathreplacing,patterns}

\newcommand{\aBH}{\mathrm{a}} 
\newcommand{\QBH}{\mathrm{Q}}
\newcommand{\qBH}{\mathrm{q}}

\newcommand{\alphaHeun}{\upalpha}
\newcommand{\betaHeun}{\upbeta}
\newcommand{\gammaHeun}{\upgamma}
\newcommand{\deltaHeun}{\updelta}
\newcommand{\epsilonHeun}{\upepsilon}

\newcommand{\HeunG}[1]{
\operatorname{HeunG}\!
\left(
#1
\right)%
}

\global\long\def\FV#1#2#3#4#5#6#7#8#9{\mathfrak{F}\left(\begin{array}{c}
#1\\
#2
\end{array}#3\begin{array}{c}
#4\\
\\\end{array}#5\begin{array}{c}
#6\\
#7
\end{array};#8,#9\right)}%

\global\long\def\FIV#1#2#3#4#5#6{\mathfrak{F}\left(\begin{array}{c}
#1\\
#2
\end{array}#3\begin{array}{c}
#4\\
#5
\end{array};#6\right)}%

\newcommand{\QL}{Q}

\title{Quasinormal modes of Kerr--Newman--de Sitter black holes from Heun connection formulae}

\author{Hideo Furugori
\email{hideo@toyota-ti.ac.jp}}
\affil{Mathematical Physics Laboratory, Toyota Technological Institute, Nagoya 468-8511, Japan}
\author{Naotaka Kubo
\email{naotaka.kubo@riken.jp}}
\affil{RIKEN Center for Interdisciplinary Theoretical and Mathematical Sciences (iTHEMS), RIKEN,
Wako 351-0198, Japan}
\author{Hayato Motohashi\email{motohashi@tmu.ac.jp}}
\affil{Department of Physics, Tokyo Metropolitan University, Hachioji, Tokyo 192-0397, Japan}
\author{Daisuke Yoshida \email{yoshida.daisuke.k9@f.mail.nagoya-u.ac.jp}}
\affil{Institute for Advanced Research, Nagoya University, Nagoya 464-8602, Japan}

\date{}

\begin{document}

\preprintnumber{RIKEN-iTHEMS-Report-26, TTI-MATHPHYS-45}

\maketitle

\begin{abstract}
We study quasinormal modes governed by the Teukolsky equation in Kerr--Newman--de Sitter spacetime, formally treating the spin weight and field charge as arbitrary parameters.
We use exact Heun connection formulae expressed through classical conformal blocks.
We analytically derive the quasinormal mode spectra in the small cosmological constant, near-extremal, and near-Nariai limits.
In the small cosmological constant limit, we obtain the leading charged de Sitter spectrum together with its subleading corrections.
In both the small cosmological constant and near-extremal limits, solving the full connection condition beyond the leading pole approximation gives rise to generically noninteger-power corrections to the quasinormal mode frequencies, whereas no analogous correction appears in the near-Nariai limit.
We also analyze the angular eigenvalue problem as an illustration of the same connection-formula method.
\end{abstract}

\setcounter{tocdepth}{2}
\tableofcontents

\section{Introduction}

Quasinormal modes (QNMs) play a central role in the study of black hole perturbations, characterizing the dissipative response of a black hole under linear perturbations~\cite{Berti:2025hly}.
For many black hole perturbation problems, the separated radial and angular equations belong to the Heun class.
In asymptotically flat Kerr and Reissner--Nordstr\"{o}m backgrounds, the relevant equations are typically of confluent Heun type, whereas the presence of a cosmological horizon in de Sitter backgrounds leads naturally to equations with several regular singular points and, in particular, to the general Heun equation.
The Teukolsky equations in Kerr--de Sitter and Kerr--Newman--de Sitter spacetimes provide important examples of this structure~\cite{Suzuki:1998vy,Suzuki:1999nn,Suzuki:1999pa}.

From the viewpoint of the Heun equation, the QNM problem is naturally formulated as a connection problem between local solutions around different singular points.
The QNM boundary conditions require that the solution that is ingoing at the black hole horizon (in-mode) coincide with the one that is outgoing at the cosmological horizon (up-mode).
Therefore, the QNM spectrum can be characterized by the vanishing of an appropriate connection coefficient.
Although this formulation provides an exact analytic characterization of the spectral problem, the connection coefficients of the general Heun equation are highly nontrivial, and the Heun representation alone does not usually lead to explicit analytic expressions for the QNM frequencies.
It has been used for numerical computations of Kerr--de Sitter QNMs~\cite{Hatsuda:2020sbn} and for an exact Heun function formulation of wave scattering by Kerr--Newman--de Sitter black holes, including QNMs, greybody factors, absorption/reflection rates, and Green functions~\cite{Motohashi:2021zyv}.

Recent developments relating Heun equations to four-dimensional supersymmetric gauge theories and two-dimensional conformal field theories have substantially changed this situation.
In particular, connection formulae for Heun functions and their confluent limits can be expressed in terms of semiclassical conformal blocks and Nekrasov partition functions~\cite{Bonelli:2021uvf,Bonelli:2022ten}.
Such techniques have already been applied to black hole perturbations, including the computation of QNM conditions, scattering amplitudes, greybody factors, and Love numbers~\cite{Bonelli:2021uvf}
(using the same four-dimensional framework, Ref.~\cite{Aminov:2020yma} performed an analysis based on quantization conditions instead of connection formulae).
They have since been extended to holographic thermal correlators and QNM quantization conditions in asymptotically AdS black holes~\cite{Jia:2024zes}, to one-loop effective actions and thermodynamic corrections in Kerr-(A)dS backgrounds~\cite{Arnaudo:2024rhv,Arnaudo:2025btb}, and to Green-function and late-time dynamics in asymptotically de Sitter black holes~\cite{Arnaudo:2025uos,Arnaudo:2025kit}. Related developments have also addressed angular spectral problems beyond the standard Heun class, such as the extremal charged C-metric~\cite{Yang:2026dpt}.

In the present work, we apply this framework to the Teukolsky equation in Kerr--Newman--de Sitter spacetime, characterized by four parameters: the mass $M$, the charge $\QBH$, the specific angular momentum $\aBH$, and the cosmological constant $\Lambda \eqqcolon 3/L^2$. We formally treat the spin $s$ and field charge $\qBH$ appearing in the Teukolsky equation as arbitrary independent parameters.
Both the radial and angular Teukolsky equations can be mapped to the general Heun equation, and the corresponding boundary or regularity conditions can be formulated in terms of Heun connection coefficients~\cite{Suzuki:1998vy,Suzuki:1999nn,Suzuki:1999pa,Hatsuda:2020sbn,Motohashi:2021zyv}.

The four horizon singularities of the radial Teukolsky equation can be assigned to the four regular singular points of the general Heun equation in different ways.
These assignments are equivalent at the exact level but lead to different cross-ratio parameters and hence to different useful perturbative expansions.
In the following, we choose the singular-point assignment separately in each physical regime so that the corresponding cross-ratio parameter $t$ becomes small. This provides a common perturbative setting for the angular problem and for the three radial limits considered below.

The small-$t$ expansion of Heun connection data has also been used to analyze angular spectral problems in black hole perturbation theory~\cite{Arnaudo:2024rhv,Arnaudo:2025btb,Yang:2026dpt}. In the present setting, we apply this approach to the angular Teukolsky equation in the limit $\aBH/L\to0$. The regularity condition can then be expressed as a small-$t$ expansion of the connection formula, reproducing the spin-weighted spherical harmonic eigenvalue at leading order and yielding systematic corrections in $\aBH/L$. This angular problem provides a simple illustration of the quantization procedure that will subsequently be applied to the radial QNM condition.

For the radial equation, we study three controlled limits.
The first is the small cosmological constant limit $L \to \infty$.
In this limit, the cosmological horizons move to infinity, while the de Sitter family of QNMs has frequencies that scale with the inverse de Sitter radius.
The resulting modes can be regarded as deformations of the QNMs of the pure de Sitter static patch.
For the massless conformally coupled scalar field considered here, the four-dimensional pure
de Sitter frequencies are purely imaginary and scale with the inverse de Sitter radius~\cite{Lopez-Ortega:2006aal}.
Recently, the small cosmological constant behavior of de Sitter QNMs in Schwarzschild--de Sitter spacetime has been studied in connection with Green-function and late-time dynamics, including the emergence of Price's law~\cite{Arnaudo:2025uos,Arnaudo:2025kit}.
In the present work, we extend the spectral analysis to Kerr--Newman--de Sitter spacetime.
The leading spectrum already depends nontrivially on the combination $\qBH \QBH$, and the subleading structure includes a generically noninteger-power correction.

The second regime is the black hole near-extremal limit.
The relevant QNMs in this limit approach the superradiant threshold and are commonly referred to as zero-damping or near-horizon modes.
These modes have previously been studied using matched asymptotic expansions between the near- and far-region solutions~\cite{Detweiler:1980gk,Cardoso:2004hh,Yang:2012pj,Yang:2013uba,Zimmerman:2015trm}, as well as by analyses based on the near-horizon geometry~\cite{Davey:2024xvd}.
More recently, the near-extremal limit has also been analyzed directly in the Heun framework for massless perturbations in Kerr-(A)dS spacetime by identifying the corresponding degeneration of the differential equation~\cite{Arnaudo:2025btb}.
In the present work, we extend this analysis to Kerr--Newman--de Sitter backgrounds and charged fields.
By choosing a singular point assignment for which the extremal limit corresponds to $t\to0$, we analyze the QNM condition directly through the small-$t$ expansion of the exact Heun connection formula.
This reproduces the leading near-extremal spectrum without an independent near/far matching construction.
More importantly, by solving the full connection condition beyond the leading pole approximation, we determine the associated frequency shift, which is generically of noninteger power in the near-extremal expansion.

The third regime is the near-Nariai limit, in which the outer black hole horizon and the cosmological horizon approach each other.
Near-Nariai QNMs have been obtained previously by reducing the radial perturbation equation to simpler effective equations, such as the P\"oschl--Teller problem in special cases~\cite{Cardoso:2003sw}.
The spectrum has also been obtained analytically by directly analyzing the perturbation equations in the near-Nariai geometry~\cite{Churilova:2021nnc}; see also Refs.~\cite{Yoshida:2003zz,Yoshida:2010zzb} for numerical studies.
More recently, the Nariai limit of Kerr-(A)dS has been analyzed in the Heun framework for an uncharged field~\cite{Arnaudo:2025btb}.
Here we consider the corresponding problem for the Teukolsky equation in Kerr--Newman--de Sitter spacetime with nonzero field charge. With an appropriate singular point assignment, the Nariai limit is represented by $t\to0$, and the QNM spectrum follows directly from poles of the gamma functions appearing in the exact connection coefficient.
This yields an analytic near-Nariai spectrum while keeping rotation, black hole charge, field charge, and the cosmological constant simultaneously.
In contrast to the small cosmological constant and near-extremal limits, the QNM condition reduces directly to a gamma-function pole condition and no analogous noninteger-power correction appears.
We note that higher-order near-Nariai expansions for Kerr--de Sitter quasinormal frequencies have also been obtained using the isomonodromic accessory parameter expanison~\cite{Novaes:2018fry}.

Taken together, these results extend the Heun connection approach to charged Kerr--Newman--de Sitter spectra and clarify how the structure of the QNM condition differs among the small cosmological constant, near-extremal, and near-Nariai limits.
The different physical regimes are described within the same general-Heun connection framework through appropriate assignments of the regular singular points.

This paper is organized as follows.
In Sec.~\ref{sec:Heun}, we review the Heun equation and the connection formula used in this work.
In Sec.~\ref{sec:Teukolsky}, we introduce the Kerr--Newman--de Sitter geometry and reduce the radial and angular Teukolsky equations to the Heun form.
In addition, we express the QNM condition in terms of the connection coefficient.
In Sec.~\ref{sec:QNM}, we analyze the small-$\aBH/L$ angular eigenvalue spectrum and derive the QNM spectra in the small cosmological constant, near-extremal, and near-Nariai limits.
Sec.~\ref{sec:Summary} is devoted to a summary and discussion.

\section{The Heun equation and its connection formulae}
\label{sec:Heun}
In this section, we review the general Heun equation and the connection formulae used in this work, following Ref.~\cite{Bonelli:2022ten}.

\subsection{Fuchsian form of the Heun equation}
\label{subsec:HeunEq}
The Heun equation is a second-order Fuchsian differential equation with four regular singular points on the Riemann sphere. A M\"{o}bius transformation maps these points to \(z=0,1,t,\infty\), where \(t\neq0,1\) is the cross-ratio parameter. After a suitable redefinition of the dependent variable, the equation takes the standard form
\begin{align}
\left[ \frac{d^2}{d z^2} + \left( \frac{\gammaHeun}{z} + \frac{\deltaHeun}{z-1} + \frac{\epsilonHeun}{z - t} \right) \frac{d}{dz} + \frac{\alphaHeun \betaHeun z - q}{z(z - 1)(z - t)} \right] y(z) = 0. \label{Heun eq Fuchs}
\end{align}
The five parameters $\alphaHeun, \betaHeun, \gammaHeun, \deltaHeun, \epsilonHeun$ are not independent and satisfy
the Fuchs relation
\begin{align}
\gammaHeun + \deltaHeun + \epsilonHeun = \alphaHeun + \betaHeun + 1. \label{Fuchs relation}
\end{align}
The remaining parameter $q$ is the accessory parameter, which is not fixed by the local exponents at the singular points.

\subsubsection{Local solutions}
\label{subsec:LocalSolsF}
Following the general theory of Fuchsian differential equations, one can construct local solutions around each regular singular point.
In particular, the Frobenius series solution around $z = 0$ satisfying the boundary condition $y(0)=1$ is called the local Heun function, or the general Heun function, and is expressed as 
\begin{align}
\HeunG{t,q;\alphaHeun,\betaHeun,\gammaHeun,\deltaHeun;z} = 1 + \frac{q}{t \gammaHeun} z + \mathcal{O}(z^2).
\label{eq:Heun-Taylor}
\end{align}
The independent solutions around each regular singular point can be expressed in terms of this local Heun function as follows.

Suppose $\gammaHeun \notin \mathbb{Z}$.
A linearly independent set of series solutions to Eq.~\eqref{Heun eq Fuchs} that converge in the region $|z| < \min\{ 1, |t| \}$ is given by
\begin{align}
y_{0_{-}}(z) &\coloneqq \HeunG{t,q;\alphaHeun,\betaHeun,\gammaHeun,\deltaHeun;z} = 1 + \mathcal{O}(z),\\
y_{0_{+}}(z) &\coloneqq z^{1 - \gammaHeun} \HeunG{t,q_{0_{+}};\alphaHeun - \gammaHeun + 1, \betaHeun - \gammaHeun + 1,2 - \gammaHeun,\deltaHeun;z} = z^{1 - \gammaHeun} ( 1 + \mathcal{O}(z) ),
\end{align}
with
\begin{align}
q_{0_{+}} &= q - (\gammaHeun - 1)(\epsilonHeun + t \deltaHeun ).
\end{align}
For the second solution $y_{0_{+}}$, a branch of $z^{1 - \gammaHeun}$ must be fixed.

Suppose $\epsilonHeun \notin \mathbb{Z}$. A linearly independent set of series solutions that converge in the region $|z - t| < \min\{|t|, |1 - t|\}$ is given by
\begin{align}
y_{t_-}(z) &\coloneqq \HeunG{ \frac{t}{t-1}, q_{t_{-}}; \alphaHeun, \betaHeun, \epsilonHeun, \deltaHeun; \frac{z - t}{1 - t} }, \\
y_{t_+}(z) &\coloneqq \left( t-z \right)^{1 - \epsilonHeun} \HeunG{\frac{t}{t-1}, q_{t_{+}}; \alphaHeun - \epsilonHeun + 1, \betaHeun - \epsilonHeun + 1, 2 - \epsilonHeun, \deltaHeun; \frac{z - t}{1 - t} },
\end{align}
with
\begin{align}
q_{t_{-}} &= \frac{q - t \alphaHeun \betaHeun}{1 - t},\\
q_{t_{+}} &= \frac{q - t \alphaHeun \betaHeun}{1 - t} - (\epsilonHeun - 1) \left(\gammaHeun + \frac{t}{t-1} \deltaHeun\right). 
\end{align}
A branch of $\left( t-z \right)^{1 - \epsilonHeun}$ must be fixed.

Suppose $\deltaHeun \notin \mathbb{Z}$.
A linearly independent set of series solutions that converge in the region $|t(1 - z)/(t-z)| < \min\{ 1 , |t| \} $ is given by 
\begin{align}
y_{1_{-}}(z) &\coloneqq 
\left( \frac{z-t}{1-t}\right)^{-\alphaHeun}
\HeunG{t, q_{1_{-}} ;\alphaHeun, \deltaHeun + \gammaHeun - \betaHeun,\deltaHeun, \gammaHeun; t \frac{1-z}{t-z}}, \\
y_{1_{+}}(z) &\coloneqq \left( \frac{z-t}{1-t}\right)^{-\alphaHeun - 1 + \deltaHeun} (1 - z)^{1 - \deltaHeun}
\HeunG{t, q_{1_{+}};\alphaHeun - \deltaHeun + 1,  - \betaHeun + \gammaHeun + 1, 2 - \deltaHeun, \gammaHeun ; t \frac{1-z}{t- z}},
\end{align}
with
\begin{align}
q_{1_{-}} &= q + \alphaHeun (\deltaHeun - \betaHeun),\\
q_{1_{+}} &= q - \alphaHeun(\betaHeun + \deltaHeun - 2) + (\deltaHeun - 1)(\alphaHeun + \betaHeun - 1 - t \gammaHeun).
\end{align}
Branches of powers of $z-t$, $1-t$, and $1-z$
must be fixed consistently.

Suppose $\alphaHeun - \betaHeun \notin \mathbb{Z}$. 
A linearly independent set of series solutions that converge in the region $|z| > \max\{1, |t|\}$ is given by
\begin{align}
y_{\infty_{-}}(z) &\coloneqq z^{-\betaHeun}\HeunG{t,q_{\infty_{-}};\betaHeun, \betaHeun - \gammaHeun + 1, \betaHeun - \alphaHeun + 1, \epsilonHeun;\frac{t}{z}},\\
y_{\infty_{+}}(z) &\coloneqq z^{-\alphaHeun}\HeunG{t,q_{\infty_{+}};\alphaHeun, \alphaHeun - \gammaHeun + 1,\alphaHeun - \betaHeun + 1, \epsilonHeun; \frac{t}{z}},
\end{align}
with
\begin{align}
    q_{\infty_{-}} = q - \alphaHeun \betaHeun (1+t) + \betaHeun (\deltaHeun + t \epsilonHeun), \\ 
    q_{\infty_{+}} = q - \alphaHeun \betaHeun (1+t) + \alphaHeun (\deltaHeun + t \epsilonHeun).
\end{align}
Branches of $z^{-\alphaHeun}$ and $z^{-\betaHeun}$ have to be chosen.

\subsection{Schr\"{o}dinger form of the Heun equation and conformal blocks}
\label{sec:Heun-Sdg}
We can also write the Heun equation in Schr\"{o}dinger form.
By introducing the function
\begin{equation}
\hat{\mathcal{F}}(z)\coloneqq P_{4}\left(z\right)^{-1}y(z),\label{eq:Heun-SCB}
\end{equation}
where
\begin{equation}
P_{4}\left(z\right)=z^{-\frac{\gammaHeun}{2}}\left(1-z\right)^{-\frac{\deltaHeun}{2}}\left(t-z\right)^{-\frac{\epsilonHeun}{2}},\label{eq:P4Def}
\end{equation}
the Heun equation can be expressed in the Schr\"{o}dinger form as 
\begin{align}
\left[ \frac{d^2}{dz^2} + \left( \frac{\frac{1}{4} - a_{0}^2}{z^2} + \frac{\frac{1}{4} - a_{1}^2}{(z - 1)^2} + \frac{\frac{1}{4} - a_{t}^2}{(z - t)^2}
- \frac{\frac{1}{2} - a_{1}^2 - a_{t}^2 - a_{0}^2 + a_{\infty}^2 + u}{z(z-1)}
+ \frac{u}{z(z-t)}
\right) \right] \hat{\mathcal{F}}(z) = 0, \label{Heun eq Schrodinger}
\end{align}
where the five parameters $a_{0}, a_{1}, a_{t}, a_{\infty}, u$ are given by
\begin{align}
a_{0} &\coloneqq \frac{1-\gammaHeun}{2}
, \quad 
a_{1} \coloneqq \frac{1 - \deltaHeun}{2}, \quad 
a_{t} \coloneqq \frac{1 - \epsilonHeun}{2}, \quad 
a_{\infty} \coloneqq \frac{\alphaHeun - \betaHeun}{2},
\end{align}
and
\begin{align}
    u \coloneqq \frac{\gammaHeun \epsilonHeun (1 - t) - \deltaHeun \epsilonHeun t - 2 q + 2 t \alphaHeun \betaHeun}{2(t-1)}.
\end{align}
Conversely, the inverse relations are given by
\begin{align}
 & \alphaHeun=1-a_{0}-a_{1}-a_{t}+a_{\infty},\quad\betaHeun=1-a_{0}-a_{1}-a_{t}-a_{\infty},\nonumber \\
 & \gammaHeun=1-2a_{0},\quad\deltaHeun=1-2a_{1},\quad\epsilonHeun=1-2a_{t},\nonumber \\
 & q=\frac{1}{2}+t\left(a_{0}^{2}+a_{1}^{2}+a_{t}^{2}-a_{\infty}^{2}\right)-a_{1}t-a_{t}+a_{0}(2a_{t}-1+(2a_{1}-1)t)+(1-t)u.
\end{align}

\subsubsection{Semiclassical conformal blocks}
\label{subsec:BPZ-SCB}
Following Ref.~\cite{Bonelli:2022ten}, we express the local solutions of the Heun equation in Schr\"{o}dinger form as suitably normalized semiclassical limits of Liouville conformal blocks. A conformal block is a universal building block of a correlation function in a conformal field theory (CFT), determined by conformal symmetry for a specified channel and fixed external and intermediate representations. In the case relevant here, the conformal block with four nondegenerate primary insertions and one level-two degenerate insertion satisfies the Belavin–Polyakov–Zamolodchikov (BPZ) equation \cite{Belavin:1984vu}:
\begin{align}
 & \left(b^{-2}\partial_{z}^{2}+\frac{\Delta_{1}}{\left(z-1\right)^{2}}-\frac{\Delta_{1}+t\partial_{t}+\Delta_{t}+z\partial_{z}+\Delta_{2,1}+\Delta_{0}-\Delta_{\infty}}{z\left(z-1\right)}\right.\nonumber \\
 & \quad\left.+\frac{\Delta_{t}}{\left(z-t\right)^{2}}+\frac{t}{z\left(z-t\right)}\partial_{t}-\frac{1}{z}\partial_{z}+\frac{\Delta_{0}}{z^{2}}\right)\FV{\alpha_{1}}{\alpha_{\infty}}{\alpha}{\alpha_{t}}{\alpha_{0\theta}}{\alpha_{2,1}}{\alpha_{0}}t{\frac{z}{t}}=0.\label{eq:BPZeq}
\end{align}
Here, $b$ is a Liouville coupling, $\alpha_{0},\alpha_{t},\alpha_{1},\alpha_{\infty}$
are momenta of the four nondegenerate primary fields, and $\alpha$ is the intermediate momentum. The momentum of the level-two degenerate field is
\begin{equation}
\alpha_{2,1}=-b-\frac{1}{2b}.
\end{equation}
For $\theta=\pm$, we define the shifted momenta by
\begin{equation}
\alpha_{i\theta}=\alpha_{i}-\theta\frac{b}{2},\quad \left(i=0,t,1,\infty\right).
\end{equation}
$\Delta_{2,1}$ and $\Delta_{i}$ are scaling dimensions defined by
\begin{equation}
\Delta_{2,1}=\frac{\QL^{2}}{4}-\alpha_{2,1}^{2}=-\frac{3b^{2}}{4}-\frac{1}{2},\quad
\Delta_{i}=\frac{\QL^{2}}{4}-\alpha_{i}^{2},\quad
\left(i=0,t,1,\infty\right),
\end{equation}
with the Liouville background charge
\begin{equation}
\QL=b+b^{-1}.
\end{equation}

The conformal block in Eq.~\eqref{eq:BPZeq} can be regarded as a local solution of the BPZ equation around $z=0$. The solutions around the other singular points can be obtained via M\"obius transformations; see Eq.~\eqref{eq:CB5-Mobius}.

To recover the Heun equation in Schr\"odinger form \eqref{Heun eq Schrodinger},
we take the semiclassical limit
\begin{equation}
b\rightarrow0,\quad
\alpha,\alpha_{i}\rightarrow\infty,\quad 
b\alpha=a\text{ fixed},\quad 
b\alpha_{i}=a_{i}\text{ fixed}.
\label{eq:SCLim}
\end{equation}
Although the five-point conformal block itself diverges, we can obtain a finite function by dividing it by a four-point conformal block. 
For each local solution, the semiclassical conformal block is defined as follows \cite{Bonelli:2022ten}:
\begin{subequations}
\label{eq:SCB-Def}
\begin{align}
\mathcal{F}_{0_{\theta}}\left(a_{i},a;t,z\right) & =\lim_{b\rightarrow0}\frac{\FV{\alpha_{1}}{\alpha_{\infty}}{\alpha}{\alpha_{t}}{\alpha_{0\theta}}{\alpha_{2,1}}{\alpha_{0}}t{\frac{z}{t}}}{\FIV{\alpha_{1}}{\alpha_{\infty}}{\alpha}{\alpha_{t}}{\alpha_{0}}t},\label{eq:SCB0-Def}\\
\mathcal{F}_{t_{\theta}}\left(a_{i},a;t,z\right) & =\lim_{b\rightarrow0}\left(t-1\right)^{-\Delta_{2,1}}\frac{\FV{\alpha_{1}}{\alpha_{\infty}}{\alpha}{\alpha_{0}}{\alpha_{t\theta}}{\alpha_{2,1}}{\alpha_{t}}{\frac{t}{t-1}}{\frac{t-z}{t}}}{\FIV{\alpha_{1}}{\alpha_{\infty}}{\alpha}{\alpha_{0}}{\alpha_{t}}{\frac{t}{t-1}}},\label{eq:SCBt-Def}\\
\mathcal{F}_{1_{\theta}}\left(a_{i},a;t,z\right) & =\lim_{b\rightarrow0}\left(t\left(1-t\right)\right)^{\Delta_{2,1}}\left(t-z\right)^{-2\Delta_{2,1}}\frac{\FV{\alpha_{0}}{\alpha_{t}}{\alpha}{\alpha_{\infty}}{\alpha_{1\theta}}{\alpha_{2,1}}{\alpha_{1}}t{\frac{1-z}{t-z}}}{\FIV{\alpha_{0}}{\alpha_{t}}{\alpha}{\alpha_{\infty}}{\alpha_{1}}t},\label{eq:SCB1-Def}\\
\mathcal{F}_{\infty_{\theta}}\left(a_{i},a;t,z\right) & =\lim_{b\rightarrow0}t^{\Delta_{2,1}}z^{-2\Delta_{2,1}}\frac{\FV{\alpha_{t}}{\alpha_{0}}{\alpha}{\alpha_{1}}{\alpha_{\infty\theta}}{\alpha_{2,1}}{\alpha_{\infty}}t{\frac{1}{z}}}{\FIV{\alpha_{t}}{\alpha_{0}}{\alpha}{\alpha_{1}}{\alpha_{\infty}}t}.
\label{eq:SCBi-Def}
\end{align}
\end{subequations}

The BPZ equation \eqref{eq:BPZeq} also simplifies considerably in the semiclassical limit \eqref{eq:SCLim}. 
The scaling dimensions behave as $b^{2}\Delta_{i}\rightarrow\frac{1}{4}-a_{i}^{2}$. 
The $t$-derivative of the conformal block can be read off from Eq.~\eqref{eq:CB5-Ldg} as
\begin{align}
 & t\partial_{t}\FV{\alpha_{1}}{\alpha_{\infty}}{\alpha}{\alpha_{t}}{\alpha_{0\theta}}{\alpha_{2,1}}{\alpha_{0}}t{\frac{z}{t}}\nonumber\\
 & =b^{-2}\left(-\frac{1}{4}-a^{2}+a_{t}^{2}+a_{0}^{2}+t\partial_{t}F\left(a_i,a;t\right)+\mathcal{O}\left(b^{2}\right)\right)\FV{\alpha_{1}}{\alpha_{\infty}}{\alpha}{\alpha_{t}}{\alpha_{0\theta}}{\alpha_{2,1}}{\alpha_{0}}t{\frac{z}{t}}.
\end{align}
Here \(F(a_i,a;t)\) denotes the classical four-point conformal block defined in Eq.~\eqref{eq:CCB-Def}.
Using the four-point asymptotic expansion Eq.~\eqref{eq:CB4-Ldg2}, the prefactor can be written in terms of the four-point conformal block. Thus, defining
\begin{equation}
u=\lim_{b\rightarrow0}b^{2}t\partial_{t}\log\FIV{\alpha_{1}}{\alpha_{\infty}}{\alpha}{\alpha_{t}}{\alpha_{0}}t=-\frac{1}{4}-a^{2}+a_{t}^{2}+a_{0}^{2}+t\partial_{t}F\left(a_{i},a;t\right),
\label{eq:u0-Def}
\end{equation}
we obtain Eq.~\eqref{Heun eq Schrodinger} with $\mathcal{F}(z)\propto\mathcal{F}_{0_{\theta}}\left(a_{i},a;t,z\right)$.
By applying the appropriate M\"obius transformations, one can check that the other semiclassical conformal blocks in Eq.~\eqref{eq:SCB-Def} also satisfy the Heun equation in Schr\"{o}dinger form \eqref{Heun eq Schrodinger}. 
For prescribed Heun data $(a_0,a_t,a_1,a_\infty,u,t)$, this relation must be inverted, at least perturbatively in $t$, to determine the intermediate momentum $a=a(u,t)$.

\subsubsection{Local solutions from semiclassical conformal blocks}
\label{subsec:LocalSols}
The semiclassical blocks and the corresponding Schr\"odinger-form Heun solutions have the same local Frobenius exponents. Each local Heun solution is therefore proportional to \(P_4(z)\hat{\mathcal{F}}(z)\) as in Eq.~\eqref{eq:Heun-SCB}. The proportionality factors are independent of \(z\) and are fixed by matching the leading local behavior of the solutions in Sec.~\ref{subsec:LocalSolsF} (with Eq.~\eqref{eq:Heun-Taylor}) with that of the semiclassical blocks in Eq.~\eqref{eq:SCB-Ldg}.
From Eqs.~\eqref{eq:P4Def} and \eqref{Heun eq Schrodinger}, the prefactor \(P_4(z)\) has the following local behavior:
\begin{equation}
P_{4}\left(z\right)=\begin{cases}
t^{a_{t}-\frac{1}{2}}z^{a_{0}-\frac{1}{2}}\left(1+\mathcal{O}\left(z\right)\right), & \left(z\sim0\right)\\
t^{a_{0}-\frac{1}{2}}\left(1-t\right)^{a_{1}-\frac{1}{2}}\left(t-z\right)^{a_{t}-\frac{1}{2}}\left(1+\mathcal{O}\left(z-t\right)\right), & \left(z\sim t\right)\\
\mathrm{e}^{-i\pi\left(a_{t}-\frac{1}{2}\right)}\left(1-t\right)^{a_{t}-\frac{1}{2}}\left(1-z\right)^{a_{1}-\frac{1}{2}}\left(1+\mathcal{O}\left(z-1\right)\right), & \left(z\sim1\right)\\
\mathrm{e}^{i\pi\left(1-a_{1}-a_{t}\right)}z^{a_{0}+a_{t}+a_{1}-\frac{3}{2}}\left(1+\mathcal{O}\left(\frac{1}{z}\right)\right), & \left(z\sim\infty\right)
\end{cases}.
\end{equation}
Note that we have implicitly chosen the branch cut and hence the phase signs.\footnote{We choose this phase sign so that the connection formulae can be checked numerically using Mathematica. 
More specifically, we have compared the connection coefficients generated using the Heun function provided by Mathematica with those given above. 
In practical computations, the functions $a\left(t\right)$ and $F\left(t\right)$ are expressed as Taylor series, so we must take $t$ to be small. 
When generating the numerical coefficients using the Heun function, we set $z=x+iy$ with $y>0$.}
Therefore,
\begin{subequations}
\label{eq:HeunG-SCB}
\begin{align}
y_{0_{\theta}}\left(z\right) & =P_{4}\left(z\right)t^{\frac{1}{2}-a_{t}+\theta a_{0}}\mathrm{e}^{\frac{\theta}{2}\partial_{a_{0}}F\left(a_{i},a;t\right)}\mathcal{F}_{0_{\theta}}\left(a_{i},a;t,z\right),\label{eq:HeunG-SCB0}\\
y_{t_{\theta}}\left(z\right) & =P_{4}\left(z\right)t^{\frac{1}{2}-a_{0}+\theta a_{t}}\left(1-t\right)^{\frac{1}{2}-a_{1}}\mathrm{e}^{\frac{\theta}{2}\partial_{a_{t}}F\left(a_{i},a;t\right)}\mathcal{F}_{t_{\theta}}\left(a_{i},a;t,z\right),\label{eq:HeunG-SCBt}\\
y_{1_{\theta}}\left(z\right) & =P_{4}\left(z\right)\mathrm{e}^{i\pi\left(-\theta a_{1}+a_{t}\right)}\left(1-t\right)^{\frac{1}{2}-a_{t}}\mathrm{e}^{\frac{\theta}{2}\partial_{a_{1}}F\left(a_{i},a;t\right)}\mathcal{F}_{1_{\theta}}\left(a_{i},a;t,z\right),\label{eq:HeunG-SCB1}\\
y_{\infty_{\theta}}\left(z\right) & =P_{4}\left(z\right)\mathrm{e}^{-i\pi\left(1-a_{1}-a_{t}\right)}\mathrm{e}^{\frac{\theta}{2}\partial_{a_{\infty}}F\left(a_{i},a;t\right)}\mathcal{F}_{\infty_{\theta}}\left(a_{i},a;t,z\right).\label{eq:HeunG-SCBi}
\end{align}
\end{subequations}

\subsection{Connection formulae}
\label{subsec:ConForm}
Having expressed the local solutions of the Heun equation in terms of the semiclassical conformal blocks in Eq.~\eqref{eq:HeunG-SCB}, we now present explicit connection formulae for the local solutions using those for the semiclassical conformal blocks. Because there are four singular points, there are six pairwise connection formulae. 
Throughout the remainder of this paper, we abbreviate $F\left(a_{i},a;t\right)$ as $F$ and $\mathcal{F}_{z_{\theta}}\left(a_{i},a;t,z\right)$ as $\mathcal{F}_{z_{\theta}}$.

All connection formulae are expressed in terms of the following connection matrix:
\begin{equation}
\mathcal{M}_{\theta\theta'}\left(c_{1},c_{2};c_{3}\right)\coloneqq\frac{\Gamma\left(-2\theta'c_{2}\right)\Gamma\left(1+2\theta c_{1}\right)}{\Gamma\left(\frac{1}{2}+\theta c_{1}-\theta'c_{2}+c_{3}\right)\Gamma\left(\frac{1}{2}+\theta c_{1}-\theta'c_{2}-c_{3}\right)}.\label{eq:ConMat}
\end{equation}
Here, the term ``matrix'' indicates that we also use the following matrix notation:
\begin{equation}
\mathcal{M}\left(c_{1},c_{2};c_{3}\right)\coloneqq\left(\left[\mathcal{M}_{\theta\theta'}\left(c_{1},c_{2};c_{3}\right)\right]_{\theta\theta'}^{2\times2}\right)=\left(\begin{array}{cc}
\mathcal{M}_{--}\left(c_{1},c_{2};c_{3}\right) & \mathcal{M}_{-+}\left(c_{1},c_{2};c_{3}\right)\\
\mathcal{M}_{+-}\left(c_{1},c_{2};c_{3}\right) & \mathcal{M}_{++}\left(c_{1},c_{2};c_{3}\right)
\end{array}\right).
\end{equation}
We order the values of $\theta$ as $\left\{-,+\right\}$.
A phase factor matrix also plays an important role in combining the connection formulae:
\begin{equation}
\Theta\left(c\right)\coloneqq\left(\begin{array}{cc}
\mathrm{e}^{-i\pi c} & 0\\
0 & \mathrm{e}^{i\pi c}
\end{array}\right).\label{eq:PhaseMat}
\end{equation}

In the following subsections, we list the connection formulae for the
local solutions of the Heun equation.
Details of the derivation of the connection formulae for the semiclassical conformal blocks are provided in Appendix~\ref{subsec:CF-SCB}.

\subsubsection{$0\leftrightarrow t$\label{subsec:ConForm0t}}

The connection formula relating the semiclassical conformal blocks expanded about $z=0$ and $z=t$ is given by
\begin{equation}
\left(\begin{array}{c}
\mathcal{F}_{0_{-}}\\
\mathcal{F}_{0_{+}}
\end{array}\right)=\mathcal{M}\left(a_{0},a_{t};a\right)\left(\begin{array}{c}
\mathcal{F}_{t_{-}}\\
\mathcal{F}_{t_{+}}
\end{array}\right),\quad\left(\begin{array}{c}
\mathcal{F}_{t_{-}}\\
\mathcal{F}_{t_{+}}
\end{array}\right)=\mathcal{M}\left(a_{t},a_{0};a\right)\left(\begin{array}{c}
\mathcal{F}_{0_{-}}\\
\mathcal{F}_{0_{+}}
\end{array}\right).\label{eq:SCB0tM}
\end{equation}
Using Eqs.~\eqref{eq:SCB0tM}, \eqref{eq:HeunG-SCB0}, and \eqref{eq:HeunG-SCBt}, the connection formula relating the local solutions of the Heun equation expanded about $z=0$ and $z=t$ is obtained as
\begin{equation}
y_{0_{\theta}}\left(z\right)=\sum_{\theta'=\pm}C_{0_{\theta}}{}^{t_{\theta'}}y_{t_{\theta'}}\left(z\right),\quad y_{t_{\theta}}\left(z\right)=\sum_{\theta'=\pm}C_{t_{\theta}}{}^{0_{\theta'}}y_{0_{\theta'}}\left(z\right),
\end{equation}
with
\begin{align}
C_{0_{\theta}}{}^{t_{\theta'}} & =t^{\left(1+\theta\right)a_{0}-\left(1+\theta'\right)a_{t}}\left(1-t\right)^{-\frac{1}{2}+a_{1}}\mathrm{e}^{\frac{1}{2}\left(\theta\partial_{a_{0}}-\theta'\partial_{a_{t}}\right)F}\mathcal{M}_{\theta\theta'}\left(a_{0},a_{t};a\right),\label{C 0 t}\\
C_{t_{\theta}}{}^{0_{\theta'}} & =t^{-\left(1+\theta'\right)a_{0}+\left(1+\theta\right)a_{t}}\left(1-t\right)^{\frac{1}{2}-a_{1}}\mathrm{e}^{\frac{1}{2}\left(-\theta'\partial_{a_{0}}+\theta\partial_{a_{t}}\right)F}\mathcal{M}_{\theta\theta'}\left(a_{t},a_{0};a\right).\label{C t 0}
\end{align}

\subsubsection{$0\leftrightarrow1$\label{subsec:ConForm01}}

The connection formula relating the semiclassical conformal blocks expanded about $z=0$ and $z=1$ is given by
\begin{align}
\left(\begin{array}{c}
\mathcal{F}_{0_{-}}\\
\mathcal{F}_{0_{+}}
\end{array}\right) & =i\Theta\left(a_{0}\right)\mathcal{M}\left(a_{0},a;a_{t}\right)\left(\begin{array}{cc}
0 & \mathrm{e}^{-i\pi a}t^{-a}\mathrm{e}^{\frac{1}{2}\partial_{a}F}\\
\mathrm{e}^{i\pi a}t^{a}\mathrm{e}^{-\frac{1}{2}\partial_{a}F} & 0
\end{array}\right)\mathcal{M}\left(a,a_{1};a_{\infty}\right)\Theta\left(-a_{1}\right)\left(\begin{array}{c}
\mathcal{F}_{1_{-}}\\
\mathcal{F}_{1_{+}}
\end{array}\right),\nonumber \\
\left(\begin{array}{c}
\mathcal{F}_{1_{-}}\\
\mathcal{F}_{1_{+}}
\end{array}\right) & =-i\Theta\left(a_{1}\right)\mathcal{M}\left(a_{1},a;a_{\infty}\right)\left(\begin{array}{cc}
0 & \mathrm{e}^{-i\pi a}t^{-a}\mathrm{e}^{\frac{1}{2}\partial_{a}F}\\
\mathrm{e}^{i\pi a}t^{a}\mathrm{e}^{-\frac{1}{2}\partial_{a}F} & 0
\end{array}\right)\mathcal{M}\left(a,a_{0};a_{t}\right)\Theta\left(-a_{0}\right)\left(\begin{array}{c}
\mathcal{F}_{0_{-}}\\
\mathcal{F}_{0_{+}}
\end{array}\right).\label{eq:SCB01M}
\end{align}
Using Eqs.~\eqref{eq:SCB01M}, \eqref{eq:HeunG-SCB0}, and \eqref{eq:HeunG-SCB1}, the connection formula relating the local solutions of the Heun equation expanded about $z=0$ and $z=1$ is obtained as
\begin{equation}
y_{0_{\theta}}\left(z\right)=\sum_{\theta'=\pm}C_{0_{\theta}}{}^{1_{\theta'}}y_{1_{\theta'}}\left(z\right),\quad y_{1_{\theta}}\left(z\right)=\sum_{\theta'=\pm}C_{1_{\theta}}{}^{0_{\theta'}}y_{0_{\theta'}}\left(z\right),
\end{equation}
where
\begin{align}
C_{0_{\theta}}{}^{1_{\theta'}}&=\sum_{\sigma=\pm1}\mathrm{e}^{i\pi\left(\frac{1}{2}+\theta a_{0}+\sigma a-a_{t}\right)}t^{\frac{1}{2}-a_{t}+\theta a_{0}+\sigma a}\left(1-t\right)^{-\frac{1}{2}+a_{t}}\mathrm{e}^{\frac{1}{2}\left(\theta\partial_{a_{0}}-\sigma\partial_{a}-\theta'\partial_{a_{1}}\right)F}\nonumber \\
 & \quad\times\mathcal{M}_{\theta\sigma}\left(a_{0},a;a_{t}\right)\mathcal{M}_{\left(-\sigma\right)\theta'}\left(a,a_{1};a_{\infty}\right),\label{C 0 1}\\
C_{1_{\theta}}{}^{0_{\theta'}} & =\sum_{\sigma=\pm1}\mathrm{e}^{-i\pi\left(\frac{1}{2}+\theta'a_{0}-\sigma a-a_{t}\right)}t^{-\frac{1}{2}+a_{t}-\theta'a_{0}+\sigma a}\left(1-t\right)^{\frac{1}{2}-a_{t}}\mathrm{e}^{\frac{1}{2}\left(-\theta'\partial_{a_{0}}-\sigma\partial_{a}+\theta\partial_{a_{1}}\right)F}\nonumber \\
 & \quad\times\mathcal{M}_{\theta\sigma}\left(a_{1},a;a_{\infty}\right)\mathcal{M}_{\left(-\sigma\right)\theta'}\left(a,a_{0};a_{t}\right).\label{C 1 0}
\end{align}

\subsubsection{$0\leftrightarrow\infty$\label{subsec:ConForm0i}}

The connection formula relating the semiclassical conformal blocks expanded about $z=0$ and $z=\infty$ is given by
\begin{align}
\left(\begin{array}{c}
\mathcal{F}_{0_{-}}\\
\mathcal{F}_{0_{+}}
\end{array}\right) & =\Theta\left(a_{0}\right)\mathcal{M}\left(a_{0},a;a_{t}\right)\left(\begin{array}{cc}
0 & t^{-a}\mathrm{e}^{\frac{1}{2}\partial_{a}F}\\
t^{a}\mathrm{e}^{-\frac{1}{2}\partial_{a}F} & 0
\end{array}\right)\mathcal{M}\left(a,a_{\infty};a_{1}\right)\Theta\left(a_{\infty}\right)\left(\begin{array}{c}
\mathcal{F}_{\infty_{-}}\\
\mathcal{F}_{\infty_{+}}
\end{array}\right),\nonumber \\
\left(\begin{array}{c}
\mathcal{F}_{\infty_{-}}\\
\mathcal{F}_{\infty_{+}}
\end{array}\right) & =\Theta\left(-a_{\infty}\right)\mathcal{M}\left(a_{\infty},a;a_{1}\right)\left(\begin{array}{cc}
0 & t^{-a}\mathrm{e}^{\frac{1}{2}\partial_{a}F}\\
t^{a}\mathrm{e}^{-\frac{1}{2}\partial_{a}F} & 0
\end{array}\right)\mathcal{M}\left(a,a_{0};a_{t}\right)\Theta\left(-a_{0}\right)\left(\begin{array}{c}
\mathcal{F}_{0_{-}}\\
\mathcal{F}_{0_{+}}
\end{array}\right).\label{eq:SCB0iM}
\end{align}
Using Eqs.~\eqref{eq:SCB0iM}, \eqref{eq:HeunG-SCB0}, and \eqref{eq:HeunG-SCBi}, the connection formula relating the local solutions of the Heun equation expanded about $z=0$ and $z=\infty$ is obtained as
\begin{equation}
y_{0_{\theta}}\left(z\right)=\sum_{\theta'=\pm}C_{0_{\theta}}{}^{\infty_{\theta'}}y_{\infty_{\theta'}}\left(z\right),\quad y_{\infty_{\theta}}\left(z\right)=\sum_{\theta'=\pm}C_{\infty_{\theta}}{}^{0_{\theta'}}y_{0_{\theta'}}\left(z\right),
\end{equation}
where
\begin{align}
C_{0_{\theta}}{}^{\infty_{\theta'}}&=\sum_{\sigma=\pm1}\mathrm{e}^{i\pi\left(1-a_{t}-a_{1}+\theta a_{0}+\theta'a_{\infty}\right)}t^{\frac{1}{2}-a_{t}+\theta a_{0}+\sigma a}\mathrm{e}^{\frac{1}{2}\left(\theta\partial_{a_{0}}-\sigma\partial_{a}-\theta'\partial_{a_{\infty}}\right)F}\nonumber \\
 & \quad\times\mathcal{M}_{\theta\sigma}\left(a_{0},a;a_{t}\right)\mathcal{M}_{\left(-\sigma\right)\theta'}\left(a,a_{\infty};a_{1}\right),\label{C 0 i}\\
C_{\infty_{\theta}}{}^{0_{\theta'}}&=\sum_{\sigma=\pm1}\mathrm{e}^{-i\pi\left(1-a_{t}-a_{1}+\theta'a_{0}+\theta a_{\infty}\right)}t^{-\frac{1}{2}+a_{t}-\theta'a_{0}+\sigma a}\mathrm{e}^{\frac{1}{2}\left(-\theta'\partial_{a_{0}}-\sigma\partial_{a}+\theta\partial_{a_{\infty}}\right)F}\nonumber \\
 & \quad\times\mathcal{M}_{\theta\sigma}\left(a_{\infty},a;a_{1}\right)\mathcal{M}_{\left(-\sigma\right)\theta'}\left(a,a_{0};a_{t}\right).\label{C i 0}
\end{align}

\subsubsection{$t\leftrightarrow1$\label{subsec:ConFormt1}}

The connection formula relating the semiclassical conformal blocks expanded about $z=t$ and $z=1$ is given by
\begin{align}
\left(\begin{array}{c}
\mathcal{F}_{t_{-}}\\
\mathcal{F}_{t_{+}}
\end{array}\right) & =\Theta\left(-a_{t}\right)\mathcal{M}\left(a_{t},a;a_{0}\right)\left(\begin{array}{cc}
0 & t^{-a}\mathrm{e}^{\frac{1}{2}\partial_{a}F}\\
t^{a}\mathrm{e}^{-\frac{1}{2}\partial_{a}F} & 0
\end{array}\right)\mathcal{M}\left(a,a_{1};a_{\infty}\right)\Theta\left(-a_{1}\right)\left(\begin{array}{c}
\mathcal{F}_{1_{-}}\\
\mathcal{F}_{1_{+}}
\end{array}\right),\nonumber \\
\left(\begin{array}{c}
\mathcal{F}_{1_{-}}\\
\mathcal{F}_{1_{+}}
\end{array}\right) & =\Theta\left(a_{1}\right)\mathcal{M}\left(a_{1},a;a_{\infty}\right)\left(\begin{array}{cc}
0 & t^{-a}\mathrm{e}^{\frac{1}{2}\partial_{a}F}\\
t^{a}\mathrm{e}^{-\frac{1}{2}\partial_{a}F} & 0
\end{array}\right)\mathcal{M}\left(a,a_{t};a_{0}\right)\Theta\left(a_{t}\right)\left(\begin{array}{c}
\mathcal{F}_{t_{-}}\\
\mathcal{F}_{t_{+}}
\end{array}\right).\label{eq:SCBt1M}
\end{align}
Using Eqs.~\eqref{eq:SCBt1M}, \eqref{eq:HeunG-SCBt}, and \eqref{eq:HeunG-SCB1}, the connection formula relating the local solutions of the Heun equation expanded about $z=t$ and $z=1$ is obtained as
\begin{equation}
y_{t_{\theta}}\left(z\right)=\sum_{\theta'=\pm}C_{t_{\theta}}{}^{1_{\theta'}}y_{1_{\theta'}}\left(z\right),\quad y_{1_{\theta}}\left(z\right)=\sum_{\theta'=\pm}C_{1_{\theta}}{}^{t_{\theta'}}y_{t_{\theta'}}\left(z\right),
\end{equation}
where
\begin{align}
C_{t_{\theta}}{}^{1_{\theta'}}&=\sum_{\sigma=\pm1}\mathrm{e}^{-i\pi\left(1+\theta\right)a_{t}}t^{\frac{1}{2}-a_{0}+\theta a_{t}+\sigma a}\left(1-t\right)^{a_{t}-a_{1}}\mathrm{e}^{\frac{1}{2}\left(\theta\partial_{a_{t}}-\sigma\partial_{a}-\theta'\partial_{a_{1}}\right)F}\nonumber \\
 & \quad\times\mathcal{M}_{\theta\sigma}\left(a_{t},a;a_{0}\right)\mathcal{M}_{\left(-\sigma\right)\theta'}\left(a,a_{1};a_{\infty}\right),\label{C t 1}\\
C_{1_{\theta}}{}^{t_{\theta'}}&=\sum_{\sigma=\pm1}\mathrm{e}^{i\pi\left(1+\theta'\right)a_{t}}t^{-\frac{1}{2}+a_{0}-\theta'a_{t}+\sigma a}\left(1-t\right)^{-a_{t}+a_{1}}\mathrm{e}^{\frac{1}{2}\left(-\theta'\partial_{a_{t}}-\sigma\partial_{a}+\theta\partial_{a_{1}}\right)F}\nonumber \\
 & \quad\times\mathcal{M}_{\theta\sigma}\left(a_{1},a;a_{\infty}\right)\mathcal{M}_{\left(-\sigma\right)\theta'}\left(a,a_{t};a_{0}\right).\label{C 1 t}
\end{align}

\subsubsection{$t\leftrightarrow\infty$\label{subsec:ConFormti}}

The connection formula relating the semiclassical conformal blocks expanded about $z=t$ and $z=\infty$ is given by
\begin{align}
\left(\begin{array}{c}
\mathcal{F}_{t_{-}}\\
\mathcal{F}_{t_{+}}
\end{array}\right) & =-i\Theta\left(-a_{t}\right)\mathcal{M}\left(a_{t},a;a_{0}\right)\left(\begin{array}{cc}
0 & \mathrm{e}^{i\pi a}t^{-a}\mathrm{e}^{\frac{1}{2}\partial_{a}F}\\
\mathrm{e}^{-i\pi a}t^{a}\mathrm{e}^{-\frac{1}{2}\partial_{a}F} & 0
\end{array}\right)\mathcal{M}\left(a,a_{\infty};a_{1}\right)\Theta\left(a_{\infty}\right)\left(\begin{array}{c}
\mathcal{F}_{\infty_{-}}\\
\mathcal{F}_{\infty_{+}}
\end{array}\right),\nonumber \\
\left(\begin{array}{c}
\mathcal{F}_{\infty_{-}}\\
\mathcal{F}_{\infty_{+}}
\end{array}\right) & =i\Theta\left(-a_{\infty}\right)\mathcal{M}\left(a_{\infty},a;a_{1}\right)\left(\begin{array}{cc}
0 & \mathrm{e}^{i\pi a}t^{-a}\mathrm{e}^{\frac{1}{2}\partial_{a}F}\\
\mathrm{e}^{-i\pi a}t^{a}\mathrm{e}^{-\frac{1}{2}\partial_{a}F} & 0
\end{array}\right)\mathcal{M}\left(a,a_{t};a_{0}\right)\Theta\left(a_{t}\right)\left(\begin{array}{c}
\mathcal{F}_{t_{-}}\\
\mathcal{F}_{t_{+}}
\end{array}\right).\label{eq:SCBtiM}
\end{align}
Using Eqs.~\eqref{eq:SCBtiM}, \eqref{eq:HeunG-SCBt}, and \eqref{eq:HeunG-SCBi}, the connection formula relating the local solutions of the Heun equation expanded about $z=t$ and $z=\infty$ is obtained as
\begin{equation}
y_{t_{\theta}}\left(z\right)=\sum_{\theta'=\pm}C_{t_{\theta}}{}^{\infty_{\theta'}}y_{\infty_{\theta'}}\left(z\right),\quad y_{\infty_{\theta}}\left(z\right)=\sum_{\theta'=\pm}C_{\infty_{\theta}}{}^{t_{\theta'}}y_{t_{\theta'}}\left(z\right),
\end{equation}
where
\begin{align}
C_{t_{\theta}}{}^{\infty_{\theta'}}&=\sum_{\sigma=\pm1}\mathrm{e}^{i\pi\left(\frac{1}{2}-a_{t}-a_{1}-\theta a_{t}-\sigma a+\theta'a_{\infty}\right)}t^{\frac{1}{2}-a_{0}+\theta a_{t}+\sigma a}\left(1-t\right)^{\frac{1}{2}-a_{1}}\mathrm{e}^{\frac{1}{2}\left(\theta\partial_{a_{t}}-\sigma\partial_{a}-\theta'\partial_{a_{\infty}}\right)F}\nonumber \\
 & \quad\times\mathcal{M}_{\theta\sigma}\left(a_{t},a;a_{0}\right)\mathcal{M}_{\left(-\sigma\right)\theta'}\left(a,a_{\infty};a_{1}\right),\label{C t i}\\
C_{\infty_{\theta}}{}^{t_{\theta'}}&=\sum_{\sigma=\pm1}\mathrm{e}^{-i\pi\left(\frac{1}{2}-a_{t}-a_{1}-\theta'a_{t}+\sigma a+\theta a_{\infty}\right)}t^{-\frac{1}{2}+a_{0}-\theta'a_{t}+\sigma a}\left(1-t\right)^{-\frac{1}{2}+a_{1}}\mathrm{e}^{\frac{1}{2}\left(-\theta'\partial_{a_{t}}-\sigma\partial_{a}+\theta\partial_{a_{\infty}}\right)F}\nonumber \\
 & \quad\times\mathcal{M}_{\theta\sigma}\left(a_{\infty},a;a_{1}\right)\mathcal{M}_{\left(-\sigma\right)\theta'}\left(a,a_{t};a_{0}\right).\label{C i t}
\end{align}

\subsubsection{$1\leftrightarrow\infty$\label{subsec:ConForm1i}}

The connection formula relating the semiclassical conformal blocks expanded about $z=1$ and $z=\infty$ is given by
\begin{align}
\left(\begin{array}{c}
\mathcal{F}_{1_{-}}\\
\mathcal{F}_{1_{+}}
\end{array}\right)  =-\mathcal{M}\left(a_{1},a_{\infty};a\right)\left(\begin{array}{c}
\mathcal{F}_{\infty_{-}}\\
\mathcal{F}_{\infty_{+}}
\end{array}\right),\quad
\left(\begin{array}{c}
\mathcal{F}_{\infty_{-}}\\
\mathcal{F}_{\infty_{+}}
\end{array}\right)  =-\mathcal{M}\left(a_{\infty},a_{1};a\right)\left(\begin{array}{c}
\mathcal{F}_{1_{-}}\\
\mathcal{F}_{1_{+}}
\end{array}\right).\label{eq:SCB1iM}
\end{align}
Using Eqs.~\eqref{eq:SCB1iM}, \eqref{eq:HeunG-SCB1}, and \eqref{eq:HeunG-SCBi}, the connection formula relating the local solutions of the Heun equation expanded about $z=1$ and $z=\infty$ is obtained as
\begin{equation}
y_{1_{\theta}}\left(z\right)=\sum_{\theta'=\pm}C_{1_{\theta}}{}^{\infty_{\theta'}}y_{\infty_{\theta'}}\left(z\right),\quad y_{\infty_{\theta}}\left(z\right)=\sum_{\theta'=\pm}C_{\infty_{\theta}}{}^{1_{\theta'}}y_{1_{\theta'}}\left(z\right),
\end{equation}
with
\begin{align}
C_{1_{\theta}}{}^{\infty_{\theta'}} & =\mathrm{e}^{-i\pi\left(1+\theta\right)a_{1}}\left(1-t\right)^{\frac{1}{2}-a_{t}}\mathrm{e}^{\frac{1}{2}\left(\theta\partial_{a_{1}}-\theta'\partial_{a_{\infty}}\right)F}\mathcal{M}_{\theta\theta'}\left(a_{1},a_{\infty};a\right),\label{C 1 i}\\
C_{\infty_{\theta}}{}^{1_{\theta'}} & =\mathrm{e}^{i\pi\left(1+\theta'\right)a_{1}}\left(1-t\right)^{-\frac{1}{2}+a_{t}}\mathrm{e}^{\frac{1}{2}\left(-\theta'\partial_{a_{1}}+\theta\partial_{a_{\infty}}\right)F}\mathcal{M}_{\theta\theta'}\left(a_{\infty},a_{1};a\right).\label{C i 1}
\end{align}

\section{Kerr--Newman--de Sitter spacetime and the Teukolsky equations}
\label{sec:Teukolsky}
\subsection{Kerr--Newman--de Sitter spacetime}
The Kerr--Newman--de Sitter spacetime is a solution to the Einstein--Maxwell equations with a cosmological constant $\Lambda$, and the metric and electromagnetic potential in the Boyer--Lindquist coordinates are given by 
\begin{align}
g_{\mu\nu}dx^{\mu} dx^{\nu} &= - \frac{\Delta_{r}(r)}{\Sigma(r,\vartheta) \Xi^2} \left( d \tau - \aBH \sin^2 \vartheta d \varphi \right)^2 + \Sigma(r,\vartheta) \left( \frac{dr^2}{\Delta_{r}(r)} + \frac{d \vartheta^2}{\Delta_{\vartheta}(\vartheta)} \right)
\\
&\qquad + \frac{\Delta_{\vartheta}(\vartheta) \sin^2 \vartheta}{\Sigma(r,\vartheta) \Xi^2}  \left( (r^2 + \aBH^2) d \varphi - \aBH d\tau \right)^2,
\\
A_{\mu} dx^{\mu} &= - \frac{\QBH r}{\Sigma(r,\vartheta) \Xi} \left( d\tau - \aBH \sin^2 \vartheta d\varphi \right),
\end{align}
where we introduced the functions
\begin{align}
    \Sigma(r,\vartheta) &\coloneqq r^2 + \aBH^2 \cos^2 \vartheta, \\
    \Delta_{r}(r) &\coloneqq (r^2 + \aBH^2)\left(1 - \frac{r^2}{L^2} \right) - 2 M r + \QBH^2, \label{def Delta}\\
    \Delta_{\vartheta}(\vartheta) &\coloneqq 1 + \frac{\aBH^2}{L^2} \cos^2 \vartheta,
\end{align}
and the constants
\begin{align}
\aBH \coloneqq \frac{J}{M}, \quad L \coloneqq \sqrt{\frac{3}{\Lambda}}, \quad \Xi \coloneqq 1 + \frac{\aBH^2}{L^2},
\end{align}

The spacetime is characterized by four parameters $M, J, \QBH, \Lambda$, where the first three, appearing as integration constants, represent the mass, angular momentum, and charge, while $\Lambda$ is the cosmological constant.
Throughout the paper, we focus on the case $M>0$, $\aBH^2 + \QBH^2 > 0,$ and $\Lambda >0$.
In addition, we impose the conditions~\eqref{upper bound for Q}, \eqref{condition for a}, and \eqref{Mass range} discussed in the appendix, which ensure that the four roots of $\Delta_{r}(r) = 0$ are real and nondegenerate. In this case,
one of the roots, say the negative root $r_{n}$, is negative and the other three roots, say $r_{-}, r_{+}, r_{c}$, are positive.
We adopt the following ordering:
\begin{align}
r_{n} < 0 < r_{-} < r_{+} < r_{c}.
\end{align}
Then, the physical meaning of each root is as follows:
\begin{align}
r_{c} &: \text{cosmological horizon,} \\
r_{+} &:\text{outer black hole horizon,}\\
r_{-} &:\text{inner black hole horizon,}\\
r_{n}&:\text{cosmological horizon in the negative $r$ region.}
\end{align}
Note that the negative root $r_n$ is not merely an artificial root when $\aBH \neq 0$, because
$r = 0$ is singular only on the equatorial plane $\vartheta = \pi/2$ and the spacetime can be extended to the negative $r$ region through $r=0$ away from the ring singularity. The negative $r$-region again possesses another asymptotic de Sitter region and $r = r_{n}$ is the cosmological horizon there. 

The relevant Killing vector is the null generator of the outer black hole horizon, given by
\begin{align}
\xi_{+}^{\mu} \partial_{\mu} = \partial_{\tau} + \Omega_{+} \partial_{\varphi},
\end{align}
with the horizon angular velocity $\Omega_{+}$ defined by
\begin{align}
\Omega_{+} \coloneqq  \lim_{r \to r_{+}} - \frac{g_{\tau \tau}}{g_{\tau\varphi}}  = \frac{\aBH}{r_{+}^2 + \aBH^2}.
\end{align}
The horizon Coulomb potential $\Phi_{+}$ is then defined by
\begin{align}
\Phi_{+} \coloneqq \lim_{r \to r_{+}} - A_{\mu} \xi_{+}^{\mu} = \frac{1}{\Xi}\frac{r_{+}\QBH} {r_{+}^2 + \aBH^2}.
\end{align}
The surface gravity $\kappa_{+}$ can be evaluated as
\begin{align}
\kappa_{+} = \lim_{r \to r_{+}} \frac{1}{2} \sqrt{ - \frac{g^{\mu\nu} \partial_{\mu}(|\xi_{+}|^2)\partial_{\nu}(|\xi_{+}|^2) }{|\xi_{+}|^2}}
= \frac{\Delta_{r}'(r_{+})}{2 \Xi (r_{+}^2 + \aBH^2)}
= - \frac{(r_{+} - r_{-})(r_{+} - r_{c})(r_{+} - r_{n})}{2 (\aBH^2 + L^2)  (r_{+}^2 + \aBH^2)}.
\end{align}

Similarly, one can introduce the Coulomb potential, the angular velocity and the surface gravity at each horizon $r = r_{i}$ by
\begin{align}
\Phi_{i} &\coloneqq \frac{1}{\Xi} \frac{r_{i} \QBH}{r_i^2 + \aBH^2}, \quad
\Omega_{i} \coloneqq \frac{\aBH}{r_{i}^2 + \aBH^2}, \quad 
\kappa_{i} \coloneqq \frac{\Delta_{r}'(r_{i})}{2 \Xi (r_{i}^2 + \aBH^2)}.
\end{align}

\subsection{Teukolsky equations}

On the Kerr--de Sitter background ($\QBH = 0$), the field equations for a conformally coupled massless scalar field, a massless Dirac field, an electromagnetic field, a Rarita--Schwinger field, and gravitational perturbations can be separated and written in the unified Teukolsky form, with spin weights $s=0,\pm1/2,\pm1,\pm3/2,\pm2$, respectively.
For the Kerr--Newman--de Sitter background ($\QBH \neq 0$), the corresponding formulation is established for the charged conformally coupled massless scalar field ($s=0$) and the charged massless Dirac field ($s=\pm1/2$), with the field charge parameter $\qBH$. See appendix \ref{app:units} for the units used in this paper.
The field equation for a $U(1)$ gauge field ($s =  \pm 1$) on the Kerr--Newman--de Sitter background is also included in the unified Teukolsky equation by setting $\qBH = 0$.
We note that this gauge field should be distinguished from perturbations of the electromagnetic field supporting the Kerr--Newman--de Sitter background, which are coupled to gravitational perturbations~\cite{Dudley:1977zz,Dudley:1978vd}.
Following the unified formulation of Refs.~\cite{Suzuki:1998vy,Suzuki:1999nn,Motohashi:2021zyv}, we formally treat the spin weight $s$ and field charge $\qBH$ appearing in the separated equations as arbitrary independent parameters.

In each setup explained above, the Teukolsky equations can be expressed as follows. Let $\psi$ denote a master variable constructed from the Newman--Penrose scalar of spin weight $s$. Then, the field equations admit a separation of variables of the form 
\begin{align}
\psi(\tau,r,\vartheta,\varphi) = R(r)\,S(\vartheta)\,\mathrm{e}^{-i\omega \tau}\mathrm{e}^{im\varphi}.
\end{align}
Here, the constants $\omega$ and $m$ are the frequency and azimuthal quantum number, respectively.
Then, the equations for radial and angular variables are given by
\begin{align}
&\Biggl[
\frac{d}{dx}\left( \Delta_{x}(x)(1-x^2)\frac{d}{dx}\right)
+ \lambda - s \left( 1 - \frac{\aBH^2}{L^2} \right) - 2 \frac{\aBH^2}{L^2} x^2\\
&\qquad + \frac{4 s x \Xi}{\Delta_{x}(x)} \left( \frac{\aBH^2}{L^2} m  - \aBH \Xi \omega \right)
- \frac{\Xi^2 \left(m + s x - (1-x^2)\aBH \omega \right)^2}{\Delta_{x}(x) (1-x^2)}
\Biggr] S(x) = 0, \label{Teukolsky a}
\end{align}
and
\begin{align}
&\Biggl[
\Delta_{r}^{-s}\frac{d}{dr}\Delta_{r}^{s+1}\frac{d}{dr}
+ \frac{\mathcal{K}^2 - i s \mathcal{K} \Delta_{r}'}{\Delta_{r}}
+ 2is \mathcal{K}'
- 2(s+1)\left(2s+1\right) \frac{r^2}{L^2}
+ 2s \left(1 - \frac{\aBH^2}{L^2}\right)
- \lambda
\Biggr] R(r) = 0, \label{Teukolsky r}
\end{align}
where $\lambda$ is the separation constant, $x \coloneqq \cos \vartheta$, and
\begin{align}
    \Delta_x(x) &\coloneqq \Delta_{\vartheta}(\cos^{-1}(x)) = 1 + \frac{\aBH^2}{L^2} x^2 \\
    \mathcal{K}(r) &\coloneqq \Xi (r^2 + \aBH^2) \left( \omega 
   - \frac{1}{\Xi}\frac{r \QBH}{r^2 + \aBH^2} \qBH
  - \frac{\aBH}{r^2 + \aBH^2} m
   \right) .
\end{align}
The former and latter equations are referred to as the angular and radial Teukolsky equations, respectively.

\subsection{Angular Teukolsky equation}
\label{sec:angular Teukolsky}
When $\aBH = 0$, the angular Teukolsky equation reduces to the spin-weighted spherical harmonic equation, whose spectrum is $\lambda = \ell(\ell + 1) - s (s - 1)$, with $\ell=|s|,|s|+1,\ldots$ and $m=-\ell,-\ell+1,\ldots,\ell$. In contrast, when $\aBH \neq 0$, the equation reduces to the Heun equation. In this subsection, we demonstrate this reduction and discuss how the spectrum is determined from the connection formula.

\subsubsection{Reduction to Heun equation}

The angular Teukolsky equation has regular singular points at the zeros of $(1 - x^2) \Delta_{x}(x)$ and $x = \infty$, that is, 
\begin{align}
x = \pm 1, \pm i \frac{L}{\aBH}, \infty.
\end{align}
Let us introduce a new variable,
\begin{align}
z \equiv \frac{-2}{x - 1} \frac{x + i L/\aBH}{- 1 + i L/\aBH},
\end{align}
which maps the regular singular points as 
\begin{align}
z(-1) &= 1, \\
z(1) &= \infty, \\
z(i L/\aBH) &= \frac{- 4 i L/\aBH}{(1 - i L/\aBH)^2} \equiv t, \\
z(- i L/\aBH) &= 0, \\
z(\infty) &=  \frac{2}{1 - iL/\aBH} \equiv z_{\infty}.
\end{align}
The regular singularity at $z = z_{\infty}$ can be removed by defining a new function $y(z)$ through
\begin{align}
S(x) = z^{A_{0}} (z-1)^{A_{1}} (z - t)^{A_{t}} (z - z_\infty) y(z),
\end{align}
with
\begin{align}
A_{0} &= \frac{1}{2} \left(
s + i \left( - \frac{\aBH}{L} m + \left(1 + \frac{\aBH^2}{L^2} \right) L \omega \right)
\right),
\\
A_{1} &= \frac{1}{2} (s - m),\\
A_{t} &=  \frac{1}{2} \left(
s - i \left( - \frac{\aBH}{L} m + \left(1 + \frac{\aBH^2}{L^2} \right) L \omega \right)
\right),
\end{align}
the angular Teukolsky equation reduces to the Heun equation~\eqref{Heun eq Fuchs},
\begin{align}
\left[ \frac{d^2}{d z^2} + \left( \frac{\gammaHeun}{z} + \frac{\deltaHeun}{z-1} + \frac{\epsilonHeun}{z - t} \right) \frac{d}{dz} + \frac{\alphaHeun \betaHeun z - q}{z(z - 1)(z - t)} \right] y(z) = 0.
\end{align}
The six parameters are identified as 
\begin{align}
\alphaHeun &\equiv 2s + 1\\
\betaHeun &\equiv s - m + 1\\
\gammaHeun &\equiv 1 + 2 A_{0}\\
\deltaHeun &\equiv 1 + 2 A_{1}\\
\epsilonHeun &\equiv 1 + 2 A_{t}\\
q &\equiv 
\frac{1}{\left(1 + i \frac{\aBH}{L}\right)^2} \left(- \lambda +2s + 2i(1-m+s)(1+2s) \frac{\aBH}{L} + 2 (m-s+2ms)\frac{\aBH^2}{L^2} 
- 2 \frac{\aBH}{L} \left(1 + \frac{\aBH^2}{L^2}\right)(1+2s) L \omega
\right)
.
\end{align}
Note that $\alphaHeun$ and $\betaHeun$ are chosen so that they satisfy the Fuchs relation \eqref{Fuchs relation}.
From these expressions, the parameters in the Heun equation in the Schr\"{o}dinger form can be obtained as
\begin{align}
a_{0} &= \frac{1- \gammaHeun}{2} = - A_{0} =  - \frac{1}{2} \left(
s + i \left( - \frac{\aBH}{L} m + \left(1 + \frac{\aBH^2}{L^2} \right) L \omega \right)
\right)
,\\
a_{1} &=  \frac{1- \deltaHeun}{2} = - A_{1} = \frac{m - s}{2}, \\
a_{t} &= \frac{1 - \epsilonHeun}{2} = - A_{t}=  - \frac{1}{2} \left(
s - i \left( - \frac{\aBH}{L} m + \left(1 + \frac{\aBH^2}{L^2} \right) L \omega \right)
\right),\\
a_{\infty} &= \frac{\alphaHeun - \betaHeun}{2} = \frac{m+s}{2},
\end{align}
and
\begin{align}
 u &= \frac{\gammaHeun \epsilonHeun (1 - t) - \deltaHeun \epsilonHeun t - 2 q + 2 t \alphaHeun \betaHeun}{2(t-1)}
 \notag\\
 &= \frac{1}{\left(1 - i \frac{\aBH}{L}\right)^2} \left( - \lambda
- \frac{1}{2}(1-s)^2 + i (1 + 2 m s - s^2) \frac{\aBH}{L} + \frac{1}{2}(1-2s + (2m + s)^2)\frac{\aBH^2}{L^2}
 \right)
  \notag\\
 &\qquad 
 - \frac{m^2}{2} \frac{\aBH^2}{L^2} - \frac{1}{2} \left(1 + \frac{\aBH^2}{L^2}\right)^2 L^2 \omega^2
 - \frac{\aBH}{L}\left(1 + i \frac{\aBH}{L}\right)^2 m L \omega - 2 \frac{\aBH}{L}\frac{(1+i\frac{\aBH}{L}) (s+i m \frac{\aBH}{L})}{1 - i \frac{\aBH}{L}} L \omega.
\end{align}

\subsubsection{Angular eigenvalue problem}
We now turn to the angular eigenvalue problem. The physically relevant solutions are required to be regular at both poles, $x = \pm 1$, and this regularity condition discretizes the separation constant $\lambda$.
In the present convention, regularity requires
\begin{align}
S(x) \approx 
\begin{cases}
(1-x)^{|m+s|/2} &(x \to 1)\\
(1+x)^{|m-s|/2} &(x \to -1)
\end{cases}.
\end{align}

In our mapping to the Heun equation, the poles $x=+ 1$ and $x = -1$ correspond to $z = \infty$ and $z=1$, respectively.
Using the local Heun functions, we obtain the following two local solutions around $z = \infty$:
\begin{align}
S_{\infty_{-}}(z) &= z^{A_{0}} (z-1)^{A_{1}}(z-t)^{A_{t}} (z - z_{\infty}) y_{\infty_{-}}(z), \\
S_{\infty_{+}}(z) &= z^{A_{0}} (z-1)^{A_{1}}(z-t)^{A_{t}} (z - z_{\infty}) y_{\infty_{+}}(z),
\end{align}
The corresponding solutions around $z = 1$ are
\begin{align}
S_{1_{-}}(z) &= z^{A_{0}} (z-1)^{A_{1}}(z-t)^{A_{t}} (z - z_{\infty}) y_{1_{-}}(z), \\
S_{1_{+}}(z) &= z^{A_{0}} (z-1)^{A_{1}}(z-t)^{A_{t}} (z - z_{\infty}) y_{1_{+}}(z), 
\end{align}
The solutions around $z = \infty$ have the following asymptotic behavior:
\begin{align}
S_{\infty_{-}} &\approx z^{A_{0}+A_{1}+A_{t}+1 - \betaHeun}
=z^{+\frac{m+s}{2}} \propto (1-x)^{- \frac{m+s}{2}},
\\
S_{\infty_{+}} &\approx z^{A_{0}+A_{1}+A_{t}+1 - \alphaHeun} = z^{- \frac{m+s}{2}} \propto (1 - x)^{+ \frac{m+s}{2}}.
\end{align}
Similarly, the solutions around $z = 1$ have the following asymptotic behavior:
\begin{align}
S_{1_{-}} &\propto (z-1)^{A_{1}} = (z-1)^{- \frac{m-s}{2}} \propto (x+1)^{- \frac{m-s}{2}},
\\
S_{1_{+}} &\propto (z-1)^{A_{1}+1-\deltaHeun} = (z-1)^{+\frac{m-s}{2}}
\propto (x+1)^{+\frac{m-s}{2}}.
\end{align}
Thus,  by introducing $\sigma_{\pm} \coloneqq \text{sgn}(m \pm s)$ for $m\pm s\neq0$, the regular boundary condition can be represented as
\begin{align}
S(z) \approx 
\begin{cases}
S_{\infty_{\sigma_{+}}}(z) & (z \to \infty),\\
S_{1_{\sigma_{-}}}(z) & (z \to 1)
\end{cases}.
\end{align}
In terms of the connection formula of local Heun functions between $z=1$ and $z = \infty$, this boundary condition can be expressed by the condition for the connection matrix as
\begin{align}
C_{1_{\sigma_{-}}}{}^{
\infty_{-\sigma_{+}}}  = \mathrm{e}^{-i\pi\left(1+\sigma_{-}\right)a_{1}}\left(1-t\right)^{\frac{1}{2}-a_{t}}\mathrm{e}^{\frac{1}{2}\left(\sigma_{-}\partial_{a_{1}}+\sigma_{+}\partial_{a_{\infty}}\right)F}\mathcal{M}_{\sigma_{-}(-\sigma_{+})}\left(a_{1},a_{\infty};a\right) = 0, \label{angular quantization conditon}
\end{align}
with
\begin{align}
\mathcal{M}_{\sigma_{-}(-\sigma_{+})}\left(a_{1},a_{\infty};a\right) = \frac{\Gamma\left(2\sigma_{+}a_{\infty}\right)\Gamma\left(1+2\sigma_{-} a_{1}\right)}{\Gamma\left(\frac{1}{2}+\sigma_{-} a_{1}+\sigma_{+}a_{\infty}+a\right)\Gamma\left(\frac{1}{2}+\sigma_{-} a_{1} + \sigma_{+}a_{\infty}-a\right)}. \label{Msmmsp}
\end{align}
Since $F$ and $a$ are given as a series of $t$, the condition \eqref{angular quantization conditon} provides the analytic quantization condition for the angular Heun equation.

In particular, if the prefactor in Eq.~\eqref{angular quantization conditon} is finite, the angular spectrum is determined by the poles of the gamma functions in the denominator of Eq.~\eqref{Msmmsp}. Thus, the quantization condition is
\begin{align}
\frac{1}{2} + \sigma_{-}a_{1} + \sigma_{+} a_{\infty} \pm a = - N,
\end{align}
where $N \in \mathbb{Z}_{\geq 0}$.
This condition can be expressed as 
\begin{align}
a^2 = \left( \ell + \frac{1}{2}  \right)^2, \label{angular quantization conditon 2}
\end{align}
where we have introduced $\ell$ instead of $N$, which is defined by 
\begin{align}
\ell \coloneqq N + \max(|m|, |s|) = \max(|m|, |s|), \max(|m|, |s|) + 1, \cdots, 
\end{align}
and used the expression
\begin{align}
\sigma_{-} a_{1} &= \frac{|m-s|}{2},\qquad 
\sigma_{+} a_{\infty} = \frac{|m+s|}{2},
\end{align}
and the identity
\begin{align}
\frac{|m-s| + |m + s|}{2} = \max(|m|,|s|).
\end{align} 

The above analysis is directly valid when $m-s$ and $m+s$ are non-integers, so that the two Frobenius solutions at each pole form a nonresonant local basis. In the physical Teukolsky problem, however, both $m-s$ and $m+s$ are integers, and the corresponding local exponents are resonant. For nonzero exponent differences, the solutions associated with the larger local exponents, $y_{1\sigma_-}$ and $y_{\infty\sigma_+}$, possess finite, logarithm-free limits and continue to represent the solutions regular at the two angular poles. By contrast, the smaller-exponent solutions, $y_{1,-\sigma_-}$ and $y_{\infty,-\sigma_+}$, generally become singular in the resonant limit: they develop terms proportional to the regular solutions, while their finite parts define the corresponding logarithmic companion solutions.

The resonant connection formula may therefore be obtained by approaching the integer exponent differences from the nonresonant case. In this limiting procedure, the divergent contributions proportional to the regular solution cancel between the two terms in the connection formula. This cancellation can substantially modify the coefficient multiplying the regular solution. By contrast, the coefficient of the logarithmic companion is given, up to a finite nonzero normalization, by the limiting value of the nonresonant regular-to-singular connection coefficient. Consequently, for nonzero exponent differences, regularity at both poles is still imposed by
\begin{align}
\lim_{\text{resonant}} C_{1_{\sigma_{-}}}{}^{\infty_{-\sigma_{+}}} = 0.
\end{align}
Thus, the quantization condition derived above by analytic continuation from non-integer $m-s$ and $m+s$ remains valid in the physical resonant case. When $m-s=0$ or $m+s=0$, the two local exponents coincide and the basis must first be replaced by a regular solution and a normalized logarithmic companion. Regularity then requires the coefficient of this logarithmic companion to vanish; the unrescaled limit of the connection coefficient above need not vanish. Equivalently, the Wronskian of the solutions regular at the two poles must vanish. This prescription gives the same condition \eqref{angular quantization conditon 2}.

We have therefore reduced the angular eigenvalue problem to the quantization condition~\eqref{angular quantization conditon} or~\eqref{angular quantization conditon 2}. Together with the series expression for $a$, this condition determines the separation constant $\lambda$. Although these relations give an exact implicit description of the angular spectrum, extracting an explicit expression requires solving them order by order in an appropriate expansion parameter. We postpone this analysis to Sec.~\ref{sec:angular spectrum small aoverL}, where the angular eigenvalues are obtained perturbatively in $\aBH/L$.

\subsection{Radial Teukolsky equation}
\subsubsection{Reduction to the Heun equation}
For later convenience, we denote the four roots by $r_{0}, r_{1}, r_{t}, r_{\infty}$ without yet specifying their correspondence to the physical horizons. In terms of these roots, $\Delta_r(r)$ can be factorized as
\begin{align}
\Delta_r(r) = - \frac{1}{L^2}(r - r_{0})(r - r_{1})(r - r_{t})(r - r_{\infty}). \label{Delta by poles}
\end{align}
By comparing Eqs.~\eqref{def Delta} and \eqref{Delta by poles}, we obtain
\begin{align}
r_{0} + r_{1} + r_{t} + r_{\infty} = 0,
\end{align}
and
\begin{align}
M &= - \frac{r_{0}r_{1}r_{t}r_{\infty}}{2L^2} \left(\frac{1}{r_{0}} + \frac{1}{r_{1}} +\frac{1}{r_{t}} +\frac{1}{r_{\infty}} \right), \\
\QBH^2 &= - L^2 - \frac{r_{0}r_{1}r_{t}r_{\infty}}{L^2} - r_{0} r_{1} - r_{0} r_{t} - r_{0} r_{\infty} - r_{1}r_{t} - r_{1}r_{\infty} - r_{t} r_{\infty},\\
\aBH^2 &= L^2 + r_{0} r_{1} + r_{0} r_{t} + r_{0} r_{\infty} + r_{1}r_{t} + r_{1}r_{\infty} + r_{t} r_{\infty}.
\end{align}

We introduce the variable $z$ by
\begin{align}
z(r) = \frac{r_{1} - r_{\infty}}{r - r_{\infty}} \frac{r - r_{0}}{r_{1} - r_{0}}. \label{def z(r)}
\end{align}
The four zeros of $\Delta_r(r)$, namely, $r_{0}, r_{1}, r_{t}, r_{\infty}$, together with the point $r = \infty$, are mapped to
\begin{align}
z(r_{0}) &= 0, \\
z(r_{1}) &= 1, \\
z(r_{t}) &= \frac{r_{1} - r_{\infty}}{r_{t} - r_{\infty}} \frac{r_{t} - r_{0}}{r_{1} - r_{0}} \eqqcolon t,\label{eq:tDef} \\
z(r_{\infty}) &= \infty, \\
z(\infty) & =  \frac{r_{1} - r_{\infty}}{r_{1} - r_{0}} \eqqcolon z_{\infty}.
\end{align}
Thus, the subscripts labeling the zeros of $\Delta_r(r)$ have been chosen to match their images under the map \eqref{def z(r)}.

The radial Teukolsky equation \eqref{Teukolsky r} is a Fuchsian differential equation with five regular singular points at $r = r_{0}, r_{1}, r_{t}, r_{\infty}, \infty$, which correspond to $z = 0, 1, t, \infty, z_{\infty}$, respectively.
The regular singular point at $r = \infty$ ($z = z_{\infty}$) can be removed through the transformation
\begin{align}
R =  z^{B_{0}} (z - 1)^{B_{1}} (z - t)^{B_{t}}(z - z_{\infty})^{2s + 1} y(z),
\end{align}
where
\begin{align}
B_{0} \coloneqq B(r_{0}), \quad 
B_{1} \coloneqq B(r_{1}), \quad
B_{t} \coloneqq B(r_{t}),
\end{align}
and
\begin{align}
    B(r) \coloneqq i \frac{\mathcal{K}(r)}{\Delta_r'(r)}.
\end{align}
We also introduce
\begin{align}
B_{\infty} \coloneqq B(r_{\infty}), 
\end{align}
for later convenience. These constants satisfy
\begin{align}
B_{0} + B_{1} + B_{t} + B_{\infty} = 0.
\end{align}

By using the expression for $\mathcal{K}$ and the surface gravity $\kappa_{i}$, we can express $B(r_{i})$ as 
\begin{align}
B(r_{i}) =  i \frac{\mathcal{K}(r_{i})}{\Delta_r'(r_{i})} =  \frac{i}{2} \frac{\omega - \Phi_{i} \qBH - \Omega_{i} m}{\kappa_{i}}.
\end{align}

Rewriting as the differential equation for $y(z)$, the radial Teukolsky equation reduces to the Heun equation~\eqref{Heun eq Fuchs},
\begin{align}
\left[ \frac{d^2}{d z^2} + \left( \frac{\gammaHeun}{z} + \frac{\deltaHeun}{z-1} + \frac{\epsilonHeun}{z - t} \right) \frac{d}{dz} + \frac{\alphaHeun \betaHeun z - q}{z(z - 1)(z - t)} \right] y(z) = 0,
\end{align}
with the parameters identified as 
\begin{align}
\alphaHeun &\equiv 
s + 1 - 2 B_{\infty}, \\
\betaHeun &\equiv 
2 s + 1,\\
\gammaHeun &\equiv 2B_{0} + s + 1, \\
\deltaHeun &\equiv 2B_{1} + s + 1, \\
\epsilonHeun &\equiv 2B_{t} + s + 1, \\
q &\equiv - \frac{\lambda L^2 - 2 s (L^2 - \aBH^2) - (s + 1) (2s + 1)(r_{0}r_{\infty} + r_{1} r_{t} )}{(r_{\infty} - r_{t})(r_{0} - r_{1})} \notag\\
& \qquad + i \frac{L^2(2s + 1) \left(  2 \Xi  (\omega (r_{0}r_{\infty} +  \aBH^2) - \aBH m ) - \qBH \QBH (r_{0} + r_{\infty}) \right) }{(r_{\infty} - r_{t})(r_{\infty} - r_{0})(r_{0} - r_{1})}.
\end{align}
Then, the parameters in the Heun equation in the Schr\"{o}dinger form can be obtained as
\begin{align}
a_{0} &= \frac{1- \gammaHeun}{2} = - B_{0} - \frac{1}{2} s 
,\\
a_{1} &=  \frac{1- \deltaHeun}{2} = - B_{1} - \frac{1}{2} s,\\
a_{t} &= \frac{1 - \epsilonHeun}{2} = - B_{t} - \frac{1}{2} s,\\
a_{\infty} &= \frac{\alphaHeun - \betaHeun}{2} = - B_{\infty} - \frac{1}{2} s,
\end{align}
and
\begin{align}
 u = \frac{\gammaHeun \epsilonHeun (1 - t) - \deltaHeun \epsilonHeun t - 2 q + 2 t \alphaHeun \betaHeun}{2(t-1)}.
\end{align}

Using the expressions for $B_{i}$, the combinations appearing in $a_{i}$ can be expressed as
\begin{align}
B_{i} + \frac{s}{2} = \frac{1}{2} \left( i \frac{\omega - \Phi_{i} \qBH - \Omega_{i} m}{\kappa_{i}} + s  \right). \label{ai}
\end{align}

\subsubsection{Quasinormal modes}
\label{subsec:QNM condition}
Let us express the condition for the quasinormal modes in terms of the connection coefficients.

For this purpose, let us clarify the asymptotic behavior of the function $R$ corresponding to each local solution around a pole.
Specifically, for $i \in \{0,t,1,\infty\}$, we denote the radial functions corresponding to the local solutions $y_{i_{-}}$ and $y_{i_{+}}$ by $R_{i_{-}}$ and $R_{i_{+}}$, respectively.
Thus, the asymptotics of $R_{0_{\mp}}$ are evaluated as 
\begin{align}
R_{0_{-}} &= z^{B_{0}} (z - 1)^{B_{1}} (z - t)^{B_{t}}(z - z_{\infty})^{2s + 1} y_{0_{-}}(z) \propto z^{B_{0}} \propto (r - r_{0})^{B_{0}},\\
R_{0_{+}} &= z^{B_{0}} (z - 1)^{B_{1}} (z - t)^{B_{t}}(z - z_{\infty})^{2s + 1} y_{0_{+}}(z) \propto z^{B_{0} + 1 - \gammaHeun} = z^{- B_{0} - s} \propto (r - r_{0})^{-B_{0}-s},
\end{align}
and similar calculations hold for $R_{1_{\mp}}$ and $R_{t_{\mp}}$.
$R_{\infty_{\mp}}$ can also be evaluated as 
\begin{align}
R_{\infty_{-}} &= z^{B_{0}} (z - 1)^{B_{1}} (z - t)^{B_{t}}(z - z_{\infty})^{2s + 1} y_{\infty_{-}}(z) \propto z^{B_{0} + B_{1} + B_{t} + 2s+1 - \betaHeun} = \left(\frac{1}{z}\right)^{ B_{\infty}} \propto (r-r_{\infty})^{B_{\infty}},\\
R_{\infty_{+}} &= z^{B_{0}} (z - 1)^{B_{1}} (z - t)^{B_{t}}(z - z_{\infty})^{2s + 1} y_{\infty_{+}}(z) \propto z^{B_{0} + B_{1} + B_{t} + 2 s + 1 - \alphaHeun} = \left(\frac{1}{z}\right)^{ - B_{\infty} - s} \propto (r - r_{\infty})^{-B_{\infty} - s}.
\end{align}
Thus the results can be summarized as 
\begin{align}
    R_{i_{-}} &\propto (r - r_{i})^{B_{i}},\\
    R_{i_{+}} &\propto (r - r_{i})^{- B_{i} - s}.
\end{align}

Let us define the tortoise coordinate $r_{*}$ by
\begin{align}
\frac{d r_{*}}{d r} \coloneqq \Xi \frac{r^2 + \aBH^2}{\Delta_r(r)} = \Xi \frac{r_{i}^2 + \aBH^2}{\Delta_r'(r_{i})} \frac{1}{r - r_{i}} (1 + \mathcal{O}(r - r_{i})).
\end{align}
Integrating this around $r \approx r_{i}$, we obtain
\begin{align}
    r_{*} \approx \Xi \frac{r_{i}^2 + \aBH^2}{\Delta_r'(r_{i})} \log|r - r_{i}|,
\end{align}
up to an additive constant.
Therefore, we obtain
\begin{align}
r - r_{i} \approx \mathrm{e}^{\frac{\Delta_r'(r_{i})}{\Xi (r_{i}^2 + \aBH^2)} r_{*} + \text{const.}}
\end{align}
Then, the asymptotics of $R_{i_{-}}$ are expressed in terms of $r_{*}$ as 
\begin{align}
R_{i_{-}} &\propto \mathrm{e}^{\frac{\Delta_r'(r_{i})}{\Xi (r_{i}^2 + \aBH^2)}  B_{i} r_{*} }
= \mathrm{e}^{ i \frac{\mathcal{K}(r_{i})}{\Xi (r_{i}^2 + \aBH^2)} r_{*}}
= \mathrm{e}^{ + i \left( \omega - \Phi_{i} \qBH - \Omega_{i} m \right) r_{*}}, \\
R_{i_{+}} &\propto \mathrm{e}^{- \frac{\Delta_r'(r_{i})}{\Xi (r_{i}^2 + \aBH^2)}  (B_{i} + s ) r_{*} }
= \mathrm{e}^{ - i \frac{\mathcal{K}(r_{i})}{\Xi (r_{i}^2 + \aBH^2)} r_{*} - \frac{\Delta_r'(r_{i})}{\Xi (r_{i}^2 + \aBH^2)} s r_{*}}
= \mathrm{e}^{ - i \left( \omega - \Phi_{i} \qBH - \Omega_{i} m \right) r_{*} - 2 \kappa_{i} s r_{*} }.
\end{align}
Therefore, if we identify the outer black hole horizon $r_{+}$ with $r_{i^{+}}$ and the cosmological horizon $r_{c}$ with $r_{i^{c}}$, then
\begin{align}
R_{i^{+}_{-}} &\approx \mathrm{e}^{+ i (\omega + \cdots ) r_{*}}  \text{: out mode},\\
R_{i^{+}_{+}} &\approx \mathrm{e}^{- i (\omega + \cdots ) r_{*}}  \text{: in mode},
\end{align}
and
\begin{align}
R_{i^{c}_{-}} &\approx \mathrm{e}^{+ i (\omega + \cdots ) r_{*}}  \text{: up mode},\\
R_{i^{c}_{+}} &\approx \mathrm{e}^{- i (\omega + \cdots ) r_{*}}  \text{: down mode}.
\end{align}

Quasinormal modes are defined by the boundary conditions
\begin{align}
R(r) \propto 
\begin{cases}
    \text{in mode}: R_{i^{+}_{+}} & (r \to r_{+}),\\
    \text{up mode}: R_{i^{c}_{-}} & (r \to r_{c}).
\end{cases}
\end{align}
Since the connection formula relates these modes as
\begin{align}
R_{i^{c}_{-}} =  C_{i^{c}_{-}}{}^{i^{+}_{+}} R_{i^{+}_{+}} + C_{i^{c}_{-}}{}^{i^{+}_{-}} R_{i^{+}_{-}}, 
\end{align}
or inversely,
\begin{align}
R_{i^{+}_{+}} =  C_{i^{+}_{+}}{}^{i^{c}_{+}} R_{i^{c}_{+}} + C_{i^{+}_{+}}{}^{i^{c}_{-}} R_{i^{c}_{-}}, 
\end{align}
the condition for the quasinormal modes can be expressed as 
\begin{align}
C_{i^{c}_{-}}{}^{i^{+}_{-}} = 0,\quad 
\text{or}
\quad
C_{i^{+}_{+}}{}^{i^{c}_{+}} = 0. \label{QNM conditions by C}
\end{align}

Depending on the mapping of the Teukolsky equation to the Heun equation, that is, the choice of the assignment $(r_{0}, r_{t}, r_{1}, r_{\infty})$, one can obtain apparently different expressions for the QNM condition. For example, if one identifies four horizons as $(r_{0}, r_{t}, r_{1}, r_{\infty}) = (r_{n}, r_{c}, r_{+}, r_{-})$, as we will do in Sec.~\ref{sec:flat limit}, the condition can be given by
\begin{align}
C_{t_{-}}{}^{1_{-}} = 0.
\end{align}
The connection coefficient is given explicitly by
Eq.~\eqref{C t 1} in terms of the parameters $a_{0}, a_{1}, a_{t}, a_{\infty}, a$, and $F(t)$. In principle, $a$ and $F(t)$ can be calculated from Eqs.~\eqref{eq:u0-Def} and \eqref{eq:CCB-Def}, and their expansions through $\mathcal{O}(t)$ are given by Eqs.~\eqref{eq:CCB-Ldg} and \eqref{a leading}. This provides an exact analytic quantization condition for the quasinormal spectrum.

In the following sections, we focus on controlled limits in which the condition for the quasinormal spectrum can be expressed in closed form.
The basic idea is to approximate the infinite series for $F(t)$ using the leading terms in its small-$t$ expansion.
In the limits considered below, the relevant Heun parameters remain finite as $t\to0$, so that the coalescence of two regular singular points reduces the Heun equation to a hypergeometric equation.
For the radial problem, such a small-$t$ degeneration is associated with two horizon radii either coalescing or moving to infinity.
We therefore consider separately the $\Lambda \to 0$ limit ($|r_{c}|, |r_{n}| \to \infty$), the black hole extremal limit ($r_{-} \to r_{+}$), and the Nariai limit ($r_{c} \to r_{+}$).

\section{Quasinormal Modes in Controlled Limits}
\label{sec:QNM}
In this section, we focus on four limits:
the $\aBH/L \to 0$ limit for the angular Teukolsky equation and the small cosmological constant limit $L \to \infty$, the black hole extremal limit $M \to M_{\mathrm{E}}$, and the Nariai limit $M \to M_{\mathrm{N}}$ for the radial Teukolsky equation. Under appropriate mappings of the Teukolsky equation to the Heun equation, each of these limits can be regarded as a small-$t$ limit, and the quantization conditions can be evaluated perturbatively in closed form using the small-$t$ expansion of $F(t)$.

\subsection{Angular eigenvalue spectrum in the \texorpdfstring{$\aBH/L \to 0$}{a/L → 0} limit}
\label{sec:angular spectrum small aoverL}
In Sec.~\ref{sec:angular Teukolsky}, 
we obtained the quantization condition~\eqref{angular quantization conditon} or~\eqref{angular quantization conditon 2} for the angular Teukolsky equation.
In this subsection, we solve this condition perturbatively in the small-$\aBH/L$ regime while keeping $L \omega$ finite and derive an explicit expansion of the angular eigenvalue $\lambda$.

The important point is that the small $\aBH/L$ expansion corresponds to the small $|t|$ expansion because 
\begin{align}
t = 4i \frac{\frac{\aBH}{L}}{(1 + i \frac{\aBH}{L})^2} \approx 4 i \frac{\aBH}{L} + \mathcal{O}((\aBH/L)^{2}).
\end{align}
Therefore, we can use the leading-order expressions for $F$ and $a$ in the $t$-expansion, obtained from Eqs.~\eqref{eq:CCB-Ldg} and \eqref{a leading}:
\begin{align}
F &= \mathcal{O}(\aBH/L), \\ 
a &= \sqrt{- \frac{1}{4} - u + a_{t}^2 + a_{0}^2} + \mathcal{O}(\aBH/L).
\end{align}
Using the expressions for $u$, $a_{t}$, and $a_{0}$ obtained in Sec.~\ref{sec:angular Teukolsky}, we obtain
\begin{align}
a = \sqrt{\lambda + \left( s - \frac{1}{2} \right)^2} + \mathcal{O}(\aBH/L).
\end{align}
With the resonant cases understood as described in Sec.~\ref{sec:angular Teukolsky}, we can use
the quantization condition \eqref{angular quantization conditon 2}. Therefore, we obtain the angular eigenvalue spectrum as 
\begin{align}
\lambda + \left( s - \frac{1}{2} \right)^2 + \mathcal{O}(\aBH/L) = \left(\ell + \frac{1}{2}\right)^2,
\end{align}
that is,
\begin{align}
\lambda = \ell \left( \ell + 1 \right) - s \left(s - 1 \right) + \mathcal{O}(\aBH/L), \label{lambda leading}
\end{align}
at leading order in the $\aBH/ L$ expansion.

We can straightforwardly extend the analysis to higher order. Let us derive the next-to-leading-order correction. Introducing $\delta \lambda$ by
\begin{align}
\lambda =  \ell \left( \ell + 1 \right) - s \left(s - 1 \right)  + \delta \lambda \frac{\aBH}{L} + \mathcal{O}((\aBH/L)^2),
\end{align}
and using the $\mathcal{O}(t)$ expression for $a$ in Eq.~\eqref{a leading}, we obtain
\begin{align}
a^2 = \left( \ell + \frac{1}{2} \right)^2 + \left( 2 m \frac{(s^2 + \ell(\ell + 1))}{\ell ( \ell + 1)} L \omega  + \delta \lambda \right) \frac{\aBH}{L} + \mathcal{O}((\aBH/L)^2).
\end{align} 
Therefore, the quantization condition requires
\begin{align}
\delta \lambda = - 2 m \frac{s^2 + \ell (\ell+1)}{\ell (1 + \ell)} L \omega. \label{lambda correction}
\end{align}
Equation~\eqref{lambda correction} applies to $\ell>0$; for $s=m=\ell=0$, the angular equation gives $\delta\lambda=0$. This convention also applies to Eq.~\eqref{eq:dlDef} below.
This result agrees with the expansion of the separation constant to first order in $\aBH\omega$ obtained in Ref.~\cite{Suzuki:1998vy,Novaes:2018fry,Arnaudo:2025btb}.\footnote{
The separation constant ${}_sA_{\ell m}$ used in Ref.~\cite{Arnaudo:2025btb} is related to our separation constant $\lambda$ by
$\lambda={}_sA_{\ell m}+s(1-\aBH^2/L^2)$.
}
We note that the $\mathcal{O}(\aBH/L)$ correction depends on $\omega$, which will be determined by the quantization condition for the radial Teukolsky equation. 
For example, in the next subsection, we will derive the analytic expression for $L \omega$ in the $L \to \infty$ limit.

\subsection{Quasinormal modes in the small cosmological constant limit}
\label{sec:flat limit}
In this subsection, we examine the limit $\Lambda \to 0$ in detail.
We hold $M$, $\aBH$, $\QBH$, and $\qBH$ fixed and assume $M^2>\aBH^2+\QBH^2$, so that the limiting Kerr--Newman black hole is subextremal.
Since we use the parameter $L = \sqrt{3/\Lambda}$, this limit corresponds to $L \to \infty$. In this limit, the horizon radii can be evaluated as 
\begin{align}
r_{+} &= r^{\text{KN}}_{+} + \mathcal{O}(L^{-2}), \\
r_{-} &= r^{\text{KN}}_{-} + \mathcal{O}(L^{-2}),
\end{align}
and 
\begin{align}
r_{c} &= L + \mathcal{O}(L^{0}), \\
r_{n} &= - L + \mathcal{O}(L^{0}),
\end{align}
where $r_{\pm}^{\text{KN}}$ are the horizon radii of the Kerr--Newman black hole, 
\begin{align}
r_{\pm}^{\text{KN}} &\coloneqq M \pm \sqrt{M^2 - (\QBH^2 + \aBH^2)}.
\end{align}

The behavior of the surface gravities in this limit can be obtained as 
\begin{align}
\kappa_{c} &= - \frac{1}{L} + \mathcal{O}(L^{-2}), \quad
\Phi_{c} = \frac{\QBH}{L} + \mathcal{O}(L^{-2}), \quad 
\Omega_{c} = \mathcal{O}(L^{-2})
\\ 
\kappa_{n} &= + \frac{1}{L}
+\mathcal{O}(L^{-2}), \quad  
\Phi_{n} = - \frac{\QBH}{L} + \mathcal{O}(L^{-2}), \quad 
\Omega_{n} = \mathcal{O}(L^{-2}),
\\
\kappa_{\pm} &= \kappa^{\text{KN}}_{\pm} + \mathcal{O}(L^{-2}), \quad
\Phi_{\pm} = \Phi^{\text{KN}}_{\pm} + \mathcal{O}(L^{-2}),\quad 
\Omega_{\pm} = \Omega^{\text{KN}}_{\pm} + \mathcal{O}(L^{-2}),
\end{align}
where $\kappa_{\pm}^{\text{KN}}, \Phi_{\pm}^{\text{KN}}, \Omega_{\pm}^{\text{KN}}$ are the corresponding quantities in Kerr--Newman spacetime given by
\begin{align}
\kappa_{\pm}^{\text{KN}} &= \frac{r_{\pm}^{\text{KN}} - M}{ (r_{\pm}^{\text{KN}})^2 + \aBH^2}, \quad
\Phi^{\text{KN}}_{\pm} = \frac{r^{\text{KN}}_{\pm} \QBH}{(r_{\pm}^{\text{KN}})^2 + \aBH^2}, \quad 
\Omega^{\text{KN}}_{\pm} = \frac{\aBH}{(r_{\pm}^{\text{KN}})^2 + \aBH^2}.
\end{align}

One can see that $\kappa_{c}$ and $\kappa_{n}$, which appear in the denominator in Eq.~\eqref{ai}, go to $0$ in the limit $L \to \infty$. 
Therefore, for the modes with finite $\omega$ in $L \to \infty$ limit, the parameters $a_{c}$ and $a_{n}$ diverge, and therefore the systematic small-$t$ expansion of the connection formulae is no longer valid. In the following, we focus on modes that behave as $\omega = \mathcal{O}(L^{-1})$, for which the Heun parameters $a_{i}$ remain finite in the $L \to \infty$ limit.

We express the frequency $\omega$ as a series expansion in $L^{-1}$ as 
\begin{align}
\omega &= \delta \omega L^{-1} + \mathcal{O}(L^{-2},L^{-1 - 2 \bar{a}}), \label{omega expansion large L}
\end{align}
and examine the analytic expression for the leading-order contribution $\delta \omega$.
We use $\mathcal{O}(f,g)$ as shorthand for $\mathcal{O}(|f|+|g|)$, listing the two scales separately to keep track of their respective contributions; the same convention applies to three or more terms.
Here $\bar{a} \coloneqq \lim_{L \to \infty} a$ and we focus on the case in which $\bar{a}$ is positive and real. The contribution $\mathcal{O}(L^{-1-2\bar{a}})$ in Eq.~\eqref{omega expansion large L} is necessary in the higher-order analysis. We will address it at the end of this subsection.

The angular spectrum in the large-$L$ limit, with $\aBH$ held finite, can be obtained from the analysis in the previous section. We obtain
\begin{align}
\lambda = \bar{\lambda} + \aBH \delta \lambda L^{-1} + \mathcal{O}(L^{-2},L^{-1-2\bar{a}}),
\end{align}
with
\begin{align}
\bar{\lambda} &= \ell(\ell + 1) - s(s-1),\\
\delta \lambda &=  - 2 m \frac{s^2 + \ell (\ell+1)}{\ell (1 + \ell)} \delta \omega,
\label{eq:dlDef}
\end{align}
where we have replaced $L \omega$ with $\delta \omega$ using Eq.~\eqref{omega expansion large L}.

Let us identify the zeros of $\Delta_r(r)$ as
\begin{align}
(r_{0}, r_{t}, r_{1}, r_{\infty}) \equiv (r_{n}, r_{c}, r_{+}, r_{-}),
\end{align}
so that the limit $L \to \infty$ corresponds to the small $t$ limit, which is defined in Eq.~\eqref{eq:tDef}. In fact, the parameter $t$ can be evaluated as 
\begin{align}
t &= \frac{r_{1} - r_{\infty}}{r_{t} - r_{\infty}} \frac{r_{t} - r_{0}}{r_{1} - r_{0}} = \frac{r_{+} - r_{-}}{r_{c} - r_{-}} \frac{r_{c} - r_{n}}{r_{+} - r_{n}} = \bar{t} + \mathcal{O}(L^{-2}),
\end{align}
where we define
\begin{align}
\bar{t} \coloneqq  4 \frac{\sqrt{M^2 - (\aBH^2 + \QBH^2)}}{L}. \label{tbar flat}
\end{align}
Then, the parameters in the Heun equation in the Schr\"{o}dinger form can be evaluated as 
\begin{align}
a_{1} &\equiv -\frac{1}{2} \left( i \frac{\omega - \Phi_{+} \qBH - \Omega_{+} m}{\kappa_{+}} + s  \right) = \frac{1}{2}\left( i \frac{\Phi^{\text{KN}}_{+} \qBH + \Omega^{\text{KN}}_{+} m }{\kappa^{\text{KN}}_{+}}  - s \right) + \mathcal{O}(L^{-1},L^{-1-2\bar{a}}), \\ 
a_{\infty} &\equiv - \frac{1}{2} \left( i \frac{\omega - \Phi_{-} \qBH - \Omega_{-} m}{\kappa_{-}} + s  \right) = +\frac{1}{2}\left( i \frac{\Phi^{\text{KN}}_{-} \qBH + \Omega^{\text{KN}}_{-} m }{\kappa^{\text{KN}}_{-}}  - s \right)
+ \mathcal{O}(L^{-1},L^{-1-2\bar{a}}), \\
a_{t} &\equiv -\frac{1}{2} \left( i \frac{\omega - \Phi_{c} \qBH - \Omega_{c} m}{\kappa_{c}} + s  \right) =  \frac{   i (\delta\omega - \qBH \QBH) - s}{2} + \mathcal{O}(L^{-1},L^{-2\bar{a}}), \\
a_{0} &\equiv -\frac{1}{2} \left( i \frac{\omega - \Phi_{n} \qBH - \Omega_{n} m}{\kappa_{n}} + s  \right) = \frac{ i( - \delta\omega -  \qBH \QBH)  - s}{2} + \mathcal{O}(L^{-1},L^{-2\bar{a}}),
\end{align}
and
\begin{align}
u = \frac{1}{2} \left( (\qBH \QBH + i s)^2 - 1 + 2 s - 2 \bar{\lambda} - \delta \omega^2 \right)
+ \mathcal{O}(L^{-1},L^{-2\bar{a}}).
\end{align}
Since the expansion in $L^{-1}$ corresponds to the small-$t$ expansion, we can approximate
$a$ using its leading-order expression in Eq.~\eqref{a leading},
\begin{equation}
a = \sqrt{-\frac{1}{4}-u+a_{t}^{2}+a_{0}^{2}}+\mathcal{O}\left(t\right) = \bar{a} + \mathcal{O}(L^{-1}, L^{-2 \bar{a}}),
\end{equation}
where we obtain the expression
\begin{align}
\bar{a} &= \sqrt{\left( \frac{1}{2} - s \right)^2 + \bar{\lambda} - \qBH^2 \QBH^2} = \sqrt{\left(\ell + \frac{1}{2}\right)^2 - \qBH^2 \QBH^2}.
\label{eq:baraDef}
\end{align}

Since the assignment of the regular singular points in this subsection corresponds to $i^{+} = 1$ and $i^{c} = t$ in the notation in Sec.~\ref{subsec:QNM condition}, the quantization condition \eqref{QNM conditions by C} for the
quasinormal modes can be expressed as 
\begin{align}
C_{t_{-}}{}^{1_{-}} &=  t^{\frac{1}{2} - a_{0} - a_{t}-a}\left(1-t\right)^{a_{t}-a_{1}}\mathrm{e}^{\frac{1}{2}\left(-\partial_{a_{t}}+\partial_{a}+\partial_{a_{1}}\right)F} 
\notag\\
&\quad \times \left(
\mathcal{M}_{--}\left(a_{t},a;a_{0}\right)\mathcal{M}_{+-}\left(a,a_{1};a_{\infty}\right) 
+
 t^{2a} \mathrm{e}^{-\partial_{a}F} \mathcal{M}_{-+}\left(a_{t},a;a_{0}\right)\mathcal{M}_{--}\left(a,a_{1};a_{\infty}\right) 
 \right) = 0, \label{QNM condition C=0 flat}
\end{align}
where we used the expression for the connection coefficient given in Eq.~\eqref{C t 1}. Here,  $\mathcal{M}_{\theta\theta'}\left(c_{1},c_{2};c_{3}\right)$ is defined in Eq.~\eqref{eq:ConMat}. The relevant components are explicitly given by
\begin{align}
\mathcal{M}_{--}\left(a_{t},a;a_{0}\right) &= 
\frac{\Gamma(2 a) \Gamma(1-2 a_{t})}{\Gamma\left(\frac{1}{2} - a_{t} + a + a_{0} \right) \Gamma\left(\frac{1}{2} - a_{t} + a - a_{0} \right)},
\\
\mathcal{M}_{+-}\left(a,a_{1};a_{\infty}\right) 
&=
\frac{\Gamma(2 a_{1}) \Gamma(1 + 2 a) }{\Gamma\left(\frac{1}{2} + a + a_{1} + a_{\infty} \right)
\Gamma\left(\frac{1}{2} + a + a_{1} - a_{\infty} \right)},\\
\mathcal{M}_{-+}\left(a_{t},a;a_{0}\right) &= \frac{\Gamma\left(-2a\right)\Gamma\left(1-2 a_{t}\right)}{\Gamma\left(\frac{1}{2}- a_{t}-a+a_{0}\right)\Gamma\left(\frac{1}{2} - a_{t}- a-a_{0}\right)},\\
\mathcal{M}_{--}\left(a,a_{1};a_{\infty}\right) &= \frac{\Gamma\left(2a_{1}\right)\Gamma\left(1-2 a\right)}{\Gamma\left(\frac{1}{2}- a+a_{1}+a_{\infty}\right)\Gamma\left(\frac{1}{2}- a+a_{1}-a_{\infty}\right)}.
\end{align}
The requirement that $a$ be positive and real in the $L \to \infty$ limit can be satisfied if
\begin{align}
     \ell  > |\qBH \QBH| - \frac{1}{2}.
\end{align}
In deriving the pole condition and its corrections below, we also assume generic nonresonant parameters: the remaining gamma-function factors are finite and nonzero at the selected simple zero, and $2\bar a\notin\mathbb Z$. Resonant cases require taking limits of the full connection condition after resolving simultaneous poles.
Under these assumptions,
the second term in the bracket in Eq.~\eqref{QNM condition C=0 flat} is suppressed by the factor $t^{2\bar{a}}$, and hence the leading-order QNM condition is determined by the zeros of the first term, whose zero arises from a pole of one of the gamma functions in the denominator of $\mathcal{M}_{--}\left(a_{t},a;a_{0}\right)$ or $\mathcal{M}_{+-}\left(a,a_{1};a_{\infty}\right) $.
At leading order, the frequency $\delta \omega$ appears only in the combination $a_{0} - a_{t}$.
Actually, it can be expanded as  
\begin{align}
a_{0} - a_{t} = - i \delta \omega + \mathcal{O}(L^{-1}, L^{-2\bar{a}}).
\end{align}
Therefore, the QNM frequency must correspond to the pole of $\Gamma(\frac{1}{2}-a_{t} + a + a_{0})$, that is,
\begin{align}
\frac{1}{2} - a_{t} + a + a_{0}
=
\frac{1}{2} - i \delta\omega + \bar{a}
+ \mathcal{O}(L^{-1}, L^{-2 \bar{a}})
\equiv  - n + \mathcal{O}(L^{-2 \bar{a}}),
\end{align}
with $n \in \mathbb{Z}_{\geq 0}$.
Solving this condition, we obtain the leading-order QNM spectrum as 
\begin{align}
\delta\omega =  - i \left( n + \frac{1}{2} + \bar{a} \right),
\label{eq:doDef}
\end{align}
that is, the frequency $\omega$ is given by
\begin{align}
\omega 
& = - \frac{i}{L} \left( n + \frac{1}{2} + \bar{a} \right) + \mathcal{O}(L^{-2}, L^{-1 - 2 \bar{a}}) \\
& = - \frac{i}{L} \left( n + \frac{1}{2} + \sqrt{\left( \frac{1}{2} + \ell \right)^2 - \qBH^2 \QBH^2} \right)
 + \mathcal{O}(L^{-2}, L^{-1 - 2 \bar{a}})
.  \label{omega dS leading}
\end{align}
This is the leading-order expression for the QNM spectrum in the large-$L$ (small-$\Lambda$) limit. We note that in the uncharged case $\qBH \QBH = 0$, we recover the result in Ref.~\cite{Arnaudo:2025kit}, 
$
\omega = -i L^{-1} \left( n + \ell + 1 \right) + \mathcal{O}(L^{-2}, L^{-1-2\bar{a}})$.

Let us now proceed to higher-order corrections. Higher-order terms that do not involve the factor $L^{-2\bar{a}}$ can be obtained straightforwardly by expanding $a_{0}, a_{t}$ and $a$ to the required order and solving the quantization condition $1/2 - a_{t} + a + a_{0} = - n$ order by order in $L^{-1}$. We note that the expansion of $a$ in powers of $t$ is given in Eq.~\eqref{a leading} in the appendix.
For example, the $\mathcal{O}(L^{-2})$ terms in $\omega$, introduced as 
\begin{align}
\omega = \delta \omega L^{-1} + \frac{1}{2} \delta^2 \omega L^{-2} + \mathcal{O}(L^{-3}, L^{-1-2\bar{a}}) ,
\end{align}
can be obtained by repeating the above calculations to next-to-leading order. A direct calculation gives the following expressions:
\begin{align}
a_{0} - a_{t} = - i \delta \omega 
+ i \left( - \frac{1}{2} \delta^2 \omega
+\aBH m + 3 M \QBH \qBH
\right)
L^{-1}
+
\mathcal{O}(L^{-2}, L^{-2\bar{a}}),
\end{align}
and
\begin{align}
a &= 
\bar{a} + \delta a L^{-1} + \mathcal{O}(L^{-2},L^{-2 \bar{a}}),
\end{align}
with
\begin{align}
\delta a &=  \frac{4 M \QBH \qBH(2 \qBH^2 \QBH^2 - s^2 - 3 \ell(1+\ell)) - 4 \aBH m (s^2 + \ell(\ell+1))}{ (1 - 4\bar{a}^2) \bar{a}}
\delta \omega + \frac{1}{2 \bar{a}} \aBH \delta \lambda,
\end{align}
where $\bar{a}$, $\delta \lambda$ and $\delta \omega$ are defined in Eqs.~\eqref{eq:baraDef}, \eqref{eq:dlDef} and \eqref{eq:doDef}, respectively.
Therefore,
the quantization condition $1/2 - a_{t} + a + a_{0} = - n$ can be solved as 
\begin{align}
\frac{1}{2} \delta^2 \omega =  \aBH m + 3 M \QBH \qBH - i \delta a. \label{delta 2 omega dS}
\end{align}
We note that, since $\delta a$ is pure imaginary, the correction $\delta^2 \omega$ is real and therefore it represents the deviation of the de Sitter mode spectrum from the imaginary axis.

On the other hand, the correction of order $\mathcal{O}(L^{-1-2\bar{a}})$, which we explicitly introduce as 
\begin{align}
\omega = \delta \omega L^{-1} + \Delta \omega L^{-1} \bar{t}^{2\bar{a}} + \mathcal{O}(L^{-2}, L^{-2-2\bar{a}}, L^{-1 - 4 \bar{a}}) ,
\label{eq:frequency-noninteger-flat}
\end{align}
is obtained by requiring the two terms in Eq.~\eqref{QNM condition C=0 flat} to cancel. 
We note that $\bar{t}$ is defined by Eq.~\eqref{tbar flat}.

Before proceeding, we clarify the remainder notation in the frequency expansion above and in the quantization condition below.
Logarithmic factors arise from $t^{2a}=t^{2\bar a}\exp[2(a-\bar a)\log t]$ because $a$ depends on $L$.
In particular, the ordinary $\mathcal{O}(L^{-1})$ correction to $a$ gives a contribution of order $L^{-1-2\bar a}\log(L/M)$.
Since $\bar a$ is independent of $L\omega$, the noninteger-power frequency shift changes $a$ only at order $L^{-1-2\bar a}$; the resulting logarithmic contribution is of order $L^{-1-4\bar a}\log(L/M)$ and is absorbed into $\mathcal{O}(L^{-4\bar a})$.
In Eq.~\eqref{eq:frequency-noninteger-flat}, the additional factor $L^{-1}$ relating the frequency correction to the shift in the quantization condition gives a logarithmic factor multiplying $L^{-2-2\bar a}$, which is already covered by the $\mathcal{O}(L^{-2})$ remainder.
In the modified quantization condition below, for the nonresonant parameters considered here, the corresponding remainder is more precisely estimated as
\begin{equation}
\mathcal{O}\!\left(L^{-1-2\bar a}(1+|\log(L/M)|),L^{-4\bar a}\right).
\label{eq:logarithmic-remainder-flat}
\end{equation}
For brevity, in the remainder estimates for the noninteger-power correction below, we suppress this logarithmic factor and write $\mathcal{O}(L^{-1-2\bar a},L^{-4\bar a})$, retaining only the power counting in $L^{-1}$.

For the two terms in the brackets in Eq.~\eqref{QNM condition C=0 flat} to cancel, the argument of the gamma function $\Gamma(\frac{1}{2}-a_{t} + a + a_{0})$ must deviate from a negative integer by an amount proportional to $L^{-2\bar{a}}$, or equivalently to $\bar{t}^{2\bar{a}}$.
We therefore impose the following modified quantization condition,
\begin{align}
\frac{1}{2}-a_{t} + a + a_{0} = - n - \Delta n ~ \bar{t}^{2\bar{a}} + \mathcal{O}(L^{-1 -2 \bar{a}}, L^{-4 \bar{a}}).
\label{modified quantization condition}
\end{align}
The gamma function can be expanded as 
\begin{align}
\frac{1}{\Gamma\left( \frac{1}{2}-a_{t} + a + a_{0} \right)} = - (-1)^{n} n! \Delta n \bar{t}^{2 \bar{a}} + \mathcal{O}(L^{-1 -2 \bar{a}}, L^{-4 \bar{a}}).
\end{align}
Therefore, we obtain
\begin{align}
&\mathcal{M}_{--}\left(a_{t},a;a_{0}\right)\mathcal{M}_{+-}\left(a,a_{1};a_{\infty}\right) 
+
 \bar{t}^{2a} \mathrm{e}^{-\partial_{a}F} \mathcal{M}_{-+}\left(a_{t},a;a_{0}\right)\mathcal{M}_{--}\left(a,a_{1};a_{\infty}\right) \notag\\
&\quad
=
- \frac{\Gamma(2 \bar{a}) \Gamma(1 - 2 \bar{a}_{t})}{\Gamma(\frac{1}{2} - \bar{a}_{t}+\bar{a} - \bar{a}_{0})} \mathcal{M}_{+-}\left(\bar{a},\bar{a}_{1};\bar{a}_{\infty}\right)  (-1)^{n} n! \Delta n \bar{t}^{2 \bar{a}}
+
 \bar{t}^{2\bar{a}} \mathcal{M}_{-+}\left(\bar{a}_{t},\bar{a};\bar{a}_{0}\right)\mathcal{M}_{--}\left(\bar{a},\bar{a}_{1};\bar{a}_{\infty}\right)
 \notag\\
& \qquad + \mathcal{O}(L^{-1-2\bar{a}}, L^{-4 \bar{a}}).
\end{align}
The cancellation at order $L^{-2 \bar{a}}$ therefore determines the deviation as
\begin{align}
\Delta n = \frac{(-1)^{n}}{n!} \frac{\Gamma(\frac{1}{2} - \bar{a}_{t}+\bar{a} - \bar{a}_{0})}{\Gamma(2 \bar{a}) \Gamma(1 - 2 \bar{a}_{t})} \frac{\mathcal{M}_{-+}\left(\bar{a}_{t},\bar{a};\bar{a}_{0}\right)\mathcal{M}_{--}\left(\bar{a},\bar{a}_{1};\bar{a}_{\infty}\right)}{ \mathcal{M}_{+-}\left(\bar{a},\bar{a}_{1};\bar{a}_{\infty}\right)}.
\end{align}
The left-hand side of Eq.~\eqref{modified quantization condition} can be evaluated as 
\begin{align}
\frac{1}{2}-a_{t} + a + a_{0}= - n - i \Delta \omega \bar{t}^{2\bar{a}} + \mathcal{O}(L^{-1}, L^{-1-2\bar{a}}, L^{-4\bar{a}}).
\end{align}
The QNM condition is satisfied when $\Delta \omega = - i \Delta n$.
Thus, including this correction, the QNM frequency $\omega$ is given by
\begin{align}
\omega 
& = - \frac{i}{L} \left( n + \frac{1}{2} + \bar{a} + \Delta n \bar{t}^{2 \bar{a}} \right) + \mathcal{O}(L^{-2}, L^{-2 - 2 \bar{a}}, L^{-1 - 4 \bar{a}}).
\label{Delta omega dS}
\end{align}
A subleading contribution to the connection coefficient with an analogous $\bar{t}^{2\bar{a}}$-type scaling also appears in Ref.~\cite{Arnaudo:2025kit}. In the present analysis, by solving the full QNM condition, we determine how this contribution shifts the quasinormal mode frequency.
Since $\bar{t}=\mathcal{O}(L^{-1})$, this correction scales as $L^{-1-2\bar a}$ and is therefore generically of noninteger order in $1/L$.
In the uncharged limit $\qBH\QBH=0$, however, $\bar a=\ell+1/2$, so that this contribution appears at an integer order and overlaps with the ordinary integer-power expansion.
This limit is resonant: the gamma functions in the correction coefficient can have simultaneous poles. The uncharged leading spectrum quoted above agrees with the formal limit of Eq.~\eqref{omega dS leading}, but its derivation and higher-order corrections require resolving these poles in the full connection condition before taking the limit; direct substitution into the nonresonant correction formula is not justified.

\subsection{Quasinormal modes in the near-extremal limit}
Let us consider the case in which the mass parameter is slightly greater than that of the extremal solution, that is,
\begin{align}
\frac{M - M_{\mathrm{E}}(\aBH,\QBH,L)}{M_{\mathrm{E}}(\aBH,\QBH,L)} \coloneqq \epsilon^2,
\end{align}
with $0< \epsilon \ll 1$.
Here, $M_{\mathrm{E}}(\aBH,\QBH,L)$ is the mass of the extremal Kerr--Newman--de Sitter black hole with the spin parameter $\aBH$, charge $\QBH$, and the cosmological constant $\Lambda = 3/L^2$, given by Eq.~\eqref{ME}.
The parameter $\epsilon$ measures the deviation from extremality, and the extremal limit corresponds to $\epsilon \to 0$.
In this limit, the inner and outer black hole horizons coalesce at the radius $r_{\mathrm{E}}(\aBH, \QBH, L)$ given by Eq.~\eqref{rE}.
Solving $\Delta_{r} = 0$ order by order in $\epsilon$, the radii of these horizons can be obtained as 
\begin{align}
r_{+} &= r_{\mathrm{E}}(\aBH,\QBH,L) + \delta r + \mathcal{O}(\epsilon^2), \\
r_{-} &= r_{\mathrm{E}}(\aBH,\QBH,L) - \delta r + \mathcal{O}(\epsilon^2),
\end{align}
with
\begin{align}
\delta r = \epsilon \, r_{\mathrm{E}} \sqrt{ 
\frac{2}{3} + \frac{4}{3} \frac{\left( 1 - \frac{\aBH^2}{L^2} \right)}{\sqrt{\left(1 - \frac{\aBH^2}{L^2} \right)^2 - 12 \frac{\aBH^2 + \QBH^2}{L^2}}}
}
= \epsilon r_{\mathrm{E}} \sqrt{2 \frac{L^2 - \aBH^2 - 2 r_{\mathrm{E}}^2}{L^2 - \aBH^2 - 6 r_{\mathrm{E}}^2}}
,
\end{align}
where we have used the relation
\begin{align}
\aBH^2 + \QBH^2 = \frac{r_{\mathrm{E}}^2}{L^2} (- \aBH^2 + L^2 - 3 r_{\mathrm{E}}^2 ),
\end{align}
and the inequality \eqref{ineq rE and rN}, that is,
\begin{align}
L^2 - \aBH^2 - 6 r_{\mathrm{E}}^2 > 0.
\end{align}

The surface gravities at these horizons can be expanded as
\begin{align}
\kappa_{+} = - \kappa_{-} + \mathcal{O}(\epsilon^2) = \frac{(\bar{r}_{c} - r_{\mathrm{E}})(r_{\mathrm{E}} - \bar{r}_{n})}{(\aBH^2 + L^2)(\aBH^2 + r_{\mathrm{E}}^2)} \delta r + \mathcal{O} (\epsilon^2) \eqqcolon \kappa + \mathcal{O}(\epsilon^2),
\end{align}
where we define $\kappa$ as the small expansion parameter, instead of $\epsilon$ or $\delta r$, by
\begin{align}
\kappa = \frac{(\bar{r}_{c} - r_{\mathrm{E}})(r_{\mathrm{E}} - \bar{r}_{n})}{(\aBH^2 + L^2)(\aBH^2 + r_{\mathrm{E}}^2)} \delta r = \frac{L^2 - \aBH^2 - 6 r_{\mathrm{E}}^2}{(\aBH^2 + L^2)(\aBH^2 + r_{\mathrm{E}}^2)} \delta r.
\end{align}
We note that we have used the relations
\begin{align}
\bar{r}_{c} &= - r_{\mathrm{E}} + \sqrt{L^2 - \aBH^2 - 2 r_{\mathrm{E}}^2}, \\
\bar{r}_{n} &= - r_{\mathrm{E}} - \sqrt{L^2 - \aBH^2 - 2 r_{\mathrm{E}}^2}.
\end{align}
We express each quantity as a series expansion in $\kappa$ and use notation such as
\begin{align}
\omega = \bar{\omega} + \kappa \delta \omega + \mathcal{O}(\kappa^2, \kappa^{1 + 2 \bar{a}}).
\end{align}
We note that, as in the small cosmological constant case, the correction of $\mathcal{O}(t^{2\bar{a}})$ is necessary in the higher-order analysis.

The horizon Coulomb potentials and angular velocities can be expanded as 
\begin{align}
\Phi_{\pm} = \bar{\Phi}_{\pm} + \kappa \delta \Phi_{\pm} + \mathcal{O}(\kappa^2),\quad 
\Omega_{\pm} = \bar{\Omega}_{\pm} + \kappa \delta \Omega_{\pm} + \mathcal{O}(\kappa^2),
\end{align}
with 
\begin{align}
\bar{\Phi}_{+}& = \bar{\Phi}_{-} = \frac{r_{\mathrm{E}} \QBH}{\Xi (r_{\mathrm{E}}^2 + \aBH^2)} \eqqcolon \Phi_{\mathrm{E}} \\
\bar{\Omega}_{+} &= \bar{\Omega}_{-} = \frac{\aBH}{r_{\mathrm{E}}^2 + \aBH^2} \eqqcolon \Omega_{\mathrm{E}},
\end{align}
and 
\begin{align}
   \delta \Phi_{+} &= - \delta \Phi_{-} = - \frac{(\aBH^2 + L^2) (r_{\mathrm{E}}^2 - \aBH^2)}{(\bar{r}_{c} - r_{\mathrm{E}})(r_{\mathrm{E}} - \bar{r}_{n})r_{\mathrm{E}}} \Phi_{\mathrm{E}} \eqqcolon  \delta \Phi_{\mathrm{E}}\\
   \delta \Omega_{+} &= - \delta \Omega_{-} = 
   - \frac{2 (\aBH^2 + L^2) r_{\mathrm{E}}}{(\bar{r}_{c} - r_{\mathrm{E}})(r_{\mathrm{E}} - \bar{r}_{n})} \Omega_{\mathrm{E}} \eqqcolon \delta \Omega_{\mathrm{E}}.
\end{align}

We employ the identification
\begin{align}
(r_{0}, r_{t}, r_{1}, r_{\infty}) \equiv
(r_{-}, r_{+}, r_{c}, r_{n}).
\end{align}
Thus, the outer black hole horizon is labeled $i^{+} = t$, while the cosmological horizon is labeled $i^{c} = 1$, in the notation in Sec.~\ref{subsec:QNM condition}. 
With this mapping, the extremal limit corresponds to the small-$t$ limit,
\begin{align}
t = \bar{t} + \mathcal{O}(\kappa^2),
\end{align}
where we define $\bar{t}$ by
\begin{align}
\bar{t} \coloneqq \frac{2 (\aBH^2 + L^2) (\aBH^2 + r_{\mathrm{E}}^2) (\bar{r}_{c} - \bar{r}_{n})}{(\bar{r}_{c} - r_{\mathrm{E}})^2(r_{\mathrm{E}}-\bar{r}_{n})^2} \kappa. \label{tbar extremal} 
\end{align}
However, since the surface gravities $\kappa_{\pm}$ appear in the denominator in the expression for $a_{t}$ and $a_{0}$, 
\begin{align}
a_{0} &= - \frac{1}{2} \left( i \frac{\omega - \Phi_{-} \qBH - \Omega_{-} m}{\kappa_{-}} + s  \right) =  + \frac{1}{2} \left( i \frac{\bar{\omega} - \Phi_{\mathrm{E}} \qBH - \Omega_{\mathrm{E}} m}{\kappa} \right) + \mathcal{O}(\kappa^{0}),\\
a_{t} &= - \frac{1}{2} \left( i \frac{\omega - \Phi_{+} \qBH - \Omega_{+} m}{\kappa_{+}} + s  \right) =  - \frac{1}{2} \left( i \frac{\bar{\omega} - \Phi_{\mathrm{E}} \qBH - \Omega_{\mathrm{E}} m}{\kappa} \right) + \mathcal{O}(\kappa^{0}),
\end{align}
a controlled small-$t$ limit exists only if 
\begin{align}
\bar{\omega} = \Phi_{\mathrm{E}} \qBH + \Omega_{\mathrm{E}} m.
\end{align}
Thus, in our approach, the QNMs with the frequency approaching the superradiant threshold are selected. Since $\bar{\omega}$ is real, these modes correspond to the zero-damping modes, where the imaginary part of the frequency disappears in the extremal limit.
With this choice, we obtain finite parameters $a_{i}$ as 
\begin{align}
a_{0} &= \frac{1}{2} \left( i (\delta \omega + \delta \Phi_{\mathrm{E}} \qBH + \delta \Omega_{\mathrm{E}} m ) - s  \right) + \mathcal{O}(\kappa, \kappa^{2 \bar{a}}),\\
a_{t} &= \frac{1}{2} \left( i (- \delta \omega +\delta \Phi_{\mathrm{E}} \qBH +\delta \Omega_{\mathrm{E}} m ) - s  \right) + \mathcal{O}(\kappa, \kappa^{2 \bar{a}}).
\end{align}
Similarly, $a_{1}$ and $a_{\infty}$, which correspond to the poles at $r = r_{c}$ and $r = r_{n}$, are finite,
\begin{align}
a_{1} &= - \frac{1}{2} \left( i \frac{\omega - \Phi_{c}\qBH - \Omega_{c}m}{\kappa_{c}} + s \right) = - \frac{1}{2}\left( i \frac{(\Phi_{\mathrm{E}} - \bar{\Phi}_{c})\qBH + (\Omega_{\mathrm{E}}-\bar{\Omega}_{c})m}{\bar{\kappa}_{c}} + s \right) + \mathcal{O}(\kappa),\\
a_{\infty} &=  - \frac{1}{2} \left( i \frac{\omega - \Phi_{n}\qBH - \Omega_{n}m}{\kappa_{n}} + s \right) = - \frac{1}{2}\left( i \frac{(\Phi_{\mathrm{E}} - \bar{\Phi}_{n})\qBH + (\Omega_{\mathrm{E}}-\bar{\Omega}_{n})m}{\bar{\kappa}_{n}} + s \right) + \mathcal{O}(\kappa).
\end{align}

By a straightforward but lengthy calculation, the leading-order expression for the parameter $a$ can be evaluated as $a = \bar{a} + \mathcal{O}(\kappa, \kappa^{2 \bar{a}})$ with
\begin{align}
\bar{a} = \sqrt{- (\delta \Phi_{\mathrm{E}} \qBH + \delta \Omega_{\mathrm{E}} m)^2 + \frac{L^2}{ L^2 - \aBH^2 - 6 r_{\mathrm{E}}^2} \left( \lambda
+ \frac{ L^2 - \aBH^2 - 2r_{\mathrm{E}}^2}{L^2}\left(s - \frac{1}{2}\right) \left(s - \frac{1}{2} \frac{L^2 - \aBH^2 + 2 r_{\mathrm{E}}^2}{L^2 - \aBH^2 - 2 r_{\mathrm{E}}^2} \right) \right)
}. \label{a ext}
\end{align}
We note that, contrary to the $L \to \infty$ limit, the angular eigenvalue $\lambda$ is not obtained analytically in the near-extremal limit in our formulation. We just treat $\lambda$ as a  parameter for the radial Teukolsky equation here.
In Eq.~\eqref{a ext}, $\lambda$ denotes the angular eigenvalue evaluated at $\omega=\Phi_{\mathrm{E}}\qBH+\Omega_{\mathrm{E}} m$. Higher-order frequency corrections must also include the corresponding variation of $\lambda(\omega)$.

Since the labels for the black hole outer horizon and the cosmological horizon are assigned as $i^{+} = t$ and $i^{c} = 1$, respectively, the quasinormal mode is determined from the condition \eqref{QNM conditions by C}
\begin{align}
C_{1_{-}}{}^{t_{-}} = 0 \qquad \text{or} \qquad 
C_{t_+}{}^{1_+} = 0.
\end{align}
Since these two conditions are equivalent, we focus on the latter condition,  $C_{t_+}{}^{1_+} = 0$. 
The connection coefficient $C_{t_+}{}^{1_+}$ is given in Eq.~\eqref{C t 1},
which can be expressed as 
\begin{align}
C_{t_+}{}^{1_+} &= \mathrm{e}^{-2\pi i a_{t}} t^{\frac{1}{2} - a_{0} + a_{t} - a} (1 - t)^{a_{t} - a_{1}} \mathrm{e}^{\frac{1}{2}\left(\partial_{a_{t}} + \partial_{a} - \partial_{a_{1}}\right)F}\notag\\
& \times \left(\mathcal{M}_{+-}(a_{t},a;a_{0})\mathcal{M}_{++}(a,a_{1};a_{\infty}) + t^{2a} \mathrm{e}^{- \partial_{a} F}  \mathcal{M}_{++}(a_{t},a;a_{0}) \mathcal{M}_{-+}(a,a_{1};a_{\infty}) \right) = 0. \label{QNM condition C=0 extremal}
\end{align} 
Here, the connection matrix $\mathcal{M}_{\theta \theta'}(c_{1},c_{2};c_{3})$ is given by Eq.~\eqref{eq:ConMat}, and the relevant expressions are given by
\begin{align}
\mathcal{M}_{+ -}(a_{t},a;a_{0}) &= \frac{\Gamma\left(2 a \right)\Gamma\left(1+2 a_{t} \right)}{\Gamma\left(\frac{1}{2}+a_{t}+ a + a_{0} \right)\Gamma\left(\frac{1}{2}+ a_{t} + a - a_{0}\right)}, \label{M+-ataa0}\\
\mathcal{M}_{+ +}(a,a_1;a_{\infty}) &= \frac{\Gamma\left(-2 a_{1}\right)\Gamma\left(1+2 a\right)}{\Gamma\left(\frac{1}{2}+ a - a_{1}+a_{\infty}\right)\Gamma\left(\frac{1}{2}+ a - a_{1}-a_{\infty}\right)}, \\
\mathcal{M}_{+ +}(a_{t},a;a_{0}) &= \frac{\Gamma\left(-2 a \right)\Gamma\left(1+2 a_{t} \right)}{\Gamma\left(\frac{1}{2}+ a_{t} - a + a_{0}\right)\Gamma\left(\frac{1}{2}+ a_{t} - a - a_{0}\right)}, \\
\mathcal{M}_{- +}(a, a_{1}; a_{\infty}) &= \frac{\Gamma\left(-2 a_{1} \right)\Gamma\left(1 - 2 a \right)}{\Gamma\left(\frac{1}{2} - a- a_{1} + a_{\infty}\right)\Gamma\left(\frac{1}{2} - a - a_{1} - a_{\infty}\right)}.
\end{align}

We focus on the case where $\bar{a}$ is real and positive.
We also impose the generic nonresonance and simple-zero assumptions used in Sec.~\ref{sec:flat limit}; resonant parameters must be treated using the full connection condition.
In this case the second term in Eq.~\eqref{QNM condition C=0 extremal} is suppressed by a factor $t^{2 \bar{a}}$.
Therefore, at leading order in the $t$ expansion, the QNM condition can be derived from the zeros of the first term in Eq.~\eqref{QNM condition C=0 extremal}, which correspond to the poles of the gamma function in the denominator of Eq.~\eqref{M+-ataa0}. Since the frequency $\delta \omega$ appears in the combination $a_{0} - a_{t}$ at leading order, the QNM condition is identified with the pole of
$\Gamma(\frac{1}{2}+a_t + a-a_0)$.
Therefore, the QNM condition can be obtained as
\begin{align}
\frac{1}{2} + a_t + a-a_0 = \frac{1}{2} - i \delta \omega + \bar{a} + \mathcal{O}(\kappa, \kappa^{2 \bar{a}}) = - n + \mathcal{O}(\kappa^{2\bar{a}}). \label{quantization condition extremal leading}
\end{align}
The deviation of $\mathcal{O}(\kappa^{2\bar{a}})$ from the pole is necessary to cancel the second term in Eq.~\eqref{QNM condition C=0 extremal} as we will discuss later.

From the quantization condition \eqref{quantization condition extremal leading}, we obtain the expression for the first-order correction to the QNM spectrum 
\begin{align}
    \delta \omega = - i \left( n + \frac{1}{2} + \bar{a} \right). \label{QNM spectrum extremal leading}
\end{align}
Thus, the QNM spectrum up to $\mathcal{O}(\kappa)$ is given by 
\begin{align}
\omega = \Phi_{\mathrm{E}} \qBH + \Omega_{\mathrm{E}} m - i \kappa \left( n + \frac{1}{2} + \bar{a} \right) + \mathcal{O}(\kappa^2, \kappa^{1 + 2\bar{a}}), \label{QNM ext}
\end{align}
with $\bar{a}$ given by Eq.~\eqref{a ext}.

The algorithm for calculating the higher-order corrections to the QNM spectrum is the same as in the small cosmological constant case. Thus, the correction at $\mathcal{O}(\kappa^{2})$ can be obtained by evaluating the pole condition \eqref{quantization condition extremal leading} at $\mathcal{O}(\kappa)$, that is, $\delta a_{t} + \delta a - \delta a_{0} = 0$, where $\delta$ represents the coefficients in the expansion in $\kappa$; for example, $a = \bar{a} + \delta a \kappa + \cdots$. Note that since $\delta a_{t} - \delta a_{0} = - \frac{i}{2} \delta^2 \omega + \cdots$, this condition in fact determines the $\mathcal{O}(\kappa^2)$ correction to the QNM spectrum.

The correction of $\mathcal{O}(\kappa^{1 + 2 \bar{a}})$ can also be obtained in the same manner as in the small cosmological constant case. Repeating the discussion at the end of Sec.~\ref{sec:flat limit}, we find that the cancellation of the second term in Eq.~\eqref{QNM condition C=0 extremal} at $\mathcal{O}(\kappa^{2\bar{a}})$, that is, at $\mathcal{O}(\bar{t}^{2\bar{a}})$, requires a deviation from the pole condition of the form
\begin{align}
 \frac{1}{2} + a_{t} + a - a_{0} = - n - \Delta n \bar{t}^{2 \bar{a}} + \mathcal{O}(\kappa^{1 + 2\bar{a}}, \kappa^{4 \bar{a}}),
\end{align}
with
\begin{align}
\Delta n = \frac{(- 1)^{n}}{n!} \frac{\Gamma(\frac{1}{2} + \bar{a}_{t}+\bar{a}+\bar{a}_{0})}{\Gamma(2\bar{a}) \Gamma(1 + 2 \bar{a}_{t})} \frac{\mathcal{M}_{++}(\bar{a}_{t},\bar{a};\bar{a}_{0})\mathcal{M}_{-+}(\bar{a}, \bar{a}_{1}; \bar{a}_{\infty})}{\mathcal{M}_{++}(\bar{a},\bar{a}_{1};\bar{a}_{\infty})}.
\end{align}
Here, $\bar{t}$ is defined by Eq.~\eqref{tbar extremal}.
Analogously to Eq.~\eqref{eq:logarithmic-remainder-flat}, the remainder here includes a logarithmic factor: the abbreviated notation $\mathcal{O}(\kappa^{1+2\bar a},\kappa^{4\bar a})$ is understood as $\mathcal{O}\!\left(\kappa^{1+2\bar a}(1+|\log(\kappa r_{\mathrm{E}})|),\kappa^{4\bar a}\right)$.
The $\mathcal{O}(\kappa^{1 + 2\bar{a}})$ correction to the QNM spectrum, introduced as
\begin{align}
\omega = \bar{\omega} + \kappa \delta \omega + \kappa \bar{t}^{2 \bar{a}} \Delta \omega + \mathcal{O}(\kappa^{2}, \kappa^{2 + 2 \bar{a}}, \kappa^{1 + 4 \bar{a}}),
\end{align}
can be obtained by
\begin{align}
  \Delta \omega = - i \Delta n = - i \frac{(- 1)^{n}}{n!} \frac{\Gamma(\frac{1}{2} + \bar{a}_{t}+\bar{a}+\bar{a}_{0})}{\Gamma(2\bar{a}) \Gamma(1 + 2 \bar{a}_{t})} \frac{\mathcal{M}_{++}(\bar{a}_{t},\bar{a};\bar{a}_{0})\mathcal{M}_{-+}(\bar{a}, \bar{a}_{1}; \bar{a}_{\infty})}{\mathcal{M}_{++}(\bar{a},\bar{a}_{1};\bar{a}_{\infty})}. \label{Delta omega ext}
\end{align} 

\subsection{Quasinormal modes in the near-Nariai limit}
Let us consider the case where the mass parameter $M$ is slightly smaller than that of the Nariai solution $M_{\mathrm{N}}(\aBH,\QBH,L)$ given in Eq.~\eqref{MN}. We parameterize the deviation from the Nariai limit by $\epsilon > 0$, defined by
\begin{align}
\frac{M - M_{\mathrm{N}}(\aBH,\QBH,L)}{M_{\mathrm{N}}(\aBH,\QBH,L)} \coloneqq - \epsilon^2.
\end{align}
We investigate the QNM spectrum as an expansion in $\epsilon$.

The radii of the outer horizon $r_{+}$ and the cosmological horizon $r_{c}$ coalesce at the Nariai radius $r_{\mathrm{N}}$ given in Eq.~\eqref{rN} in the Nariai limit $\epsilon \to 0$.
The $\mathcal{O}(\epsilon)$ corrections can be derived as 
\begin{align}
r_{+} &= r_{\mathrm{N}} - \delta r + \mathcal{O}(\epsilon^2), \\
r_{c} &= r_{\mathrm{N}} + \delta r + \mathcal{O}(\epsilon^2),
\end{align}
with 
\begin{align}
\delta r = \epsilon \, r_{\mathrm{N}} \sqrt{ 
-\frac{2}{3} + \frac{4}{3} \frac{\left( 1 - \frac{\aBH^2}{L^2} \right)}{\sqrt{\left(1 - \frac{\aBH^2}{L^2} \right)^2 - 12 \frac{\aBH^2 + \QBH^2}{L^2}}}
} = \epsilon r_{\mathrm{N}}\sqrt{2 \frac{L^2 - \aBH^2 - 2 r_{\mathrm{N}}^2}{6 r_{\mathrm{N}}^2 - (L^2 -\aBH^2)}}.
\end{align}
Here we have used the relation
\begin{align}
\aBH^2 + \QBH^2 = \frac{r_{\mathrm{N}}^2}{L^2}\left(L^2 - \aBH^2 -3 r_{\mathrm{N}}^2 \right),
\end{align}
and the inequality \eqref{ineq rE and rN} given by
\begin{align}
6r_{\mathrm{N}}^2 - (L^2 - \aBH^2) > 0.
\end{align}

The surface gravities at these horizons can be expanded as
\begin{align}
\kappa_{+} = 
- \kappa_{c} + \mathcal{O}(\epsilon^2)
=
\frac{(r_{\mathrm{N}} - \bar{r}_{-})(r_{\mathrm{N}} - \bar{r}_{n})}{(\aBH^2 + L^2)(\aBH^2 + r_{\mathrm{N}}^2)} \delta r + \mathcal{O}(\epsilon^2) = \kappa + \mathcal{O}(\epsilon^2).
\end{align}
Here we have introduced $\kappa$ by
\begin{align}
\kappa \coloneqq
 \frac{(r_{\mathrm{N}} - \bar{r}_{-})(r_{\mathrm{N}} - \bar{r}_{n})}{(\aBH^2 + L^2)(\aBH^2 + r_{\mathrm{N}}^2)} \delta r
= \frac{6 r_{\mathrm{N}}^2 - (L^2 - \aBH^2)}{(\aBH^2 + L^2)(\aBH^2 + r_{\mathrm{N}}^2)} \delta r.
\end{align}
We use $\kappa$ as the expansion parameter instead of $\epsilon$.

The horizon Coulomb potentials $\Phi_{+}, \Phi_{c}$ and the angular velocities $ \Omega_{+}, \Omega_{c}$
can be expanded as 
\begin{align}
\Phi_{+} = \Phi_{\mathrm{N}} + \kappa \delta \Phi_{\mathrm{N}} + \mathcal{O}(\kappa^2), \\
\Phi_{c} = \Phi_{\mathrm{N}} - \kappa \delta \Phi_{\mathrm{N}} + \mathcal{O}(\kappa^2), \\
\Omega_{+} = \Omega_{\mathrm{N}} + \kappa \delta \Omega_{\mathrm{N}} + \mathcal{O}(\kappa^2), \\
\Omega_{c} = \Omega_{\mathrm{N}} - \kappa \delta \Omega_{\mathrm{N}} + \mathcal{O}(\kappa^2),
\end{align}
where
\begin{align}
\Phi_{\mathrm{N}} &= \frac{r_{\mathrm{N}} \QBH}{\Xi (r_{\mathrm{N}}^2 + \aBH^2)},\\
\Omega_{\mathrm{N}} &=\frac{\aBH}{r_{\mathrm{N}}^2 + \aBH^2},
\end{align}
and
\begin{align}
\delta \Phi_{\mathrm{N}} &= \frac{(\aBH^2 + L^2)(r_{\mathrm{N}}^2 - \aBH^2)}{( r_{\mathrm{N}}-\bar{r}_{-})(r_{\mathrm{N}}-\bar{r}_{n})r_{\mathrm{N}}} \Phi_{\mathrm{N}} = \frac{L^2 (r_{\mathrm{N}}^2 -\aBH^2 ) \QBH }{(\aBH^2 + r_{\mathrm{N}}^2)(6 r_{\mathrm{N}}^2 - (L^2 - \aBH^2))}, \\
\delta \Omega_{\mathrm{N}} &= \frac{2 (\aBH^2 + L^2) r_{\mathrm{N}}}{(r_{\mathrm{N}} - \bar{r}_{-})(r_{\mathrm{N}}- \bar{r}_{n})} \Omega_{\mathrm{N}} = \frac{2 \aBH (\aBH^2 + L^2) r_{\mathrm{N}}}{(\aBH^2 + r_{\mathrm{N}}^2)(6 r_{\mathrm{N}}^2 - (L^2 - \aBH^2)) }.
\end{align}

Let us identify 
\begin{align}
(r_{0}, r_{t}, r_{1}, r_{\infty}) \equiv
(r_{c}, r_{+}, r_{-}, r_{n}).
\end{align}
Thus, we assign $i^{+} \equiv t$ and  $i^{c} \equiv 0$.
With this assignment, the parameter $t$ can be expressed as 
\begin{align}
t &= 
\sqrt{L^2 -\aBH^2 -2 r_{\mathrm{N}}^2}\frac{4(\aBH^2 + L^2)(\aBH^2 + r_{\mathrm{N}}^2)}{(6 r_{\mathrm{N}}^2 - (L^2 - \aBH^2))^2} \kappa + \mathcal{O}(\kappa^2).
\end{align}
Therefore, the Nariai limit $\epsilon \to 0$ corresponds to the small-$t$ limit as desired.

We expand the frequency $\omega$ as 
\begin{align}
\omega = \bar{\omega} + \kappa \delta \omega + \mathcal{O}(\kappa^2).
\end{align}
We note that the correction of $\mathcal{O}(\kappa^{1 + 2 \bar{a}})$ is not needed.
As in the extremal case, $a_{0}$ and $a_{t}$ diverge unless $\bar{\omega}$ is 
\begin{align}
\bar{\omega} = 
\Phi_{\mathrm{N}} \qBH + 
\Omega_{\mathrm{N}} m.
\end{align}
In this case, the parameters $a_{0}$ and $a_{t}$ can be expanded as 
\begin{align}
a_{0} & =   \frac{1}{2} \left( i (\delta \omega +\delta \Phi_{\mathrm{N}} \qBH +\delta \Omega_{\mathrm{N}} m ) - s  \right) + \mathcal{O}(\kappa),\\
a_{t} & =  \frac{1}{2} \left( i (- \delta \omega +\delta \Phi_{\mathrm{N}} \qBH +\delta \Omega_{\mathrm{N}} m ) - s  \right) + \mathcal{O}(\kappa).
\end{align}
The other parameters $a_{1}$ and $a_{\infty}$ can be evaluated as 
\begin{align}
a_{1} = -\frac{1}{2} \left( i \frac{(\Phi_{\mathrm{N}} - \bar{\Phi}_{-}) \qBH + (\Omega_{\mathrm{N}} - \bar{\Omega}_{-})m}{\bar{\kappa}_{-}} + s \right) + \mathcal{O}(\kappa), \\
a_{\infty} =  - \frac{1}{2} \left( i \frac{(\Phi_{\mathrm{N}} - \bar{\Phi}_{n}) \qBH + (\Omega_{\mathrm{N}} - \bar{\Omega}_{n})m}{\bar{\kappa}_{n}} + s \right) + \mathcal{O}(\kappa).
\end{align}
As in the extremal case, a straightforward but lengthy calculation gives $a = \bar{a} + \mathcal{O}(\kappa)$, with
\begin{align}
\bar{a} = \sqrt{- (\delta \Phi_{\mathrm{N}}\qBH + \delta \Omega_{\mathrm{N}} m)^2 - \frac{L^2}{6r_{\mathrm{N}}^2 - (L^2 - \aBH^2)}
\left(
\lambda + \frac{L^2 - \aBH^2 - 2 r_{\mathrm{N}}^2}{L^2} \left(s - \frac{1}{2} \right) \left( s - \frac{1}{2} \frac{L^2 - \aBH^2 + 2 r_{\mathrm{N}}^2}{L^2 - \aBH^2 - 2 r_{\mathrm{N}}^2}\right)
\right)
}. \label{a Nariai}
\end{align}
Here, $\lambda$ denotes the angular eigenvalue evaluated at $\omega=\Phi_{\mathrm{N}}\qBH+\Omega_{\mathrm{N}} m$. Higher-order frequency corrections must also include the corresponding variation of $\lambda(\omega)$.

Given our assignment $i^{+} \equiv t$ and $i^{c} \equiv 0$, Eq.~\eqref{QNM conditions by C} shows that the QNM spectrum can be derived from either of the following two equivalent conditions:
\begin{align}
C_{0_{-}}{}^{t_{-}} = 0, \qquad \text{or} \qquad C_{t_{+}}{}^{0_{+}} = 0.
\end{align}
We consider the former condition below.
The connection coefficient $C_{0_{-}}{}^{t_{-}}$ is given by Eq.~\eqref{C 0 t}, that is,
\begin{align}
C_{0_{-}}{}^{t_{-}} & = \left(1-t\right)^{-\frac{1}{2}+a_{1}}\mathrm{e}^{\frac{1}{2}\left(- \partial_{a_{0}} + \partial_{a_{t}}\right)F} \frac{\Gamma\left(2a_{t}\right)\Gamma\left(1-2 a_{0}\right)}{\Gamma\left(\frac{1}{2}- a_{0} +a_{t}+a\right)\Gamma\left(\frac{1}{2}- a_{0}+a_{t}-a\right)}.
\end{align}
Contrary to the small cosmological constant or the near-extremal expansion, the QNM condition in the near-Nariai expansion is simply characterized by the exact poles of the gamma functions in the denominator. Thus, the QNM spectrum can be determined from the condition
\begin{align}
\frac{1}{2} - a_{0} + a_{t} \pm a  = \frac{1}{2} - i \delta \omega \pm \bar{a} + \mathcal{O}(\kappa) = - n, \label{quantization condition Nariai}
\end{align}
with $n \in \mathbb{Z}_{\geq 0}$. The coefficient of the $\mathcal{O}(\kappa)$ correction to the QNM frequency is then given by 
\begin{align}
\delta \omega =  - i \left(n + \frac{1}{2} \pm \bar{a} \right).
\end{align}
Therefore, we obtain the QNM spectrum in the near-Nariai expansion as 
\begin{align}
\omega = \Omega_{\mathrm{N}} m + \Phi_{\mathrm{N}} \qBH - i \kappa \left(n + \frac{1}{2} \pm \bar{a} \right) + \mathcal{O}(\kappa^2). \label{QNM Nariai}
\end{align}
Here, $\bar{a}$ is given by Eq.~\eqref{a Nariai}.
Higher-order corrections can also be derived by expanding $a_{0}, a_{t}$ and $a$ up to an appropriate order in $\kappa$ and imposing the quantization condition \eqref{quantization condition Nariai} order by order.

\section{Summary and Discussion}
\label{sec:Summary}

In this paper, we studied the angular eigenvalue problem and quasinormal modes governed by the Teukolsky equation in Kerr--Newman--de Sitter spacetime, treating the spin weight $s$ and field charge $\qBH$ as arbitrary parameters, using connection formulae for the general Heun equation developed in Refs.~\cite{Bonelli:2021uvf,Bonelli:2022ten}.
We summarized the connection coefficients between all pairs of regular singular points in Sec.~\ref{subsec:ConForm}.
We formulated the angular regularity condition and the quasinormal mode boundary conditions in terms of the Heun connection coefficients in 
Eqs.~\eqref{angular quantization conditon 2}
and \eqref{QNM conditions by C}, respectively.
We then analyzed these conditions in several limits that can be mapped to the small-$t$ limit, where the classical conformal blocks can be systematically evaluated using their combinatorial expansion.

For the angular equation, we reproduced the eigenvalue of the spin-weighted spherical harmonics in the small $\aBH/L$ limit in Eq.~\eqref{lambda leading} and obtained its next-to-leading-order correction in Eq.~\eqref{lambda correction},
in agreement with previous results for the $\mathcal{O}(\aBH\omega)$ correction~\cite{Suzuki:1998vy,Novaes:2018fry,Arnaudo:2025btb}. For the radial equation, we derived the de Sitter mode spectrum in the small cosmological constant limit $L\to\infty$ in Eq.~\eqref{omega dS leading}, including the corrections summarized in Eqs.~\eqref{delta 2 omega dS} and \eqref{Delta omega dS}. We also obtained the near-extremal quasinormal mode spectrum in Eq.~\eqref{QNM ext}, together with the noninteger-power correction of order $\mathcal{O}(\kappa^{1+2\bar{a}})$ in Eq.~\eqref{Delta omega ext}, and the near-Nariai spectrum in Eq.~\eqref{QNM Nariai}, for which such a noninteger-power correction is absent. For both the near-extremal and near-Nariai expansions, the $\mathcal{O}(\kappa^2)$ and higher-order integer-power corrections can be systematically obtained, although we do not present their explicit expressions. Taken together, these results provide analytic expressions for several distinct quasinormal mode families within a common Heun connection framework.

Let us first discuss the small cosmological constant limit.
In the uncharged case, our leading-order result reduces to the de Sitter mode spectrum obtained in previous studies~\cite{Du:2004jt,Lopez-Ortega:2006aal,Arnaudo:2025kit}.
For a charged field, the leading-order spectrum instead acquires a nontrivial dependence on the combination $\qBH\QBH$, providing a charged generalization of the de Sitter modes.
At higher order, the spectrum is not described by a simple expansion in integer powers of $1/L$.
In addition to the ordinary integer-power corrections, the full Heun connection condition gives rise to a contribution proportional to the generically noninteger power $L^{-1-2\bar a}$.
This term originates from the second contribution to the connection coefficient, which is subleading in the small-$t$ expansion.
Such a contribution was already identified in the uncharged case in Ref.~\cite{Arnaudo:2025kit}.
Here, however, we explicitly solve the full two-term quasinormal mode condition and show that this subleading term shifts the frequency away from the leading gamma-function pole by a generically noninteger power.
Thus, even in the small cosmological constant limit, the exact connection formula contains information beyond that captured by imposing the leading pole condition alone.

We next consider the near-extremal limit.
At leading order in the near-extremal expansion, our result reproduces the superradiant threshold.
The first nontrivial correction agrees with the near-horizon result obtained from the corresponding hypergeometric equation in Ref.~\cite{Davey:2024xvd} in the regime where $\bar a$ is real and positive.
\footnote{
This comparison is for $s=0$, with the scalar mass parameter of Ref.~\cite{Davey:2024xvd} set to $\mu^2=2/L^2$, corresponding to conformal coupling in our massless scalar equation.
At first order away from extremality, the expression given in Ref.~\cite{Davey:2024xvd} takes a different form from ours.
For real positive $\bar a$, we find that the two results agree after reorganizing the expansion so that the first-order variations of the horizon angular velocity and electric potential are included in the reference frequency.
This interpretation is also consistent with the Kerr--de Sitter result of Ref.~\cite{Arnaudo:2025btb}, where the corresponding first-order contribution is combined with the expansion of the horizon angular velocity and cancels when the frequency is expressed relative to its extremal value.
}
In the Kerr--de Sitter limit, the resulting gamma function pole condition agrees with that obtained in Ref.~\cite{Arnaudo:2025btb}.
In the present Kerr--Newman--de Sitter setting, the corresponding near-extremal spectrum includes both the black hole charge and the field charge.
Beyond the leading pole approximation, however, the full Heun connection condition contains a second contribution of order $t^{2\bar a}$.
Solving the resulting two-term condition shifts the quasinormal mode frequency by an amount of order $\kappa^{1+2\bar a}$, which is generically a noninteger power of the surface gravity.
This correction is therefore not captured by imposing the leading gamma function pole condition alone.

We finally turn to the near-Nariai limit.
In this case, the quasinormal mode condition again reduces to a gamma function pole condition, yielding the near-Nariai spectrum in Eq.~\eqref{QNM Nariai}.
Analytic expressions for the near-Nariai quasinormal mode spectrum of Kerr--Newman--de Sitter black holes were previously obtained for neutral fields in Ref.~\cite{Churilova:2021nnc}, and our result reproduces their spectrum when the field charge is set to zero.
The corresponding Kerr--de Sitter problem was also analyzed in Ref.~\cite{Arnaudo:2025btb}.
Our result extends these analyses to a nonzero field charge.
Unlike in the small cosmological constant and near-extremal limits, however, the full connection condition does not generate an additional generically noninteger-power correction to the quasinormal mode frequency.

The comparison between the near-extremal and near-Nariai limits also suggests a possible interpretation of the different structures of their quasinormal mode conditions.
In the near-extremal limit, the near-horizon region does not include the cosmological horizon, where the second quasinormal mode boundary condition is imposed.
The additional contribution proportional to $t^{2\bar a}$ may therefore be interpreted as encoding the global connection between the near-horizon region and the cosmological horizon, which is not captured by the leading near-horizon approximation~\cite{Davey:2024xvd}.
This interpretation also clarifies the restriction to real positive $\bar a$ in our expansion.
When $\bar a$ is purely imaginary,
the suppression of the factor $t^{2\bar a}$ is not guaranteed by the smallness of $\kappa$ itself,
and the two terms in the connection condition must in general be retained at the same order.
This differs from the near-horizon analysis of Ref.~\cite{Davey:2024xvd}, where an analytic near-horizon spectrum is also obtained in this regime.
Since the near-horizon analysis does not fully resolve the connection to the cosmological horizon, our result indicates that such modes cannot in general be determined from the near-horizon problem alone.
The global connection to the cosmological horizon becomes essential already at leading order.
We note that such a contribution is discussed in Ref.~\cite{Yang:2013uba} in the near-extremal Kerr case using the matched asymptotic expansion method.
By contrast, in the near-Nariai limit, the black hole and cosmological horizons belong to the same limiting region, and 
the leading radial problem acquires a hypergeometric-type spectral structure.
Both quasinormal mode boundary conditions can therefore be encoded within this limiting problem.
This provides a natural explanation for why the Heun connection condition reduces directly to a gamma function pole condition and why no analogous noninteger-power correction appears.

The freedom in assigning the four regular singular points provides a useful way to organize different perturbative limits within the same general-Heun connection framework.
Although different assignments are equivalent at the exact level, they lead to different cross-ratio parameters and hence to different useful small-$t$ expansions.
Permutations of the four regular singular points generate the six equivalent cross-ratio representations
\begin{align}
t,\qquad
1-t,\qquad
\frac{1}{t},\qquad
\frac{1}{1-t},\qquad
\frac{t}{t-1},\qquad
\frac{t-1}{t}.
\end{align}
It may be useful to investigate systematically whether these different representations provide convenient expansions for other physically relevant limits of black hole spectral problems.

The general strategy developed in this work may also be applicable to other black hole perturbation problems governed by the general Heun equation.
Whenever a physically interesting regime can be mapped to a controlled degeneration of the Heun problem through an appropriate singular-point assignment, the corresponding spectral problem may become analytically tractable.
This may provide access to other black hole spacetimes, different perturbing fields, and other spectral problems.
It would also be interesting to extend this viewpoint to confluent limits relevant to asymptotically flat black holes and to combine the exact connection conditions with numerical methods in parameter regions where no simple analytic degeneration is available.

To facilitate such applications, we have collected the Heun connection formulae in an explicit form for all pairs of regular singular points.
We have also described in detail the combinatorial evaluation of the semiclassical conformal blocks, including the Young-diagram sums entering the Nekrasov expansion.
Since these techniques originate from conformal field theory and supersymmetric gauge theory and may be less familiar in black hole perturbation theory, we hope that the present formulation will provide a useful reference for further applications of Heun connection formulae to gravitational spectral problems.

Finally, the relation between Heun connection formulae and classical conformal blocks provides a direct link between black hole perturbation theory and developments in conformal field theory and supersymmetric gauge theory.
Further progress in the analytic evaluation of Nekrasov partition functions and semiclassical conformal blocks, such as new representations, expansion schemes, analytic continuation formulae, or more efficient combinatorial methods, may provide new tools for analytic calculations of black hole spectra and their higher-order corrections.
In this way, developments on the gauge-theory and conformal-field-theory sides may broaden the range of black hole spectral problems that can be studied analytically.
Conversely, it would be interesting to explore whether black hole spectral problems can provide useful information about particular asymptotic regimes of the corresponding conformal blocks.

\section*{Acknowledgment}
We thank Yasuyuki Hatsuda and Hajime Kobayashi for helpful discussion.
This work was supported by RIKEN Special Postdoctoral Researchers Program (N.K.) and JSPS KAKENHI Grant No.~JP22K03639, No.~JP26K21813 (H.M.), and No.~JP26K07084~(D.Y.).

\appendix

\section{Conformal blocks and semiclassical conformal blocks\label{sec:CBandSCB}}

In this appendix, we collect the definitions and properties of the conformal blocks and their semiclassical limits that are used in the main text.

\subsection{Combinatorial formulae for the regular four-point block and the degenerate five-point block\label{subsec:CB}}

This section presents the Young-diagram representations of the regular four-point conformal block and the five-point block with one level-two degenerate insertion. We define the fixed-point factors explicitly and illustrate their evaluation in the zero-, one-, and two-instanton sectors of the four-point block.

Under the AGT correspondence, the four-point block is obtained from the $\Omega$-deformed partition function of four-dimensional $\mathcal{N}=2$ $SU(2)$ gauge theory with four flavors \cite{Nekrasov:2002qd,Alday:2009aq}. Here, the term “partition function” refers to an equivariant supersymmetric path integral rather than to a thermal partition function. Schematically, the full gauge theory partition function factorizes as $Z_{{\rm full}}=Z_{{\rm cl}}Z_{{\rm 1-loop}}^{{\rm pert}}Z_{{\rm inst}}$. In what follows, we require only the fixed-point expansion of the instanton partition function $Z_{{\rm inst}}$. Under the AGT correspondence, $Z_{{\rm inst}}$, together with an elementary $U(1)$ factor, gives the normalized conformal block. The classical factor $Z_{{\rm cl}}$ accounts for the leading OPE power, whereas $Z_{{\rm 1-loop}}^{{\rm pert}}$ is associated with the three-point functions and normalization factors entering the full correlator. Equivariant localization labels the $U(2)$ fixed points by ordered pairs of Young diagrams, whose total number of boxes is the instanton number \cite{Nekrasov:2003rj}. We therefore begin with a brief review of Young diagrams.

\subsubsection{Young diagrams, arms, and legs\label{app:young-diagrams}}

A Young diagram $Y$ is specified by a partition
\begin{equation}
Y=(\nu_{1},\nu_{2},\cdots),\quad\nu_{1}\geq\nu_{2}\geq\cdots\geq0.
\end{equation}
The total number of boxes is denoted by
\begin{equation}
|Y|=\sum_{i\geq1}\nu_{i}.
\end{equation}
Here, \(\nu'_j\) is the number of boxes in the \(j\)-th column of \(Y\).
\begin{figure}[t]
\centering \begin{tikzpicture}[x=0.56cm,y=0.56cm,font=\small]
  \begin{scope}[shift={(0,0)}]
    \node[font=\bfseries] at (3,1.0) {$Y=(6,5,3,1)$};
    \foreach \r/\len in {1/6,2/5,3/3,4/1}{
      \foreach \c in {1,...,\len}{
        \draw (\c-1,-\r+1) rectangle (\c,-\r);
      }
    }
    \fill[gray!28] (1,-1) rectangle (2,-2);
    \draw[very thick] (1,-1) rectangle (2,-2);
    \node at (1.5,-1.5) {$s$};
    \foreach \c in {3,4,5}{
      \fill[pattern=north east lines,pattern color=gray]
        (\c-1,-1) rectangle (\c,-2);
      \draw (\c-1,-1) rectangle (\c,-2);
    }
    \draw[decorate,decoration={brace,amplitude=4pt,mirror}]
      (2,-2.1) -- (5,-2.1)
      node[midway,below=3pt,xshift=16pt] {$A_Y(s)=3$};
    \draw[->] (-0.55,0.35) -- (1.0,0.35) node[right] {$j$};
    \draw[->] (-0.55,0.35) -- (-0.55,-1.35) node[below] {$i$};
  \end{scope}

  \begin{scope}[shift={(8.2,0)}]
    \node[font=\bfseries] at (2,1.0) {$W=(4,3,3,1)$};
    \foreach \r/\len in {1/4,2/3,3/3,4/1}{
      \foreach \c in {1,...,\len}{
        \draw (\c-1,-\r+1) rectangle (\c,-\r);
      }
    }
    \fill[gray!28] (1,-1) rectangle (2,-2);
    \draw[very thick] (1,-1) rectangle (2,-2);
    \node at (1.5,-1.5) {$s$};
    \fill[pattern=north east lines,pattern color=gray]
      (1,-2) rectangle (2,-3);
    \draw (1,-2) rectangle (2,-3);
    \draw[decorate,decoration={brace,amplitude=4pt}]
      (2.15,-2) -- (2.15,-3)
      node[midway,right=10pt] {$L_W(s)=1$};
  \end{scope}
\end{tikzpicture} \caption{Arm and leg lengths are elementary box counts. For the marked site $s=(2,2)$, the left diagram has three boxes to the right, so $A_{Y}(s)=3$; at the same coordinate the right diagram has one box below, so $L_{W}(s)=1$. The fixed-point products introduced below appear elaborate only because these elementary counts are repeated for every box and for each ordered pair of labels $k,l=1,2$.}
\label{fig:app-arm-leg}
\end{figure}
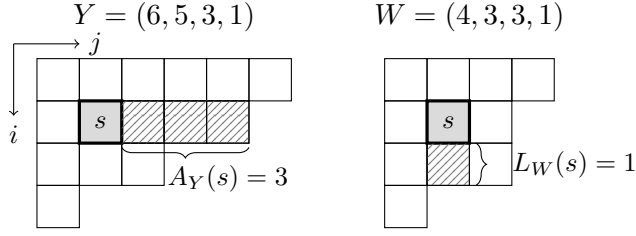
Fig.~\ref{fig:app-arm-leg} gives a graphical illustration of the Young diagram and the quantities introduced below. To define the leg length, we denote the transpose of $Y$ by
\begin{equation}
Y^{T}=(\nu'_{1},\nu'_{2},\cdots).
\end{equation}
A box is labeled by $s=(i,j)$, with rows counted from the top and columns from the left. Its arm and leg lengths relative to $Y$ are 
\begin{equation}
A_{Y}((i,j))=\nu_{i}-j,\quad L_{Y}((i,j))=\nu'_{j}-i.\label{eq:arm-leg}
\end{equation}
Thus, $A_{Y}(s)$ is the number of boxes to the right of $s$, while $L_{Y}(s)$ is the number below it. Note that, according to Eq.~\eqref{eq:arm-leg}, the lengths can be negative when $\nu_{i}<j$ or $\nu'_{j}<i$.

The empty diagram is denoted by $\varnothing$. Empty products are equal to one, but mixed lengths relative to the empty diagram are not zero: 
\begin{equation}
\prod_{s\in\varnothing}f(s)=1,\quad A_{\varnothing}((i,j))=-j,\quad L_{\varnothing}((i,j))=-i.
\end{equation}
This distinction is already needed for a one-box fixed point. A $U(2)$ fixed point is labeled by an ordered pair of Young diagrams
\begin{equation}
\vec{Y}=(Y_{1},Y_{2}),\quad|\vec{Y}|=|Y_{1}|+|Y_{2}|,\quad\vec{0}\equiv(\varnothing,\varnothing).
\end{equation}

\subsubsection{Universal fixed-point factors\label{app:fixed-point-factors}}

We now introduce the universal fixed-point factors entering the instanton partition function.

The two Coulomb parameters are collected into the vector
\begin{equation}
\vec{\alpha}=(\alpha_{1},\alpha_{2})=(\alpha,-\alpha).
\end{equation}
This is the traceless $SU(2)$ specialization of the $U(2)$ Coulomb parameters. Here, the subscripts $k=1,2$ on the components $\alpha_{k}$ are color labels and should not be confused with the external Liouville momentum $\alpha_{1}$.

For two diagrams $Y_{1},Y_{2}$, a site $(i,j)$, and a complex number $\beta$, we define 
\begin{equation}
E(\beta,Y_{1},Y_{2},(i,j))=\beta-b^{-1}L_{Y_{2}}((i,j))+b\bigl(A_{Y_{1}}((i,j))+1\bigr).
\end{equation}
If $(i,j)\in Y_{1}$, the arm is measured in $Y_{1}$ and the leg in $Y_{2}$. The dependence of \(E\) on the diagrams is therefore determined entirely by the elementary box counts illustrated in Fig.~\ref{fig:app-arm-leg}.

The fixed-point contribution of a fundamental hypermultiplet with mass parameter \(\mu\) is
\begin{equation}
z_{\mathrm{hyp}}(\vec{\alpha},\vec{Y},\mu)=\prod_{k=1,2}\prod_{(i,j)\in Y_{k}}\left[\alpha_{k}+\mu+b^{-1}\left(i-\frac{1}{2}\right)+b\left(j-\frac{1}{2}\right)\right].\label{eq:zhyp}
\end{equation}
Each box contributes one linear factor to $z_{\mathrm{hyp}}$. The corresponding vector-multiplet contribution is
\begin{equation}
z_{\mathrm{vec}}(\vec{\alpha},\vec{Y})
=\prod_{k,l=1,2}\left[
\prod_{(i,j)\in Y_{k}}E(\alpha_{k}-\alpha_{l},Y_{k},Y_{l},(i,j))^{-1}\prod_{(i',j')\in Y_{l}}\left[\QL-E(\alpha_{l}-\alpha_{k},Y_{l},Y_{k},(i',j'))\right]^{-1}
\right].
\label{eq:zvec}
\end{equation}
The functions $z_{\mathrm{hyp}}$ and $z_{\mathrm{vec}}$ are the standard finite products entering the equivariant fixed-point expansion \cite{Flume:2002az,Bruzzo:2002xf,Bonelli:2022ten}. In particular, when both Young diagrams are empty $\vec{0}=\left(\varnothing,\varnothing\right)$, all products are empty and therefore
\begin{equation}
z_{\mathrm{hyp}}(\vec{\alpha},\vec{0},\mu)=1,\quad z_{\mathrm{vec}}(\vec{\alpha},\vec{0})=1.
\end{equation}

\subsubsection{The regular four-point block\label{app:regular-four-point-block}}

Using the AGT correspondence \cite{Alday:2009aq} and the conventions of Ref.~\cite{Bonelli:2022ten}, the regular four-point conformal block expanded about $t=0$ admits the following fixed-point representation:
\begin{align}
\FIV{\alpha_{1}}{\alpha_{\infty}}{\alpha}{\alpha_{t}}{\alpha_{0}}t & =t^{\Delta-\Delta_{t}-\Delta_{0}}\left(1-t\right)^{-2\left(\frac{\QL}{2}+\alpha_{1}\right)\left(\frac{\QL}{2}+\alpha_{t}\right)}\sum_{\vec{Y}}t^{\left|\vec{Y}\right|}\mathcal{Z}_{\vec{Y}}^{\left(4\right)}\left(\alpha_{i},\alpha\right),
\label{eq:CB4-Def}
\end{align}
where $\alpha_{i}=\left(\alpha_{0},\alpha_{t},\alpha_{1},\alpha_{\infty}\right)$, and the fixed-point weight is 
\begin{equation}
\mathcal{Z}_{\vec{Y}}^{\left(4\right)}\left(\alpha_{i},\alpha\right)=z_{\mathrm{vec}}\left(\vec{\alpha},\vec{Y}\right)\prod_{\sigma=\pm}z_{\mathrm{hyp}}\left(\vec{\alpha},\vec{Y},\alpha_{t}+\sigma\alpha_{0}\right)z_{\mathrm{hyp}}\left(\vec{\alpha},\vec{Y},\alpha_{1}+\sigma\alpha_{\infty}\right).\label{eq:Z4-Def}
\end{equation}
$\Delta$ is the scaling dimension corresponding to the intermediate momentum
\begin{equation}
\Delta=\frac{\QL^{2}}{4}-\alpha^{2}.
\end{equation}
The factor $t^{|\vec{Y}|}$ is the classical weight of the $|\vec{Y}|$-instanton sector. The factor $\left(1-t\right)^{-2\left(\frac{\QL}{2}+\alpha_{1}\right)\left(\frac{\QL}{2}+\alpha_{t}\right)}$ is the abelian $U(1)$ factor that converts the natural $U(2)$ fixed-point sum into the Virasoro block.

We next spell out the first three instanton sectors, corresponding to $\left|\vec{Y}\right|=0,1,2$.

\paragraph{Zero-instanton sector: $\left|\vec{Y}\right|=0$\mbox{}\\}

The only fixed point is $\vec{Y}=\vec{0}$. Since all products are empty,
\begin{equation}
\mathcal{Z}_{\vec{0}}^{\left(4\right)}\left(\alpha_{i},\alpha\right)=1.
\end{equation}
Consequently, the conformal block begins as
\begin{equation}
\FIV{\alpha_{1}}{\alpha_{\infty}}{\alpha}{\alpha_{t}}{\alpha_{0}}t=t^{\Delta-\Delta_{t}-\Delta_{0}}\left(1+\mathcal{O}\left(t\right)\right).\label{eq:CB4-Ldg}
\end{equation}

\paragraph{One-instanton sector: $\left|\vec{Y}\right|=1$\mbox{}\\}

The two one-box pairs are 
\begin{equation}
\vec{Y}^{(1)}=(\Box,\varnothing),\quad\vec{Y}^{(2)}=(\varnothing,\Box).
\end{equation}
At $s=(1,1)$, $A_{\Box}(s)=L_{\Box}(s)=0$ and $A_{\varnothing}(s)=L_{\varnothing}(s)=-1$. Hence
\begin{equation}
z_{\mathrm{vec}}(\vec{\alpha},\vec{Y}^{(1)})=-\frac{1}{2\alpha(\QL+2\alpha)},\quad z_{\mathrm{vec}}(\vec{\alpha},\vec{Y}^{(2)})=\frac{1}{2\alpha(\QL-2\alpha)},
\end{equation}
and
\begin{equation}
z_{\mathrm{hyp}}(\vec{\alpha},\vec{Y}^{(1)},\mu)=\frac{\QL}{2}+\alpha+\mu,\quad z_{\mathrm{hyp}}(\vec{\alpha},\vec{Y}^{(2)},\mu)=\frac{\QL}{2}-\alpha+\mu.
\end{equation}
The coefficient of \(t\) in the instanton sum in Eq.~\eqref{eq:CB4-Def} is the sum of the two fixed-point contributions
$\mathcal{Z}_{\vec{Y}^{(1)}}^{\left(4\right)}+\mathcal{Z}_{\vec{Y}^{(2)}}^{\left(4\right)}$, where
\begin{align}
\mathcal{Z}_{\vec{Y}^{(1)}}^{\left(4\right)}\left(\alpha_{i},\alpha\right) & =-\frac{{\displaystyle \prod_{\sigma=\pm}\left(\frac{\QL}{2}+\alpha+\alpha_{t}+\sigma\alpha_{0}\right)\left(\frac{\QL}{2}+\alpha+\alpha_{1}+\sigma\alpha_{\infty}\right)}}{2\alpha(\QL+2\alpha)},\nonumber \\
\mathcal{Z}_{\vec{Y}^{(2)}}^{\left(4\right)}\left(\alpha_{i},\alpha\right) & =\frac{{\displaystyle \prod_{\sigma=\pm}\left(\frac{\QL}{2}-\alpha+\alpha_{t}+\sigma\alpha_{0}\right)\left(\frac{\QL}{2}-\alpha+\alpha_{1}+\sigma\alpha_{\infty}\right)}}{2\alpha(\QL-2\alpha)}.
\end{align}

\paragraph{Two-instanton sector: $\left|\vec{Y}\right|=2$\mbox{}\\}

There are five ordered pairs, 
\begin{equation}
([2],\varnothing),\quad([1,1],\varnothing),\quad(\Box,\Box),\quad(\varnothing,[2]),\quad(\varnothing,[1,1]).
\end{equation}
Here, $[2]$ is a row of length two, and $[1,1]$ is a column of height two. The coefficient of $t^{2}$ in the instanton sum is simply obtained by summing the fixed-point weights in Eq.~\eqref{eq:Z4-Def} over these five pairs. For completeness, the vector-multiplet contributions are
\begin{align}
z_{\mathrm{vec}}\left(\vec{\alpha},([2],\varnothing)\right) & =\frac{1}{4\alpha(1-b^{2})(2\alpha+b)(2\alpha+\QL)(2\alpha+\QL+b)},\nonumber \\
z_{\mathrm{vec}}\left(\vec{\alpha},([1,1],\varnothing)\right) & =\frac{1}{4\alpha(1-b^{-2})(2\alpha+b^{-1})(2\alpha+\QL)(2\alpha+\QL+b^{-1})},\nonumber \\
z_{\mathrm{vec}}\left(\vec{\alpha},(\varnothing,[2])\right) & =\frac{1}{4\alpha(1-b^{2})(2\alpha-b)(\QL-2\alpha)(\QL+b-2\alpha)},\nonumber \\
z_{\mathrm{vec}}\left(\vec{\alpha},(\varnothing,[1,1])\right) & =\frac{1}{4\alpha(1-b^{-2})(2\alpha-b^{-1})(\QL-2\alpha)(\QL+b^{-1}-2\alpha)},\nonumber \\
z_{\mathrm{vec}}\left(\vec{\alpha},(\Box,\Box)\right) & =\frac{1}{(b^{2}-4\alpha^{2})(b^{-2}-4\alpha^{2})}.
\end{align}
The corresponding hypermultiplet contributions are
\begin{align}
z_{\mathrm{hyp}}\left(\vec{\alpha},([2],\varnothing),\mu\right) & =\left(\frac{\QL}{2}+\alpha+\mu\right)\left(\frac{\QL}{2}+\alpha+\mu+b\right),\nonumber \\
z_{\mathrm{hyp}}\left(\vec{\alpha},([1,1],\varnothing),\mu\right) & =\left(\frac{\QL}{2}+\alpha+\mu\right)\left(\frac{\QL}{2}+\alpha+\mu+b^{-1}\right),\nonumber \\
z_{\mathrm{hyp}}\left(\vec{\alpha},(\varnothing,[2]),\mu\right) & =\left(\frac{\QL}{2}-\alpha+\mu\right)\left(\frac{\QL}{2}-\alpha+\mu+b\right),\nonumber \\
z_{\mathrm{hyp}}\left(\vec{\alpha},(\varnothing,[1,1]),\mu\right) & =\left(\frac{\QL}{2}-\alpha+\mu\right)\left(\frac{\QL}{2}-\alpha+\mu+b^{-1}\right),\nonumber \\
z_{\mathrm{hyp}}\left(\vec{\alpha},(\Box,\Box),\mu\right) & =\left(\frac{\QL}{2}+\alpha+\mu\right)\left(\frac{\QL}{2}-\alpha+\mu\right).
\end{align}

\subsubsection{The degenerate five-point block\label{app:degenerate-five-point-block}}

We now insert the level-two degenerate field $\Phi_{2,1}(z)$ between the operators at $0$ and $t$, and work in the OPE region $z\ll t\ll1$. On the gauge theory side, the degenerate insertion corresponds to a surface operator. Its fixed-point partition function can equivalently be represented by a two-node quiver with its parameters specialized according to the degenerate fusion rule \cite{Alday:2009fs,Awata:2010bz,Bonelli:2022ten}. In that quiver presentation, $t$ and $z/t$ grade the two fixed-point sums, and the new factor coupling the two nodes is a bifundamental determinant.

Since we have two nodes, we introduce an independent ordered pair of diagrams 
\begin{equation}
\vec{W}=(W_{1},W_{2}),\quad|\vec{W}|=|W_{1}|+|W_{2}|.
\end{equation}
For two Coulomb vectors $\vec{\alpha}=(\alpha_{1},\alpha_{2})$ and $\vec{\beta}=(\beta_{1},\beta_{2})$, the bifundamental hypermultiplet contribution is given by
\begin{align}
z_{\mathrm{bifund}}(\vec{\alpha},\vec{Y},\vec{\beta},\vec{W};\alpha_{t}) & =\prod_{k,l=1,2}\left\{ \prod_{(i,j)\in Y_{k}}\left[E(\alpha_{k}-\beta_{l},Y_{k},W_{l},(i,j))-\left(\frac{\QL}{2}+\alpha_{t}\right)\right]\right.\nonumber \\
 & \quad\left.\times\prod_{(i',j')\in W_{l}}\left[\QL-E(\beta_{l}-\alpha_{k},W_{l},Y_{k},(i',j'))-\left(\frac{\QL}{2}+\alpha_{t}\right)\right]\right\} .
 \label{eq:zbifund}
\end{align}
The degenerate five-point conformal block can then be written as \cite{Bonelli:2022ten}
\begin{align}
 & \FV{\alpha_{1}}{\alpha_{\infty}}{\alpha}{\alpha_{t}}{\alpha_{0\theta}}{\alpha_{2,1}}{\alpha_{0}}t{\frac{z}{t}}\nonumber \\
 & =t^{\Delta-\Delta_{t}-\Delta_{0\theta}}z^{\frac{b\QL}{2}+\theta b\alpha_{0}}\left(1-t\right)^{-2\left(\frac{\QL}{2}+\alpha_{1}\right)\left(\frac{\QL}{2}-\alpha_{t}\right)}\left(1-\frac{z}{t}\right)^{-2\left(\frac{\QL}{2}+\alpha_{t}\right)\left(\frac{\QL}{2}+\alpha_{2,1}\right)}\left(1-z\right)^{-2\left(\frac{\QL}{2}+\alpha_{1}\right)\left(\frac{\QL}{2}+\alpha_{2,1}\right)}\nonumber \\
 & \quad\times\sum_{\vec{Y},\vec{W}}t^{\left|\vec{Y}\right|}\left(\frac{z}{t}\right)^{\left|\vec{W}\right|}\mathcal{Z}_{\vec{Y},\vec{W},\theta}^{\left(5\right)}\left(\alpha_{i},\alpha\right),\label{eq:CB5-Def}
\end{align}
where the weight associated with the pair \((\vec Y,\vec W)\) is
\begin{align}
\mathcal{Z}_{\vec{Y},\vec{W},\theta}^{\left(5\right)}\left(\alpha_{i},\alpha\right) & =z_{\mathrm{vec}}\left(\vec{\alpha},\vec{Y}\right)z_{\mathrm{vec}}\left(\vec{\alpha_{0\theta}},\vec{W}\right)z_{\mathrm{bifund}}\left(\vec{\alpha},\vec{Y},\vec{\alpha_{0\theta}},\vec{W};\alpha_{t}\right)\nonumber \\
 & \quad\times\prod_{\sigma=\pm}\left[z_{\mathrm{hyp}}\left(\vec{\alpha},\vec{Y},\alpha_{1}+\sigma\alpha_{\infty}\right)z_{\mathrm{hyp}}\left(\vec{\alpha_{0\theta}},\vec{W},\alpha_{2,1}+\sigma\alpha_{0}\right)\right],
 \label{eq:Z5-Def}
\end{align}
with the Coulomb vector at the second quiver node $\vec{\alpha}_{0\theta}=(\alpha_{0\theta},-\alpha_{0\theta})$. 
The two vector-multiplet factors are associated with the two quiver nodes, the hypermultiplet factors encode the external parameters, and \(z_{\mathrm{bifund}}\) couples the nodes.
$\Delta_{i\theta}$ is the scaling dimension corresponding to the shifted momentum
\begin{equation}
\Delta_{i\theta}=\frac{\QL^{2}}{4}-\alpha_{i\theta}^{2},\quad \left(i=0,t,1,\infty\right).
\label{eq:SSD-Def}
\end{equation}

\subsection{Properties of the conformal blocks\label{subsec:CB-Prop}}

This section records several properties of the four- and five-point conformal blocks that will be used below.

In the limit $z\rightarrow0$, the degenerate five-point block reduces, up to its leading OPE factor, to the four-point block with the shifted momentum $\alpha_{0\theta}$. More precisely,
\begin{equation}
\FV{\alpha_{1}}{\alpha_{\infty}}{\alpha}{\alpha_{t}}{\alpha_{0\theta}}{\alpha_{2,1}}{\alpha_{0}}t{\frac{z}{t}}=z^{\frac{b\QL}{2}+\theta b\alpha_{0}}\left[\FIV{\alpha_{1}}{\alpha_{\infty}}{\alpha}{\alpha_{t}}{\alpha_{0\theta}}t+\mathcal{O}\left(z\right)\right].\label{eq:CB5-CB4}
\end{equation}
This relation follows from the definitions \eqref{eq:CB4-Def} and \eqref{eq:CB5-Def} and the identity
\begin{equation}
z_{\mathrm{bifund}}(\vec{\alpha},\vec{Y},\vec{\alpha_{0\theta}},\vec{0};\alpha_{t})=\prod_{\sigma=\pm}z_{\mathrm{hyp}}\left(\vec{\alpha},\vec{Y},-\alpha_{t}+\sigma\alpha_{0\theta}\right),
\end{equation}
together with
\begin{equation}
\FIV{\alpha_{1}}{\alpha_{\infty}}{\alpha}{-\alpha_{t}}{\alpha_{0\theta}}t=\FIV{\alpha_{1}}{\alpha_{\infty}}{\alpha}{\alpha_{t}}{\alpha_{0\theta}}t.
\end{equation}
The latter equality holds because the regular four-point conformal block depends on each external momentum $\alpha_{i}$ only through its conformal dimension $\Delta_{i}$.

The M\"obius transformations used below to map the other local insertion points to the origin induce the following identities for the regular four-point block:
\begin{subequations}
\label{eq:CB4-Mobius}
\begin{align}
\FIV{\alpha_{1}}{\alpha_{\infty}}{\alpha}{\alpha_{t}}{\alpha_{0}}t & =\mathrm{e}^{-i\pi\left(\Delta-\Delta_{t}-\Delta_{0}\right)}\left(1-t\right)^{\Delta_{\infty}-\Delta_{1}-\Delta_{t}-\Delta_{0}}\FIV{\alpha_{1}}{\alpha_{\infty}}{\alpha}{\alpha_{0}}{\alpha_{t}}{\frac{t}{t-1}},\label{eq:CB40t-Mobius}\\
\FIV{\alpha_{1}}{\alpha_{\infty}}{\alpha}{\alpha_{t}}{\alpha_{0}}t & =t^{\Delta_{\infty}+\Delta_{1}-\Delta_{t}-\Delta_{0}}\left(1-t\right)^{\Delta_{\infty}+\Delta_{0}-\Delta_{t}-\Delta_{1}}\FIV{\alpha_{0}}{\alpha_{t}}{\alpha}{\alpha_{\infty}}{\alpha_{1}}t,\label{eq:CB401-Mobius}\\
\FIV{\alpha_{1}}{\alpha_{\infty}}{\alpha}{\alpha_{t}}{\alpha_{0}}t & =t^{\Delta_{\infty}+\Delta_{1}-\Delta_{t}-\Delta_{0}}\FIV{\alpha_{t}}{\alpha_{0}}{\alpha}{\alpha_{1}}{\alpha_{\infty}}t.\label{eq:CB40i-Mobius}
\end{align}
\end{subequations}
Here, the prefactors depending on $t$ arise from the Jacobians associated with the M\"obius transformations, whereas the phase can be obtained by comparing the small-$t$ expansions on both sides using Eq.~\eqref{eq:CB4-Ldg}.

Applying the corresponding M\"obius transformations gives the following five-point blocks expanded about \(z=t\), \(z=1\), and \(z=\infty\):
\begin{subequations}
\label{eq:CB5-Mobius}
\begin{align}
 & \mathrm{e}^{-i\pi\left(\Delta-\Delta_{t}-\Delta_{0}-\Delta_{2,1}\right)}\left(1-t\right)^{\Delta_{\infty}-\Delta_{1}-\Delta_{t}-\Delta_{0}-\Delta_{2,1}}\FV{\alpha_{1}}{\alpha_{\infty}}{\alpha}{\alpha_{0}}{\alpha_{t\theta}}{\alpha_{2,1}}{\alpha_{t}}{\frac{t}{t-1}}{\frac{t-z}{t}},\label{eq:CB5t-Mobius}\\
 & t^{\Delta_{\infty}+\Delta_{1}-\Delta_{t}-\Delta_{0}+\Delta_{2,1}}\left(1-t\right)^{\Delta_{\infty}+\Delta_{0}-\Delta_{t}-\Delta_{1}+\Delta_{2,1}}\left(t-z\right)^{-2\Delta_{2,1}}\FV{\alpha_{0}}{\alpha_{t}}{\alpha}{\alpha_{\infty}}{\alpha_{1\theta}}{\alpha_{2,1}}{\alpha_{1}}t{\frac{1-z}{t-z}},\label{eq:CB51-Mobius}\\
 & t^{\Delta_{\infty}+\Delta_{1}-\Delta_{t}-\Delta_{0}+\Delta_{2,1}}z^{-2\Delta_{2,1}}\FV{\alpha_{t}}{\alpha_{0}}{\alpha}{\alpha_{1}}{\alpha_{\infty\theta}}{\alpha_{2,1}}{\alpha_{\infty}}t{\frac{1}{z}}.\label{eq:CB5i-Mobius}
\end{align}
\end{subequations}
Each of these expressions satisfies the same BPZ equation, Eq.~\eqref{eq:BPZeq}.

\subsection{Semiclassical conformal blocks\label{subsec:SCB}}

\subsubsection{The semiclassical limit}

We now take the semiclassical limit defined in Eq.~\eqref{eq:SCLim}. In this limit, the four-point conformal block behaves as
\cite{Besken:2019jyw,Bonelli:2022ten}
\begin{equation}
\FIV{\alpha_{1}}{\alpha_{\infty}}{\alpha}{\alpha_{t}}{\alpha_{0}}t=t^{\Delta-\Delta_{t}-\Delta_{0}}\exp\left[\frac{1}{b^{2}}F\left(a_{i},a;t\right)+W_{4}\left(a_{i},a;t\right)+\mathcal{O}\left(b^{2}\right)\right].
\label{eq:CB4-Ldg2}
\end{equation}
Here, $F$ is the classical four-point conformal block, whereas $W_{4}$ denotes the finite $\mathcal{O}\left(b^{0}\right)$ contribution. The classical conformal block admits the combinatorial representation
\begin{equation}
F\left(a_{i},a;t\right)=\lim_{b\rightarrow0}b^{2}\log\left[\left(1-t\right)^{-2\left(\frac{\QL}{2}+\alpha_{1}\right)\left(\frac{\QL}{2}+\alpha_{t}\right)}\sum_{\vec{Y}}t^{\left|\vec{Y}\right|}\mathcal{Z}_{\vec{Y}}^{\left(4\right)}\left(\alpha_{i},\alpha\right)\right],
\label{eq:CCB-Def}
\end{equation}
where $\mathcal{Z}_{\vec{Y}}^{\left(4\right)}$ is defined in Eq.~\eqref{eq:Z4-Def}.
Its expansion about $t=0$ begins as
\begin{equation}
F\left(a_{i},a;t\right)=\frac{\left(4a^{2}+4a_{1}^{2}-4a_{\infty}^{2}-1\right)\left(4a^{2}-4a_{0}^{2}+4a_{t}^{2}-1\right)}{8-32a^{2}}t+\mathcal{O}\left(t^{2}\right).\label{eq:CCB-Ldg}
\end{equation}
Eq.~\eqref{eq:u0-Def} may then be inverted order by order in $t$ to express the intermediate momentum $a$ in terms of the accessory parameter $u$ and the external momenta as
\begin{equation}
a=\sqrt{-\frac{1}{4}-u+a_{t}^{2}+a_{0}^{2}}
+ \frac{\left(\frac{1}{2}+u-a_{t}^2 -a_{0}^2 -a_{1}^2 + a_{\infty}^2\right)\left(\frac{1}{2} + u - 2 a_{t}^2\right)}{4\left(\frac{1}{2}+u - a_{t}^2 - a_{0}^2\right)\sqrt{-\frac{1}{4}-u+a_{t}^{2}+a_{0}^{2}}} t
+ \mathcal{O}\left(t^2\right). \label{a leading}
\end{equation}

M\"obius transformations generally permute the external momenta. Consequently, the classical conformal block evaluated with a permuted ordering of $\left(a_{0},a_{t},a_{1},a_{\infty}\right)$ is, in general, a different function of $t$. Nevertheless, following the convention of Ref.~\cite{Bonelli:2022ten}, we use the symbol $F\left(a_{i},a;t\right)$ throughout to denote the classical conformal block defined by the original ordering in Eq.~\eqref{eq:CCB-Def}, including when it appears in the normalization of the semiclassical blocks associated with $z=t$, $z=1$, or $z=\infty$.

In the same limit defined in Eq.~\eqref{eq:SCLim}, the degenerate five-point conformal block has the heavy--light expansion~
\cite{Bonelli:2022ten}
\begin{align}
&\FV{\alpha_{1}}{\alpha_{\infty}}{\alpha}{\alpha_{t}}{\alpha_{0\theta}}{\alpha_{2,1}}{\alpha_{0}}t{\frac{z}{t}}\nonumber\\
&=t^{\Delta-\Delta_{t}-\Delta_{0\theta}}z^{\frac{b\QL}{2}+\theta b\alpha_{0}}\exp\left[\left.\frac{1}{b^{2}}F\left(a_{i},a;t\right)\right|_{a_{0}\rightarrow a_{0\theta}}+W_{5}\left(a_{i},a;t,\frac{z}{t}\right)+\mathcal{O}\left(b^{2}\right)\right],
\label{eq:CB5-Ldg}
\end{align}
where $a_{i\theta}$ is the rescaled shifted momentum defined by
\begin{equation}
a_{i\theta}=a_i-\theta\frac{b^2}{2},\quad
\left(i=0,t,1,\infty\right).
\end{equation}
Since the shift of \(W_5\) contributes only at \(\mathcal{O}(b^2)\), whereas the shift of \(F\) gives an \(\mathcal{O}(b^0)\) contribution after multiplication by \(b^{-2}\), we have written the momentum shift explicitly only in the argument of \(F\).
$\Delta_{i\theta}$ is the scaling dimension corresponding to the shifted momentum defined in Eq.~\eqref{eq:SSD-Def}.
Importantly, the leading classical contribution $F\left(a_i,a;t\right)$ is the same function defined in Eq.~\eqref{eq:CCB-Def}. In addition,
\begin{equation}
W_{5}\left(a_{i},a;t,0\right)=W_{4}\left(a_{i},a;t\right).
\label{eq:W5-W4}
\end{equation}
This equality follows from Eqs.~\eqref{eq:CB5-CB4} and \eqref{eq:CB4-Ldg2}.

It follows from Eqs.~\eqref{eq:CB4-Ldg2} and \eqref{eq:CB5-Ldg} that the ratio of the five-point block to the corresponding four-point block has a finite semiclassical limit \eqref{eq:SCLim}. This observation motivates the definitions of the semiclassical conformal blocks in Eq.~\eqref{eq:SCB-Def}.

\subsubsection{Leading local behavior of $\mathcal{F}_{z_{\theta}}$}

The leading local behavior of $\mathcal{F}_{z_{\theta}}\left(a_{i},a;t,z\right)$ follows from Eqs.~\eqref{eq:CB4-Mobius}, \eqref{eq:CB4-Ldg2}, \eqref{eq:CB5-Ldg}, and \eqref{eq:W5-W4}. Explicitly, 
\begin{subequations}
\label{eq:SCB-Ldg}
\begin{align}
\mathcal{F}_{0_{\theta}}\left(a_{i},a;t,z\right) & =t^{-\theta a_{0}}\mathrm{e}^{-\frac{\theta}{2}\partial_{a_{0}}F\left(a_{i},a;t\right)}z^{\frac{1}{2}+\theta a_{0}}\left(1+\mathcal{O}\left(z\right)\right),\label{eq:SCB0-Ldg}\\
\mathcal{F}_{t_{\theta}}\left(a_{i},a;t,z\right) & =t^{-\theta a_{t}}\mathrm{e}^{-\frac{\theta}{2}\partial_{a_{t}}F\left(a_{i},a;t\right)}\left(t-z\right)^{\frac{1}{2}+\theta a_{t}}\left(1+\mathcal{O}\left(z-t\right)\right),\label{eq:SCBt-Ldg}\\
\mathcal{F}_{1_{\theta}}\left(a_{i},a;t,z\right) & =\mathrm{e}^{\pm i\pi\left(\frac{1}{2}-\theta a_{1}\right)}\mathrm{e}^{-\frac{\theta}{2}\partial_{a_{1}}F\left(a_{i},a;t\right)}\left(1-z\right)^{\frac{1}{2}+\theta a_{1}}\left(1+\mathcal{O}\left(z-1\right)\right),\label{eq:SCB1-Ldg}\\
\mathcal{F}_{\infty_{\theta}}\left(a_{i},a;t,z\right) & =\mathrm{e}^{-\frac{\theta}{2}\partial_{a_{\infty}}F\left(a_{i},a;t\right)}z^{\frac{1}{2}-\theta a_{\infty}}\left(1+\mathcal{O}\left(\frac{1}{z}\right)\right).\label{eq:SCBi-Ldg}
\end{align}
\end{subequations}
We now derive these expressions in turn. We begin with $\mathcal{F}_{0_{\theta}}\left(a_i,a;t,z\right)$ defined in Eq.~\eqref{eq:SCB0-Def}. Substituting the semiclassical expansions \eqref{eq:CB4-Ldg2} and \eqref{eq:CB5-Ldg} gives
\begin{align}
\mathcal{F}_{0_{\theta}}\left(a_{i},a;t,z\right) & =\exp\left[W_{5}\left(a_{i},a;t,\frac{z}{t}\right)-W_{4}\left(a_{i},a;t\right)\right]\nonumber \\
 & \quad\times\lim_{b\rightarrow0}t^{-\Delta_{0\theta}+\Delta_{0}}z^{\frac{b\QL}{2}+\theta b\alpha_{0}}\exp\left[\frac{1}{b^{2}}\left(\left.F\left(a_{i},a;t\right)\right|_{a_{0}\rightarrow a_{0\theta}}-F\left(a_{i},a;t\right)\right)\right].
\end{align}
Expanding the shifted classical conformal block $\left.F\left(a_{i},a;t\right)\right|_{a_{0}\rightarrow a_{0\theta}}$ to first order in the momentum shift and using Eq.~\eqref{eq:W5-W4}, we obtain
\begin{equation}
\mathcal{F}_{0_{\theta}}\left(a_{i},a;t,z\right)=t^{-\theta a_{0}}\mathrm{e}^{-\frac{\theta}{2}\partial_{a_{0}}F\left(a_{i},a;t\right)}z^{\frac{1}{2}+\theta a_{0}}\left(1+\mathcal{O}\left(z\right)\right).
\end{equation}
This reproduces Eq.~\eqref{eq:SCB0-Ldg}.

To derive the remaining semiclassical blocks, we use the M\"obius transformations \eqref{eq:CB4-Mobius}. Let us first consider $\mathcal{F}_{t_{\theta}}\left(a_{i},a;t,z\right)$ defined in Eq.~\eqref{eq:SCBt-Def}. Using Eq.~\eqref{eq:CB5-CB4}, we obtain
\begin{equation}
\mathcal{F}_{t_{\theta}}\left(a_{i},a;t,z\right)\sim\lim_{b\rightarrow0}\left(t-1\right)^{-\Delta_{2,1}}\left(\frac{t-z}{t-1}\right)^{\frac{b\QL}{2}+\theta b\alpha_{t}}\frac{\FIV{\alpha_{1}}{\alpha_{\infty}}{\alpha}{\alpha_{0}}{\alpha_{t\theta}}{\frac{t}{t-1}}}{\FIV{\alpha_{1}}{\alpha_{\infty}}{\alpha}{\alpha_{0}}{\alpha_{t}}{\frac{t}{t-1}}}, \quad (z\rightarrow t).
\end{equation}
Applying the M\"obius transformation \eqref{eq:CB40t-Mobius} to the ratio of four-point blocks gives
\begin{equation}
\frac{\FIV{\alpha_{1}}{\alpha_{\infty}}{\alpha}{\alpha_{0}}{\alpha_{t\theta}}{\frac{t}{t-1}}}{\FIV{\alpha_{1}}{\alpha_{\infty}}{\alpha}{\alpha_{0}}{\alpha_{t}}{\frac{t}{t-1}}}=\mathrm{e}^{\pm i\pi\left(\Delta_{t\theta}-\Delta_{t}\right)}\left(1-t\right)^{\Delta_{t\theta}-\Delta_{t}}\frac{\FIV{\alpha_{1}}{\alpha_{\infty}}{\alpha}{\alpha_{t\theta}}{\alpha_{0}}t}{\FIV{\alpha_{1}}{\alpha_{\infty}}{\alpha}{\alpha_{t}}{\alpha_{0}}t}.
\end{equation}
Combining this relation with the semiclassical expansion \eqref{eq:CB4-Ldg2}, we arrive at
\begin{align}
\mathcal{F}_{t_{\theta}}\left(a_{i},a;t,z\right) & =\left(t-1\right)^{\frac{1}{2}}\left(\frac{t-z}{t-1}\right)^{\frac{1}{2}+\theta a_{t}}\mathrm{e}^{\pm i\pi\theta a_{t}}\left(1-t\right)^{\theta a_{t}}t^{-\theta a_{t}}\mathrm{e}^{-\frac{\theta}{2}\partial_{a_{t}}F\left(a_{i},a;t\right)}\left(1+\mathcal{O}\left(z-t\right)\right)\nonumber \\
 & =t^{-\theta a_{t}}\mathrm{e}^{-\frac{\theta}{2}\partial_{a_{t}}F\left(a_{i},a;t\right)}\left(t-z\right)^{\frac{1}{2}+\theta a_{t}}\left(1+\mathcal{O}\left(z-t\right)\right).
\end{align}
This is precisely Eq.~\eqref{eq:SCBt-Ldg}.

We next consider $\mathcal{F}_{1_{\theta}}\left(a_{i},a;t,z\right)$ defined in Eq.~\eqref{eq:SCB1-Def}. Using Eq.~\eqref{eq:CB5-CB4}, we obtain
\begin{equation}
\mathcal{F}_{1_{\theta}}\left(a_{i},a;t,z\right)\sim\lim_{b\rightarrow0}\left(t\left(1-t\right)\right)^{\Delta_{2,1}}\left(t-z\right)^{-2\Delta_{2,1}}\left(t\frac{1-z}{t-z}\right)^{\frac{b\QL}{2}+\theta b\alpha_{1}}\frac{\FIV{\alpha_{0}}{\alpha_{t}}{\alpha}{\alpha_{\infty}}{\alpha_{1\theta}}t}{\FIV{\alpha_{0}}{\alpha_{t}}{\alpha}{\alpha_{\infty}}{\alpha_{1}}t}, \quad (z\rightarrow 1).
\end{equation}
Applying the M\"obius transformation \eqref{eq:CB401-Mobius} to the ratio of four-point blocks gives
\begin{equation}
\frac{\FIV{\alpha_{0}}{\alpha_{t}}{\alpha}{\alpha_{\infty}}{\alpha_{1\theta}}t}{\FIV{\alpha_{0}}{\alpha_{t}}{\alpha}{\alpha_{\infty}}{\alpha_{1}}t}=t^{-\Delta_{1\theta}+\Delta_{1}}\left(1-t\right)^{\Delta_{1\theta}-\Delta_{1}}\frac{\FIV{\alpha_{1\theta}}{\alpha_{\infty}}{\alpha}{\alpha_{t}}{\alpha_{0}}t}{\FIV{\alpha_{1}}{\alpha_{\infty}}{\alpha}{\alpha_{t}}{\alpha_{0}}t}.
\end{equation}
Combining this relation with the semiclassical expansion \eqref{eq:CB4-Ldg2}, we arrive at
\begin{align}
\mathcal{F}_{1_{\theta}}\left(a_{i},a;t,z\right) & =\left(t\left(1-t\right)\right)^{-\frac{1}{2}}\left(t-1\right)\left(t\frac{1-z}{t-1}\right)^{\frac{1}{2}+\theta a_{1}}t^{-\theta a_{1}}\left(1-t\right)^{\theta a_{1}}\mathrm{e}^{-\frac{\theta}{2}\partial_{a_{1}}F\left(a_{i},a;t\right)}\left(1+\mathcal{O}\left(z-1\right)\right)\nonumber \\
 & =\mathrm{e}^{\pm i\pi\left(\frac{1}{2}-\theta a_{1}\right)}\mathrm{e}^{-\frac{\theta}{2}\partial_{a_{1}}F\left(a_{i},a;t\right)}\left(1-z\right)^{\frac{1}{2}+\theta a_{1}}\left(1+\mathcal{O}\left(z-1\right)\right).
\end{align}
This is precisely Eq.~\eqref{eq:SCB1-Ldg}.

Finally, we consider $\mathcal{F}_{\infty_{\theta}}\left(a_{i},a;t,z\right)$ defined in Eq.~\eqref{eq:SCBi-Def}. Using Eq.~\eqref{eq:CB5-CB4}, we obtain
\begin{equation}
\mathcal{F}_{\infty_{\theta}}\left(a_{i},a;t,z\right)\sim\lim_{b\rightarrow0}t^{\Delta_{2,1}}z^{-2\Delta_{2,1}}\left(\frac{t}{z}\right)^{\frac{b\QL}{2}+\theta b\alpha_{\infty}}\frac{\FIV{\alpha_{t}}{\alpha_{0}}{\alpha}{\alpha_{1}}{\alpha_{\infty\theta}}t}{\FIV{\alpha_{t}}{\alpha_{0}}{\alpha}{\alpha_{1}}{\alpha_{\infty}}t}, \quad (z\rightarrow \infty).
\end{equation}
Applying the M\"obius transformation \eqref{eq:CB40i-Mobius} to the ratio of four-point blocks gives
\begin{equation}
\frac{\FIV{\alpha_{t}}{\alpha_{0}}{\alpha}{\alpha_{1}}{\alpha_{\infty\theta}}t}{\FIV{\alpha_{t}}{\alpha_{0}}{\alpha}{\alpha_{1}}{\alpha_{\infty}}t}=t^{\Delta_{\infty}-\Delta_{\infty\theta}}\frac{\FIV{\alpha_{1}}{\alpha_{\infty\theta}}{\alpha}{\alpha_{t}}{\alpha_{0}}t}{\FIV{\alpha_{1}}{\alpha_{\infty}}{\alpha}{\alpha_{t}}{\alpha_{0}}t}.
\end{equation}
Combining this relation with the semiclassical expansion \eqref{eq:CB4-Ldg2}, we arrive at
\begin{align}
\mathcal{F}_{\infty_{\theta}}\left(a_{i},a;t,z\right) & =t^{-\frac{1}{2}}z\left(\frac{t}{z}\right)^{\frac{1}{2}+\theta a_{\infty}}t^{-\theta a_{\infty}}\mathrm{e}^{-\frac{\theta}{2}\partial_{a_{\infty}}F\left(a_{i},a;t\right)}\left(1+\mathcal{O}\left(\frac{1}{z}\right)\right)\nonumber \\
 & =\mathrm{e}^{-\frac{\theta}{2}\partial_{a_{\infty}}F\left(a_{i},a;t\right)}z^{\frac{1}{2}-\theta a_{\infty}}\left(1+\mathcal{O}\left(\frac{1}{z}\right)\right).
\end{align}
This is precisely Eq.~\eqref{eq:SCBi-Ldg}.

\subsection{Connection formulae\label{subsec:CF}}

In this section, we derive the connection formulae for the semiclassical conformal blocks used in Sec.~\ref{subsec:ConForm}.

\subsubsection{Connection formulae for conformal blocks\label{subsec:CF-CB}}

The required relations follow from the connection formulae for the degenerate five-point conformal blocks established in Ref.~\cite{Bonelli:2022ten}. These formulae relate the local expansions represented by Eqs.~\eqref{eq:CB5-Def} and \eqref{eq:CB5-Mobius}.
These relations follow from crossing symmetry, which requires different OPE decompositions of the same correlation function to agree. Comparing the resulting expansions, together with the known three-point coefficients and the chosen block normalizations, yields the connection coefficients between the corresponding blocks.

Concretely, the three connection formulae relevant here are given in Ref.~\cite{Bonelli:2022ten}; in our conventions, they read as follows.
The connection formula relating the expansions about $z=0$ and $z=t$ is
\begin{align}
 & \FV{\alpha_{1}}{\alpha_{\infty}}{\alpha}{\alpha_{t}}{\alpha_{0\theta}}{\alpha_{2,1}}{\alpha_{0}}t{\frac{z}{t}}=\sum_{\theta'=\pm}\mathcal{M}_{\theta\theta'}\left(b\alpha_{0},b\alpha_{t};b\alpha\right)\nonumber \\
 & \quad\times \mathrm{e}^{-i\pi\left(\Delta-\Delta_{0}-\Delta_{2,1}-\Delta_{t}\right)}\left(1-t\right)^{\Delta_{\infty}-\Delta_{1}-\Delta_{t}-\Delta_{2,1}-\Delta_{0}}\FV{\alpha_{1}}{\alpha_{\infty}}{\alpha}{\alpha_{0}}{\alpha_{t\theta'}}{\alpha_{2,1}}{\alpha_{t}}{\frac{t}{t-1}}{\frac{t-z}{t}}.\label{eq:CB0t}
\end{align}
The connection formula relating the expansions about $z=0$ and $z=\infty$ is\footnote{
Ref.~\cite{Bonelli:2022ten} writes the corresponding connection formula without the branch-dependent phase shown here. With the branch convention adopted in this work, this phase is required. We have verified it directly from the underlying hypergeometric connection problem and independently by a numerical comparison with the local Heun solutions.
The shift \(b\alpha_{\theta'}\) and the additional phase \(\mathrm{e}^{-i\pi b^2/2}\) do not affect the semiclassical limit.
}
\begin{align}
\FV{\alpha_{1}}{\alpha_{\infty}}{\alpha}{\alpha_{t}}{\alpha_{0\theta}}{\alpha_{2,1}}{\alpha_{0}}t{\frac{z}{t}} & =\sum_{\theta',\theta''=\pm}\mathrm{e}^{i\pi\left(\theta a_{0}+\theta''a_{\infty}-\frac{b^2}{2}\right)}\mathcal{M}_{\theta\theta'}\left(b\alpha_{0},b\alpha;b\alpha_{t}\right)\mathcal{M}_{\left(-\theta'\right)\theta''}\left(b\alpha_{\theta'},b\alpha_{\infty};b\alpha_{1}\right)\nonumber \\
 & \quad\times t^{\Delta_{\infty}+\Delta_{1}+\Delta_{2,1}-\Delta_{0}-\Delta_{t}}z^{-2\Delta_{2,1}}\FV{\alpha_{t}}{\alpha_{0}}{\alpha_{\theta'}}{\alpha_{1}}{\alpha_{\infty\theta''}}{\alpha_{2,1}}{\alpha_{\infty}}t{\frac{1}{z}}.\label{eq:CB0i}
\end{align}
Here, $\alpha_{\theta}$ is a shifted momentum $\alpha_{\theta}=\alpha-\theta\frac{b}{2}$.
The connection formula relating the expansions about $z=1$ and $z=\infty$ is
\begin{align}
 & t^{\Delta_{\infty}+\Delta_{1}+\Delta_{2,1}-\Delta_{t}-\Delta_{0}}\left(1-t\right)^{\Delta_{\infty}+\Delta_{0}+\Delta_{2,1}-\Delta_{t}-\Delta_{1}}\left(z-t\right)^{-2\Delta_{2,1}}\FV{\alpha_{0}}{\alpha_{t}}{\alpha}{\alpha_{\infty}}{\alpha_{1\theta}}{\alpha_{2,1}}{\alpha_{1}}t{\frac{z-1}{z-t}}\nonumber \\
 & =\sum_{\theta'=\pm}\mathcal{M}_{\theta\theta'}\left(b\alpha_{1},b\alpha_{\infty};b\alpha\right)t^{\Delta_{\infty}+\Delta_{1}+\Delta_{2,1}-\Delta_{0}-\Delta_{t}}z^{-2\Delta_{2,1}}\FV{\alpha_{t}}{\alpha_{0}}{\alpha}{\alpha_{1}}{\alpha_{\infty\theta'}}{\alpha_{2,1}}{\alpha_{\infty}}t{\frac{1}{z}}.\label{eq:CB1i}
\end{align}
Here, $\mathcal{M}_{\theta\theta'}$ is the connection coefficient defined in Eq.~\eqref{eq:ConMat}.

\subsubsection{Connection formulae for semiclassical conformal blocks\label{subsec:CF-SCB}}

All connection formulae for the semiclassical conformal blocks presented in Sec.~\ref{subsec:ConForm} follow from the three basic relations \eqref{eq:CB0t}, \eqref{eq:CB0i}, and \eqref{eq:CB1i}.

First, the relations $0\rightarrow t$, $0\rightarrow\infty$, and $1\rightarrow\infty$ given in Eqs.~\eqref{eq:SCB0tM}, \eqref{eq:SCB0iM}, and \eqref{eq:SCB1iM}, respectively, are obtained directly from Eqs.~\eqref{eq:CB0t}, \eqref{eq:CB0i}, and \eqref{eq:CB1i}, together with the M\"obius relations \eqref{eq:CB4-Mobius} and the definitions \eqref{eq:SCB-Def}.\footnote{
To derive the $0\rightarrow\infty$ relation in Eq.~\eqref{eq:SCB0iM}, we also need to consider the shift of the intermediate momentum $\alpha\rightarrow \alpha_{\theta'}$ in Eq.~\eqref{eq:CB0i}.
In the semiclassical limit, this effect can be computed as
\begin{align}
\frac{\FV{\alpha_{t}}{\alpha_{0}}{\alpha_{\theta'}}{\alpha_{1}}{\alpha_{\infty\theta''}}{\alpha_{2,1}}{\alpha_{\infty}}t{\frac{1}{z}}}{\FV{\alpha_{t}}{\alpha_{0}}{\alpha}{\alpha_{1}}{\alpha_{\infty\theta''}}{\alpha_{2,1}}{\alpha_{\infty}}t{\frac{1}{z}}}  \sim t^{\Delta_{\theta'}-\Delta}\left.\exp\left[\frac{1}{b^{2}}\left(F\left(a_{i},a_{\theta'};t\right)-F\left(a_{i},a;t\right)\right)\right]\right|_{a_{\infty}\rightarrow a_{\infty\theta''}}
  \rightarrow t^{a\theta'}\mathrm{e}^{-\frac{1}{2}\theta'\partial_{a}F\left(a_{i},a;t\right)}.
\end{align}
}

Second, the reverse relations follow immediately using the matrix notation introduced in Sec.~\ref{subsec:ConForm}. Indeed, the identities $\Theta\left(c\right)^{-1}=\Theta\left(-c\right)$ and
\begin{equation}
\mathcal{M}\left(c_{1},c_{2};c_{3}\right)^{-1}=\mathcal{M}\left(c_{2},c_{1};c_{3}\right),\label{eq:CM-Inv}
\end{equation}
give the three inverse relations $t\rightarrow0$, $\infty\rightarrow0$, and $\infty\rightarrow1$ given in Eqs.~\eqref{eq:SCB0tM}, \eqref{eq:SCB0iM}, and \eqref{eq:SCB1iM}, respectively. 
Equation \eqref{eq:CM-Inv} follows from the explicit matrix elements in Eq.~\eqref{eq:ConMat}, together with the gamma-function recurrence relations and \(\det\mathcal M(c_1,c_2;c_3)=-c_1/c_2\).

The remaining connection formulae are obtained by composing these six elementary relations. A useful matrix identity is
\begin{equation}
i\Theta\left(c_{3}\right)\mathcal{M}\left(c_{3},c_{1};c_{2}\right)\Theta\left(c_{1}\right)\mathcal{M}\left(c_{1},c_{2};c_{3}\right)\Theta\left(c_{2}\right)=\mathcal{M}\left(c_{3},c_{2};c_{1}\right).\label{eq:CM-Comp}
\end{equation}
As an illustration, the $t\rightarrow1$ connection formula is obtained by composing the relations \eqref{eq:SCB0tM}, \eqref{eq:SCB0iM}, and \eqref{eq:SCB1iM}. In matrix form, this gives
\begin{align}
\left(\begin{array}{c}
\mathcal{F}_{t_{-}}\\
\mathcal{F}_{t_{+}}
\end{array}\right) & =-\mathcal{M}\left(a_{t},a_{0};a\right)\Theta\left(a_{0}\right)\mathcal{M}\left(a_{0},a;a_{t}\right)\nonumber \\
 & \quad\times\left(\begin{array}{cc}
0 & t^{-a}\mathrm{e}^{\frac{1}{2}\partial_{a}F}\\
t^{a}\mathrm{e}^{-\frac{1}{2}\partial_{a}F} & 0
\end{array}\right)\mathcal{M}\left(a,a_{\infty};a_{1}\right)\Theta\left(a_{\infty}\right)\mathcal{M}\left(a_{\infty},a_{1};a\right)\left(\begin{array}{c}
\mathcal{F}_{1_{-}}\\
\mathcal{F}_{1_{+}}
\end{array}\right).
\end{align}
Applying the identity \eqref{eq:CM-Comp}, together with
\begin{equation}
\Theta\left(-a\right)\left(\begin{array}{cc}
0 & t^{-a}\mathrm{e}^{\frac{1}{2}\partial_{a}F}\\
t^{a}\mathrm{e}^{-\frac{1}{2}\partial_{a}F} & 0
\end{array}\right)\Theta\left(-a\right)=\left(\begin{array}{cc}
0 & t^{-a}\mathrm{e}^{\frac{1}{2}\partial_{a}F}\\
t^{a}\mathrm{e}^{-\frac{1}{2}\partial_{a}F} & 0
\end{array}\right),
\end{equation}
reduces the result to Eq.~\eqref{eq:SCBt1M}.

\section{Parameters for nondegenerate horizons}
Let us clarify the conditions under which the function $\Delta_{r}(r)$ defined by Eq.~\eqref{def Delta}, namely,
\begin{align}
        \Delta_{r}(M,\aBH,\QBH,L;r) =
       - \frac{1}{L^2} r^4 + \left(1 - \frac{\aBH^2}{L^2}\right) r^2 - 2 M r + \aBH^2 + \QBH^2
\end{align}
has four real roots, assuming $M > 0$, $\aBH^2 + \QBH^2 > 0$, and $L > 0$.

First, it is necessary that $\Delta_{r}(r)$ has three real extrema, that is, three zero points of
\begin{align}
\Delta_{r}'(r) = - \frac{4}{L^2} \left( r^3 - \frac{L^2}{2} \left( 1 - \frac{\aBH^2}{L^2} \right)r + \frac{L^2 M}{2} \right).
\end{align}
This corresponds to the condition 
\begin{align}
   \frac{M^2}{L^2} < \frac{2}{27} \left(1 - \frac{\aBH^2}{L^2}\right)^3, \label{condition 1}
\end{align}
requiring $|\aBH| < L$.
Under this condition, let us denote the three real roots of $\Delta_{r}'(r)$ by $\rho_{1} < \rho_{2} < \rho_{3}$. Given $M > 0$, one can see that one root is negative and the other two are positive; thus,
\begin{align}
\rho_{1} < 0 < \rho_{2} < \rho_{3}.
\end{align}
The roots $\rho_{1}, \rho_{2}, \rho_{3}$ can be regarded as functions of the parameters $M$, $\aBH$, and $L$.

Taking into account that $\Delta_{r}(0) = \aBH^2 + \QBH^2 > 0$, the conditions under which $\Delta_{r}(r)$ has four real roots are expressed as
\begin{align}
&\Delta_{r}(M,\aBH,\QBH,L; \rho_{2}(M,\aBH,L)) < 0, \label{Delta rho2 < 0}\\
&\Delta_{r}(M,\aBH,\QBH,L; \rho_{3}(M,\aBH,L)) > 0. \label{Delta rho3 > 0}
\end{align}
In this case, one of the roots is negative, while the other three roots are positive.
Since $\Delta_{r}'(\rho_{2/3}) = 0$, we obtain
\begin{align}
\frac{d}{d M} \Delta_{r}(M,\aBH,\QBH,L; \rho_{2/3}(M,\aBH,L)) &= \frac{\partial}{\partial M} \Delta_{r}(M,\aBH,\QBH,L; \rho_{2/3}(M,\aBH,L)) \notag\\
&= - 2 \rho_{2/3}(M,\aBH,L) < 0,
\end{align}
and therefore the conditions \eqref{Delta rho2 < 0} and \eqref{Delta rho3 > 0} represent the lower and upper bounds on the mass parameter $M$.

Let us derive the lower and upper bounds on the mass parameter, corresponding to the extremal and Nariai limits and denoted by $M_{\mathrm{E}}(\aBH,\QBH,L)$ and $M_{\mathrm{N}}(\aBH,\QBH,L)$, respectively.
For these values of the mass parameter, $r = \rho_{2/3}$ simultaneously satisfies $\Delta_{r}(r) = 0$ and $\Delta_{r}'(r) = 0$ and, in particular, satisfies
\begin{align}
\Delta_{r}(r) - r \Delta_{r}'(r) = \frac{3}{L^2}r^4 - \left(1 - \frac{\aBH^2}{L^2}\right) r^2 + \aBH^2 + \QBH^2 = 0, \label{EN radius equation}
\end{align}
which can be solved as
\begin{align}
r^2 = \frac{L^2}{6} \left(1 - \frac{\aBH^2}{L^2} \pm \sqrt{\left(1 - \frac{\aBH^2}{L^2} \right)^2 - 12 \frac{\aBH^2 + \QBH^2}{L^2}} \right).
\end{align}
The right-hand side yields two distinct positive roots only when
\begin{align}
\left(1 - \frac{\aBH^2}{L^2} \right)^2 - 12 \frac{\aBH^2}{L^2} > 12 \frac{\QBH^2}{L^2}. \label{upper bound for Q}
\end{align}
Combining this condition with Eq.~\eqref{condition 1}, we obtain the necessary condition
\begin{align}
 \frac{|\aBH|}{L} < \sqrt{7 - 4 \sqrt{3}} \approx 0.268. \label{condition for a}
\end{align}
The two positive roots of Eq.~\eqref{EN radius equation} correspond to the radii of the degenerate horizons in the extremal and Nariai limits and are given by
\begin{align}
r_{\mathrm{E}}(\aBH,\QBH,L) &=  L \sqrt{\frac{1 - \frac{\aBH^2}{L^2} - \sqrt{\left( 1 - \frac{\aBH^2}{L^2}  \right)^2 - 12 \frac{\aBH^2 + \QBH^2}{L^2}}}{6}}, \label{rE}\\
r_{\mathrm{N}}(\aBH,\QBH,L) &=   L \sqrt{\frac{1 - \frac{\aBH^2}{L^2} + \sqrt{\left( 1 - \frac{\aBH^2}{L^2}  \right)^2 - 12 \frac{\aBH^2 + \QBH^2}{L^2}}}{6}}. \label{rN}
\end{align}
Note that, by definition, $r_{\mathrm{E}}$ and $r_{\mathrm{N}}$ satisfy
\begin{align}
r_{\mathrm{E}}^2 < \frac{1}{6}\left(L^2 - \aBH^2 \right),
\qquad r_{\mathrm{N}}^2 > \frac{1}{6}\left(L^2 - \aBH^2 \right). \label{ineq rE and rN}
\end{align}
Substituting these expressions into $\Delta_{r}'(r) = 0$, we obtain the mass parameters in the extremal and Nariai limits,
\begin{align}
M_{\mathrm{E}}(\aBH,\QBH,L) &=
\frac{1}{3} \left( 2 \left(1 - \frac{\aBH^2}{L^2} \right) + \sqrt{\left( 1 - \frac{\aBH^2}{L^2} \right)^2 - 12 \frac{\aBH^2 + \QBH^2}{L^2}} \right) r_{\mathrm{E}}(\aBH,\QBH,L)
, \label{ME}\\
M_{\mathrm{N}}(\aBH,\QBH,L) &=
\frac{1}{3} \left( 2\left(1 - \frac{\aBH^2}{L^2} \right) - \sqrt{\left( 1 - \frac{\aBH^2}{L^2} \right)^2 - 12 \frac{\aBH^2 + \QBH^2}{L^2}} \right) r_{\mathrm{N}}(\aBH,\QBH,L). \label{MN}
\end{align}
One can show that the extremal case corresponds to equality at $\rho_{2}$, while the Nariai case does so at $\rho_{3}$. One can also show that
\begin{align}
M_{\mathrm{N}}^2 - M_{\mathrm{E}}^2 = \frac{L^2}{27} \left( \left( 1 - \frac{\aBH^2}{L^2} \right)^2 - 12 \frac{\aBH^2 + \QBH^2}{L^2} \right)^{3/2} > 0.
\end{align}
Therefore, the conditions \eqref{Delta rho2 < 0} and \eqref{Delta rho3 > 0} are satisfied when
\begin{align}
M_{\mathrm{E}}(\aBH,\QBH,L) < M < M_{\mathrm{N}}(\aBH, \QBH,L). \label{Mass range}
\end{align}
Under the condition $|\aBH| < L$, one can also show that
\begin{align}
\frac{M_{\mathrm{N}}^2}{L^2} < \frac{2}{27} \left(1 - \frac{\aBH^2}{L^2} \right)^3.
\end{align}
Thus, Eq.~\eqref{condition 1} follows from the other inequalities.

The conditions under which the Kerr--Newman--de Sitter spacetime has four distinct horizons can be summarized as follows.
For a given de Sitter radius parameter $L$, the spin parameter $|\aBH|$ is bounded above as in Eq.~\eqref{condition for a}, which in particular requires $|\aBH| < L$.
Next, $|\QBH|$ is bounded above by Eq.~\eqref{upper bound for Q} in terms of $\aBH$ and $L$, and $M$ must lie between the extremal and Nariai masses as specified by Eq.~\eqref{Mass range}.

\section{Geometrized Units}
\label{app:units}
Here, we specify the units used in this paper. The quantities $M$, $J$, $\QBH$, and the gauge field $A_{\mu}dx^{\mu}$ are expressed in geometrized units. Their dimensions are $\text{[length]}^1$, $\text{[length]}^{2}$, $\text{[length]}^1$, and $\text{[length]}^{1}$, respectively.
Restoring Newton's constant $G$ and Coulomb's constant $k$, while keeping $c = 1$, the geometrized quantities are related to the corresponding physical quantities by
\begin{align}
M = G M^{\text{phys}},\quad J = G J^{\text{phys}}, \quad 
\QBH = \sqrt{k G} \QBH^{\text{phys}}, \quad A_{\mu} = \sqrt{\frac{G}{k}}A^{\text{phys}}_{\mu}.
\end{align}

We define the field charge $\qBH$ through the gauge-covariant derivative $D_{\mu} = \partial_{\mu} - i \qBH A_{\mu}$.
Since the gauge field $A_{\mu} dx^{\mu}$ in our units has the dimension of length, the charge $\qBH$ must have the dimension $\text{[length]}^{-1}$. Comparing this definition with the physical units expression $D_{\mu} = \partial_{\mu} - i (\qBH^{\text{phys}}/\hbar) A^{\text{phys}}_{\mu}$, we obtain
\begin{align}
\qBH = \frac{1}{\hbar} \sqrt{\frac{k}{G}}\qBH^{\text{phys}}.
\end{align}
Note that the dimensions of $\qBH$ and $\QBH$ involve opposite powers of length.

\bibliography{ref}
\bibliographystyle{JHEP.bst}

\end{document}